\documentclass[twocolumn]{aa}

\usepackage[T1]{fontenc}
\usepackage[utf8]{inputenc}

\usepackage{graphicx}
\usepackage{natbib}
\usepackage{ulem}
\usepackage{textcomp}
\usepackage{gensymb}
\usepackage{longtable}
\usepackage{threeparttable}
\usepackage{multicol}
\usepackage{multirow}
\usepackage{float}
\usepackage{todonotes}
\usepackage[Symbol]{upgreek}
\usepackage{amsmath}
\usepackage{etoolbox}
\usepackage{txfonts}
\usepackage{url}
\usepackage{xcolor}

\usepackage[most]{tcolorbox}
\usepackage{hyperref}
\usepackage{xcolor}
\hypersetup{
  colorlinks   = true, 
  urlcolor     = blue, 
  linkcolor    = blue, 
  citecolor   = blue 
}
\usepackage[all]{hypcap} 

\begin{document} 
	\title{Dissecting nonthermal emission in the complex multiple-merger galaxy cluster Abell\,2744: Radio and X-ray analysis }
	\titlerunning{Diffuse radio emission in Abell\,2744: Radio and X-ray analysis}
	\authorrunning{Rajpurohit et al.}

\author{K. Rajpurohit\inst{1,2,3}, F. Vazza\inst{1,2,4},  R. J. van Weeren\inst{5}, M. Hoeft\inst{3}, M. Brienza\inst{1,2},  E. Bonnassieux\inst{1,2}, C. J. Riseley\inst{1,2,6}, G. Brunetti\inst{2}, A. Bonafede\inst{1,2,4},  M. Br\"uggen\inst{4}, W. R. Formann\inst{7}, A. S. Rajpurohit\inst{8}, H. J. A. R\"ottgering\inst{5}, A. Drabent\inst{3}, P. Dom\'{i}nguez-Fern\'{a}ndez\inst{9,4}, D. Wittor\inst{4}, and F. Andrade-Santos\inst{7}}

\institute{Dipartimento di Fisica e Astronomia, Universit\'at di Bologna, Via P. Gobetti 93/2, 40129, Bologna, Italy\\
 {\email{kamlesh.rajpurohit@unibo.it}}
\and
INAF-Istituto di Radio Astronomia, Via Gobetti 101, 40129 Bologna, Italy
\and
Th\"uringer Landessternwarte, Sternwarte 5, 07778 Tautenburg, Germany
\and
Hamburger Sternwarte, Universit\"at Hamburg, Gojenbergsweg 112, 21029, Hamburg, Germany
\and
Leiden Observatory, Leiden University, P.O. Box 9513, 2300 RA Leiden, The Netherlands
\and
CSIRO Astronomy and Space Science, PO Box 1130, Bentley, WA 6102, Australia
\and
Harvard-Smithsonian Center for Astrophysics, 60 Garden Street, Cambridge, MA 02138, USA
\and
Astronomy \& Astrophysics Division, Physical Research Laboratory, Ahmedabad 380009, India
\and
Department of Physics, School of Natural Sciences UNIST, Ulsan 44919, Korea
}
   
     \abstract
{We present the first deep low frequency radio observations of the massive and highly disturbed galaxy cluster Abell 2744 using the upgraded Giant Metrewave Radio Telescope (uGMRT). The cluster is experiencing a very complex multiple merger and hosts a giant halo and four radio relics. The uGMRT observations, together with existing VLA (1-4\,GHz) and \textit {Chandra} observations, allow us to study the complexity of the physical mechanisms active in this system. Our new images reveal that the central halo emission is more extended toward low frequencies. We find that the integrated spectrum of the halo follows a power-law between 150\,MHz and 3\,GHz, while its subregions show significantly different spectra, also featuring high frequency spectral steepening. The halo also shows local regions in which the spectral index is significantly different from the average value. Our results highlight that an overall power-law spectrum, as observed in many radio halos, may also arise from the superposition of different subcomponents. The comparison of the radio surface brightness and spectral index with the X-ray brightness and temperature reveals for the first time different trends, indicating that the halo consists of two main components with distinct evolutionary signatures. All four relics in this system follow a power-law radio spectrum, compatible with shocks with Mach numbers in the range $3.0-4.5$. All relics are also highly polarized from 1-4\,GHz and show low Faraday dispersion measures, suggesting that they are located in the outermost regions of the cluster. The complexity in the distribution and properties of nonthermal components in Abell 2744 supports a multiple merger scenario, as also highlighted by previous X-ray and lensing studies. Our unique results demonstrate the importance of sensitive and high-resolution, multi-frequency radio observations for understanding the interplay between the thermal and non-thermal components of the ICM.}

   \keywords{Galaxies: clusters: individual (Abell\,2744) $-$ Galaxies: clusters: intracluster medium $-$ large-scale structures of universe $-$ Acceleration of particles $-$ Radiation mechanism: non-thermal: magnetic fields}

   \maketitle

%########################################################################
%  Section 1: Introduction
%######################################################################## 
\section{Introduction}
\label{sec:intro}
Galaxy clusters acquire their mass through accretion of smaller galaxy groups or mergers with other clusters. This results in turbulent plasma motions across a broad range of scales and shock waves in the hot intracluster medium \citep[ICM, e.g.,][]{Sarazin2002}. As a consequence, the dissipation of a significant fraction of the energy released during such mergers, accelerates cosmic ray particles leading to the formation of radio halos and radio relics \citep[for review, see][]{Brunetti2014,vanWeeren2019}. The radio spectra of these sources are typically steep\footnote{We define the spectral index, $\alpha$, so that $S_{\nu}\propto\nu^{\alpha}$, where $S$ is the flux density at frequency $\nu$.} ($\alpha\leq-1$).

Radio halos are megaparsec scale diffuse sources found at the centers of clusters and are usually unpolarized. The radio morphology of halos typically follows the X-ray morphology \citep{Pearce2017,Rajpurohit2018}, suggesting a direct connection between the thermal and nonthermal components of the ICM. 

The currently favored scenario for the formation of radio halos involves the re-acceleration of cosmic-ray electrons (CRe) to higher energies via turbulence induced during mergers \citep[reacceleration models;][]{Brunetti2001,Petrosian2001}. The reacceleration models predict a connection with cluster dynamics, a complex radio-halo morphology and spectral index distribution, and a high-frequency break in the halo spectrum.  

%########################################################################
%  Table 1: observational details 
%########################################################################

\setlength{\tabcolsep}{16pt}
\begin{table*}[!htbp]
\caption{Observational overview: uGMRT and VLA observations}
\centering
\begin{threeparttable} 
\begin{tabular}{ l  c  c c c c}
  \hline  \hline  
& uGMRT Band\,4  &  uGMRT Band\,3  & VLA L-band $^{\dagger}$  & VLA S-band $^{\dagger}$  \\  
  \hline  
Frequency range&550-950\,MHz&300-500\,MHz &1-2\,GHz &2-4\,GHz\\ 
Channel width & 49\,kHz & 97\,kHz &1\,MHz &1\,MHz\\ 
No of spectral windows &1 &1 &16 &16\\ 
No of channels &4096 &2048 &64 &64\\ 
On source time &8+8+8\,hrs &5\,hrs &12.3\,hrs &16.3\,hrs  \\
LAS$^{_\ddagger}$ &1020\arcsec&1920\arcsec &970 \arcsec&490\arcsec\\ 
\hline 
\end{tabular}
\begin{tablenotes}[flushleft]
\footnotesize
\item{\textbf{Notes.}} Full Stokes polarization information was recorded for the VLA L-band, S-band and uGMRT Band\,4 data;$^{\dagger}$ For VLA data reduction steps, we refer to \cite{Pearce2017}; $^{_\ddagger}$Largest angular scale that can be recovered by the mentioned observations. 
\end{tablenotes}
\end{threeparttable} 
\label{Tabel:obs}
\end{table*} 

An alternative mechanism proposes that the CRe are the secondary products of hadronic collisions between thermal ions and relativistic protons present in the ICM \citep[secondary models;][]{Dennison1980,Blasi1999,Dolag2000}. However, the current gamma ray limits imply that the energy budget of cosmic rays is too small to explain radio halos \citep{Brunetti2017,Pinzke2017}. Although, secondary models cannot play the main role, re-acceleration of secondary electrons is still viable \citep{Adam2021}.

Radio halos have been observed with a broad range of spectral indices, mainly between $-2$ to $-1$ \citep{Feretti2012,vanWeeren2019}. Although, the spectral behavior of radio halos in realistic (non homogeneous) conditions may result from a complex superposition of components, in general a spectral steepening is expected at higher frequencies due to a maximum energy of the emitting electrons that can be sustained by second order mechanisms \citep{Cassano2006}. The integrated radio spectra of most of radio halos follow a single power-law spectrum. To our best knowledge , there are only three radio halos that show high-frequency spectral steepening, namely halos in the Coma cluster \citep{Thierbach2003}, MACS\,J0717.5+3745 \citep{Rajpurohit2021b}, and Abell S1063  \citep{Xie2020}. The radio spectral index is key to understand the shape of the relativistic electron distribution, the properties of turbulence in the ICM, the magnetic field distribution, and the link between thermal and nonthermal plasma. 
 
Only a handful of galaxy clusters have been the subject of good-quality, high resolution spatially resolved spectral analysis. The spectral index maps, reported over large frequency coverage, so far point to a complex spectral index distribution. There are halos where the spectral index distribution remains roughly uniform, for example, 1RXS\,J0603.3+4214 \citep{vanWeeren2016a,Rajpurohit2020a,diGasperin2020}. But there are also halos which show significant spectral index fluctuations, for example, Abell 2255 \citep{Botteon2020} and MACS\,J0717.5+3745 \citep{Rajpurohit2021b}. Understanding difference in the individual systems is important to constrain the physics of these sources.     

There exists a point-to-point correlation between the radio and X-ray surface brightness for radio halos \citep{Govoni2001a,Botteon2020,Rajpurohit2021b,Luca2021}. The slope of this correlation has been reported to be mostly sub-linear or linear. The slope provides constraints on the distribution and transport of particles and magnetic field in the cluster \citep[e.g.,][]{Govoni2001a,Dolag2000,Pfrommer2008,Brunetti2014}.

Unlike halos, radio relics are usually located in the peripheral regions of clusters. They are believed to be associated with merger-induced shock fronts \citep[e.g.,][]{vanWeeren2019}. One striking observational feature of radio relics is their high degree of polarization (locally as high as 65\%) and aligned magnetic field vectors distribution  \citep{vanWeeren2010,Bonafede2012,Owen2014,DeGasperin2014,Kierdorf2016,DiGennaro2021}. 

It is widely accepted that the kinetic energy dissipated by shock powers the radio emission via diffusive shock acceleration \citep[DSA;][]{Ensslin1998,Hoeft2007,Brunetti2014} of cosmic-ray electrons (CRe). However, there is an ongoing debate about whether the acceleration starts from the thermal pool \citep[standard scenario;][]{Ensslin1998,Hoeft2007} or from a population of mildly relativistic electrons \citep[re-acceleration scenario;][]{Markevitch2005,Kang2011,Kang2016a}. According to the re-acceleration scenario, the shock fronts re-accelerate electrons from a pre-existing fossil population. There are a few examples, which seem to show a connection between the relic and active galactic nuclei. \citep{Bonafede2014,Shimwell2015,vanWeeren2017a,Gennaro2018,Stuardi2019}. 

The standard scenario has successfully explained many of the observed properties of relics, but major difficulties remain:  the acceleration of electrons from the thermal pool requires an unphysically large acceleration efficiency in the case of weak shocks to explain the high radio power observed in relics \citep{Botteon2020a} and the Mach numbers derived from X-ray observations are often significantly lower than those derived from the radio observations \citep{Akamatsu2012,Botteon2016b,Botteon2018}. Possible solution to reconcile -at least- the discrepancy between the radio and X-ray Mach numbers have been discussed. Cosmological simulations show that radio relics indeed trace a distribution of Mach numbers \citep{Skillman2013,Roh2019,Wittor2019}. Recently,  \cite{Paola2021} found that the Mach number depends on the initial strength of the shock front and the fluctuations in the ICM, suggesting that the difference in the X-ray and radio derived Mach numbers could arise when shock waves propagate through a turbulent ICM. They also showed that the radio emissivity is biased toward a high Mach number while the X-ray emissivity to a low Mach number.   

The new generation of radio telescopes such as LOw-Frequency ARray, Karl G. Jansky Very Large Array (VLA), and  upgraded Giant Metrewave Radio Telescope (uGMRT) are revealing new, unprecedented insights about relics and halos \citep[e.g.,][]{Owen2014,vanWeeren2016a,vanWeeren2017a,Pearce2017,Gennaro2018,Botteon2020,Rajpurohit2021a,Rajpurohit2021b,Bonafede2021,DiGennaro2021}.  Wideband multi-frequency radio observations from these telescopes have the potential to constrain the complexity of the mechanisms active in these sources and are essential to improve our understanding of particle acceleration in radio halos and relics.

In this paper, we present deep uGMRT (300-850\,MHz) observations of diffuse emission sources associated with the merging galaxy cluster Abell 2744. We complement our spectral and polarization analysis with VLA (1-4\,GHz) observations. To understand the relation between the thermal and nonthermal components of the ICM, we also use \textit{Chandra} data. The VLA and \textit{Chandra} observations were originally published by \citet{Pearce2017}. We use the 125\,ks X-ray \textit {Chandra} observation (ObsID: 7712, 2212, 7915, 8477, 8557) presented in \cite{Pearce2017}. For description of the \textit {Chandra} data reduction steps, the reader is referred to \cite{Pearce2017}. Our new uGMRT observations in combination with the published high frequency radio and X-ray data allows us to study the diffuse radio sources in this system in more detail than had been done previously.

The outline of this paper is as follows. In Sect.\,\ref{A2744intro}, we provide a brief overview of Abell 2744. The observations and data reduction processes are explained in Sect.\,\ref{obs}. In Sect.\,\ref{results} we present new radio continuum radio images of diffuse radio sources constructed at various resolutions. The results obtained from the spectral, polarization, and radio versus X-ray analysis are described in Sects.\,\ref{spectral_anaylsis} to \ref{polarization_analysis}. We summarized our findings in Sect.\ref{summary}.    

We adopt a flat $\Lambda$CDM cosmology with $H_{\rm{ 0}}=70$ km s$^{-1}$\,Mpc$^{-1}$, $\Omega_{\rm{ m}}=0.3$, and $\Omega_{\Lambda}=0.7$. At the cluster's redshift, $1\arcsec$ corresponds to a physical scale of 4.5\,kpc. All output images are in the J2000 coordinate system and are corrected for primary beam attenuation.

%########################################################################
% Section 2: known facts about Abell 2744 
%########################################################################
\section{Abell 2744 }
\label{A2744intro}
The galaxy cluster Abell\,2744 is located at a redshift $z=0.308$ \citep{Struble1999}. It has been extensively studied at optical, X-ray,  and radio wavelengths \citep{Markevitch2001,Kempner2004,Merten2011,Owers2011,Medezinski2016,Jauzac2018}. The cluster is extremely luminous in X-ray with $L_{\rm X,\,0.1-2.4\,keV}=13.6\times10^{44}\rm\, erg\,s^{-1}$ \citep{Ebeling2010} and has a global X-ray temperature of $8.53\pm0.37$\,keV \citep{Mantz2010}. 

\begin{figure}
    \centering
    \includegraphics[width=0.48\textwidth]{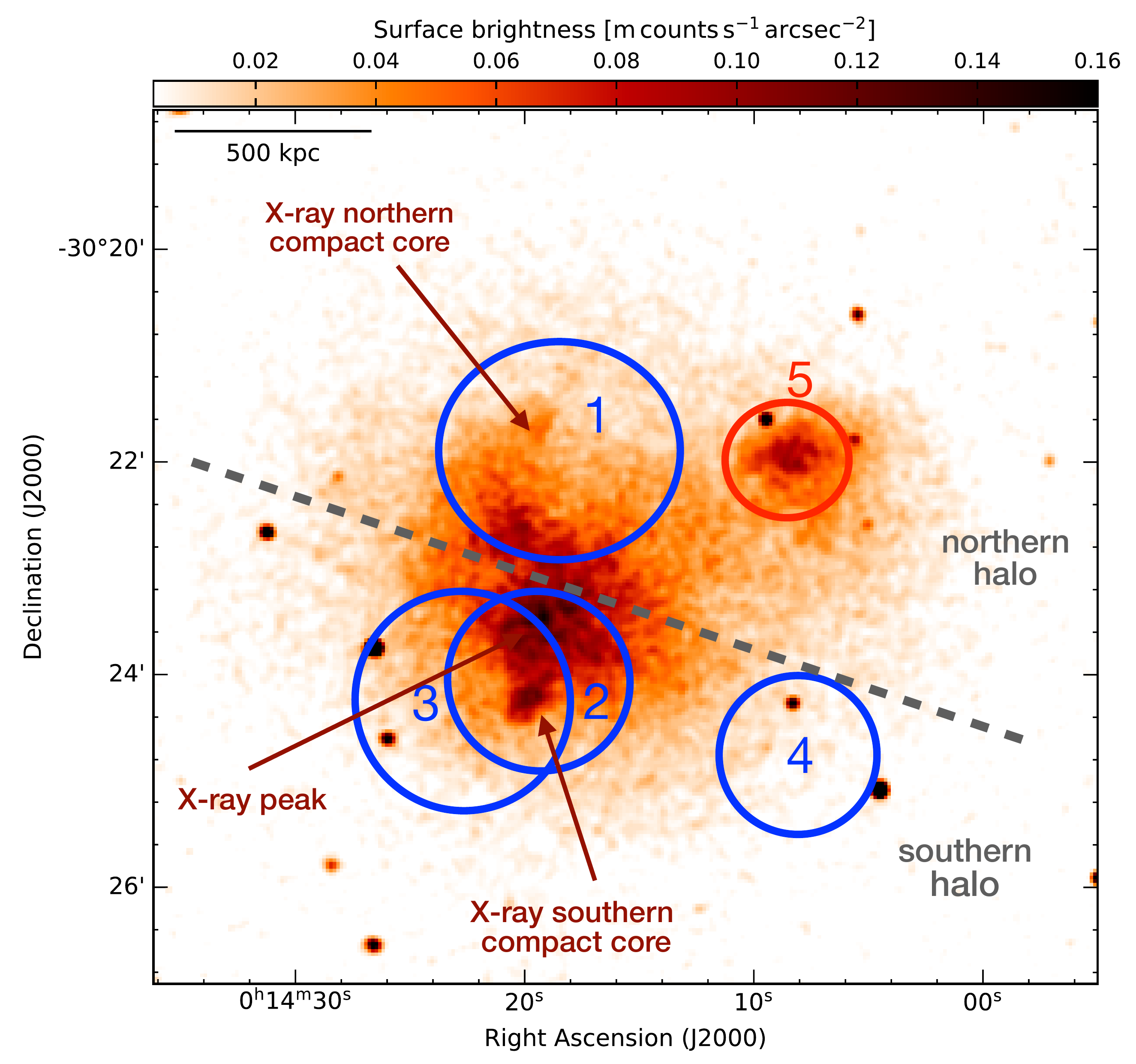}
    \vspace{-0.2cm}
 \caption{ \textit{Chandra} $0.5-2.0\rm\,keV$ band image, smoothed with a Gaussian full width at half maximum (FWHM) of $3\arcsec$. The cluster is rich in X-ray with several substructures near the center.  Blue circles marks four subclusters detected in weak/strong lensing analysis \citep{Golovich2019}. A fifth subcluster detected only in X-ray observations is shown with red circle. The dashed line represents the division between the northern (subclusters 1 and 5) and the southern (subclusters 2, 3, and 4) parts of the halo.}
      \label{subcomponents}
  \end{figure}

Optical and X-ray studies show that the cluster is in a highly dynamical disturbed state \citep{Owers2011,Merten2011}. In X-ray, the cluster shows several substructures near the center \citep{Kempner2004}. The X-ray emission concentrated on the southern compact core and extending to the northwest, see Fig.\,\ref{subcomponents}. \textit{Chandra} and \textit{XMM-Newton} observations also show evidence of density and temperature discontinuities, indicating the presence of shocks \citep{Eckert2016,Hattori2017,Pearce2017}. 

%%%%%%%%%%%%%%%%%%%%%%%%%%%%%%%%%%%%%%%%%%%%%%%%%%%%%%%%%%%%%%%%%
%  Fig 1: cluster subcomponents 

%%%%%%%%%%%%%%%%%%%%%%%%%%%%%%%%%%%%%%%%%%%%%%%%%%%%%%%%%%%%%%%%%

Gravitational lensing analysis suggests that the cluster consists of at least four individual subcomponents \citep{Golovich2019}, see Fig.\,\ref{subcomponents}. A joint analysis of strong- and weak-lensing by \cite{Jauzac2018} revealed eight substructures within the central region of the cluster. Moreover, X-ray observations shown an additional subcomponent to the northwest; shown with a red circle in Fig.\,\ref{subcomponents}.

\begin{figure*}
    \centering
    \includegraphics[width=1.0\textwidth]{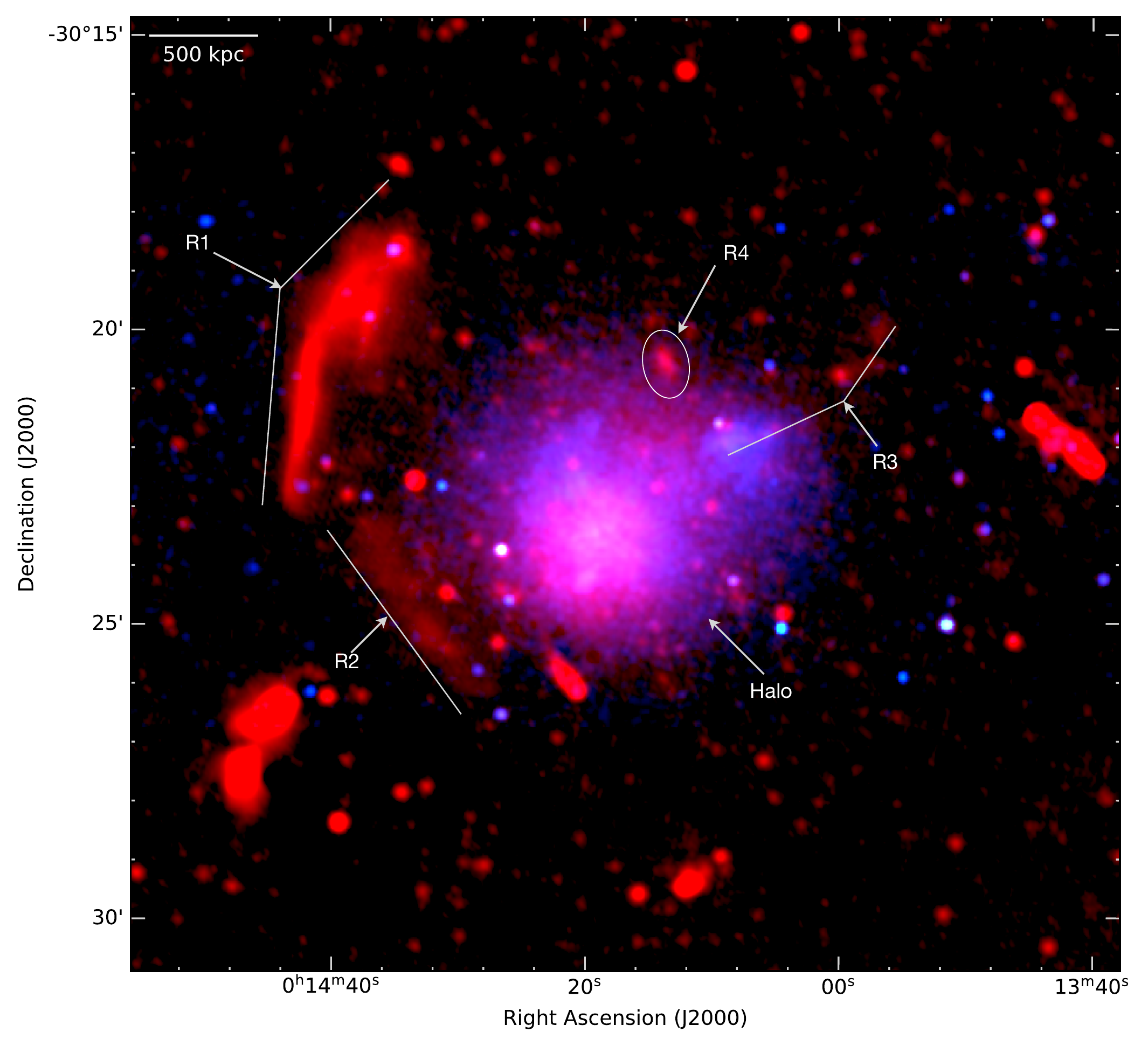}
 \caption{Xray-uGMRT (550-850 MHz) overlay Abell\,2744. The intensity in red shows the radio emission observed with uGMRT Band\,4 at a central frequency of 675\,MHz. The intensity in blue shows \textit{Chandra} X-ray emission in the $0.5-2.0$\,keV band \citep{Pearce2017}, smoothed to $3\arcsec$. The image properties are given in Table\,\ref{imaging}, IM10. The presence of four relics and the halo suggests a complex, multiple mergers.}
      \label{fig1a}
  \end{figure*}

The cluster is also rich at radio wavelength. It is known to host a giant $\sim2.1$\,Mpc large radio halo at the cluster center and a 1.5\,Mpc relic to the north-east. \citep{Giovannini1999,Govoni2001a,Govoni2001b,Orru2007,Venturi2013,Pearce2017,George2017,Paul2019}. With deep VLA observations, \cite{Pearce2017} found three new faint radio relics (namely, R2, R3, and R4) in this system, see Fig.\,\ref{fig1a}. All four relics are reported to be highly polarized at 3\,GHz \citep{Pearce2017}.

In the investigation by \cite{Pearce2017} and  \cite{Paul2019}, the radio halo was found to have a uniform spectral index distribution between 235\,MHz and 3\,GHz. A spatial correlation was also suggested between the halo regions with the flattest spectral indices and those with the highest X-ray temperature \citep{Orru2007}. Although, sensitive VLA observations revealed that the correlation is not significant \citep{Pearce2017}. \cite{Govoni2001a} studied the point-to-point correlation of the radio and X-ray emission and found that the halo shows a linear relation. 

%####################################
% imaging properties
%####################################
\setlength{\tabcolsep}{13.0pt}   
 \begin{table*}[!htbp]
\caption{Image properties }
\centering
\label{Table 2}
 \begin{threeparttable} 
\begin{tabular}{c c c c c c c r}
\hline\hline
   & Name &Restoring Beam & Weighting & uv-cut & uv-taper & RMS noise\\ % table heading
&&&&&&$\upmu\,\rm Jy\,beam^{-1}$\\
\hline
   &IM1&$10\arcsec \times 10\arcsec$&Briggs&none&10\arcsec&5\\
VLA S-band &IM2&$10\arcsec \times 10\arcsec$&Uniform&$ \geq\rm0.2\,k\uplambda$&10\arcsec&10 \\
&IM3&$15\arcsec \times 15\arcsec$&Uniform&$ \geq\rm0.2\,k\uplambda$&15\arcsec&15\\
&IM4&$20\arcsec \times 20\arcsec$&Uniform&$ \geq\rm0.2\,k\uplambda$&20\arcsec&22\\
\hline   
   &IM5&$10\arcsec \times 10\arcsec$&Briggs&none&10\arcsec&7\\
VLA L-band &IM6&$10\arcsec \times 10\arcsec$&Uniform&$ \geq\rm0.2\,k\uplambda$&10\arcsec&15 \\
   &IM7&$15\arcsec \times 15\arcsec$&Uniform&$ \geq\rm0.2\,k\uplambda$&15\arcsec&18\\
&IM8&$20\arcsec \times 20\arcsec$&Uniform&$ \geq\rm0.2\,k\uplambda$&20\arcsec&24\\
\hline
&IM9&$5\arcsec \times 5\arcsec$&Briggs&none&none&7\\
&IM10&$10\arcsec \times 10\arcsec$&Briggs&none&8\arcsec&9\\
&IM11&$10\arcsec \times 10\arcsec$&Uniform&$ \geq\rm0.2\,k\uplambda$&8\arcsec& 14\\
uGMRT Band\,4&IM12&$15\arcsec \times 15\arcsec$&Briggs&none&10\arcsec&16\\
&IM13&$15\arcsec \times 15\arcsec$&Uniform&$ \geq\rm0.2\,k\uplambda$&10\arcsec&18 \\
&IM14&$20\arcsec \times 20\arcsec$&Uniform&$ \geq\rm0.2\,k\uplambda$&15\arcsec&21\\
&IM15&$25\arcsec \times 25\arcsec$&Briggs&none&20\arcsec&24\\
 \hline 
   &IM16&$15\arcsec \times 15\arcsec$&Briggs&none&10\arcsec&41\\
  uGMRT Band\,3 &IM17&$15\arcsec \times 15\arcsec$&Uniform&$ \geq\rm0.2\,k\uplambda$&10\arcsec&48\\
 &IM18&$20\arcsec \times 20\arcsec$&Uniform&$ \geq\rm0.2\,k\uplambda$&15\arcsec&55\\
 \hline
GMRT&IM19&$15\arcsec \times 15\arcsec$&Uniform&$ \geq\rm0.2\,k\uplambda$&&381\\
 (235 MHz)&IM20&$20\arcsec \times 20\arcsec$&Uniform&$ \geq\rm0.2\,k\uplambda$&& 423\\
  \hline 
GMRT&IM21&$30\arcsec \times 30\arcsec$&Briggs&none&none&589\\
 (150 MHz)&IM22&$30\arcsec \times 30\arcsec$&Uniform&$ \geq\rm0.2\,k\uplambda$&none&612\\
 \hline  
 \end{tabular}
\begin{tablenotes}[flushleft]
  \footnotesize
   \vspace{0cm}
   \item\textbf{Notes.} Imaging was always performed using multi-scale clean, $\tt{nterms}$=2 and $\tt{wprojplanes}$=500. For all images made with {\tt Briggs} weighting we used ${\tt robust}=0$. 
    \end{tablenotes}
    \end{threeparttable} 
\label{imaging}
\end{table*}

%########################################################################
% Section 3: Observations and data reduction
%########################################################################
\section{Observations and data reduction}
\label{obs}
\subsection{uGMRT/GMRT}
Abell\,2744 was observed with the GMRT in Band\,4 using GMRT Wideband Backend (GWB) and GMRT software Backend (GSB). The GSB is the old narrow-band receiver and GWB the new wideband receiver. The pointing center was different for Band\,4 observations. All uGMRT observations were carried out in two observing runs. The total bandwidth is 200\,MHz for the Band\,3 and 240\,MHz for the Band\,4, covering the frequency range from 300-950 MHz. The observational details are summarized in Table\,\ref{Tabel:obs}. The primary calibrator 3C48 was used as a flux density calibrator.  

Both narrow and wideband GMRT data were processed using the Source Peeling and Atmospheric Modeling \citep[$\tt{SPAM}$;][]{Intema2009}, pipeline{\footnote{\url{http://www.intema.nl/doku.php?id=huibintemaspampipeline}}}.
The $\tt{SPAM}$ pipeline performs  direction-dependent calibration. The main steps are outlined below. The observations from multiple nights were combined. Each wideband data  set is split into six sub-bands, which were processed independently using the $\tt{SPAM}$ pipeline. The Band 4 data covering the frequency range 850-950\,MHz were heavily affected by radio frequency interference and were thus completely flagged out.

%%%%%%%%%%%%%%%%%%%%%%%%%%%%%%%%%%%%%%%%%%%%%%%%%%%%%%%%%%%%%%%%%
%  Fig 2: Radio-X ray overlay 

%%%%%%%%%%%%%%%%%%%%%%%%%%%%%%%%%%%%%%%%%%%%%%%%%%%%%%%%%%%%%%%%% 

The flux density of the primary calibrator 3C48 was set according to \cite{Scaife2012}. After flux density scale calibration, the data were averaged, flagged, and corrected for the bandpass. To correct the phase gains of the target field, we started from a global sky model obtained from the GMRT GSB data. $\tt{SPAM}$  measures the ionospheric phase errors toward the strongest sources in the field of view, allowing to derive direction-dependent gains for each of them, and fitting a phase-screen over the entire field of view. Within SPAM, the imaging is done with {\tt AIPS} using the wide-field imaging technique to compensate for the non-complanarity of the array. To produce deep full continuum images, the calibrated sub-bands were combined. The deconvolution was performed in {\tt CASA} using $\tt{nterms}$=2, $\tt{wprojplanes}$=500, and $\tt{Briggs}$ weighting with robust parameter 0.

%%%%%%%%%%%%%%%%%%%%%%%%%%%%%%%%%%%%%%%%%%%%%%%%%%%%%%%%%%%%%%%%%
%  Fig 3 : uGMRT band 4 images
\begin{figure*}
    \centering
    \includegraphics[width=1.00\textwidth]{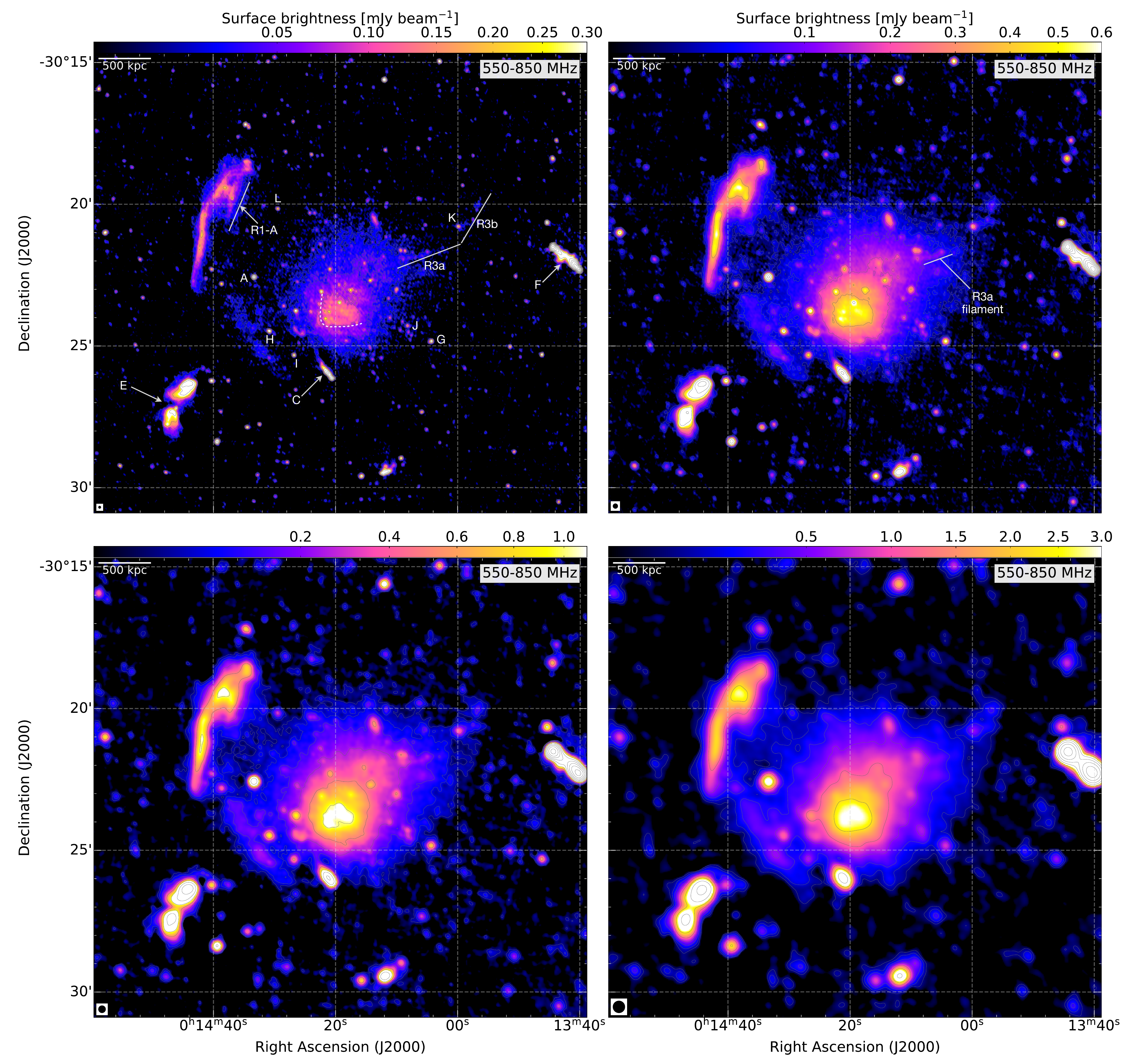}
        \vspace{-0.6cm}
 \caption{Total power uGMRT (550-850\,MHz) images of Abell\,2744 at different resolutions ($5\arcsec$, top left; $10\arcsec$, top right; $15\arcsec$, bottom left; $25\arcsec$, bottom right). The known diffuse emission sources, namely the main relic (R1), the fainter relics (R2, R3, and R4), and the extended halo are recovered in our new uGMRT observations. The radio emission at the center of the halo apparently shows a cone-like morphology. The image properties are given in Table\,\ref{imaging}. Here, panel top left, top right, bottom left and  bottom right correspond to IM9, IM10, IM12, and IM15, respectively. Contour levels are drawn at $[1,2,4,8,\dots]\,\times\,3.5\,\sigma_{{\rm{ rms}}}$. In these images there is no region below $-3\,\sigma_{{\rm{ rms}}}$. The beam size is indicated in the bottom left corner of the each image.}
      \label{fig1}
  \end{figure*}   
%%%%%%%%%%%%%%%%%%%%%%%%%%%%%%%%%%%%%%%%%%%%%%%%%%%%%%%%%%%%%%%%%

%%%%%%%%%%%%%%%%%%%%%%%%%%%%%%%%%%%%%%%%%%%%%%%%%%%%%%%%%%%%%%%%%
%  Fig 4: uGMRT band 3 and 150 MHz images
 \begin{figure*}
    \centering
    \includegraphics[width=1.0\textwidth]{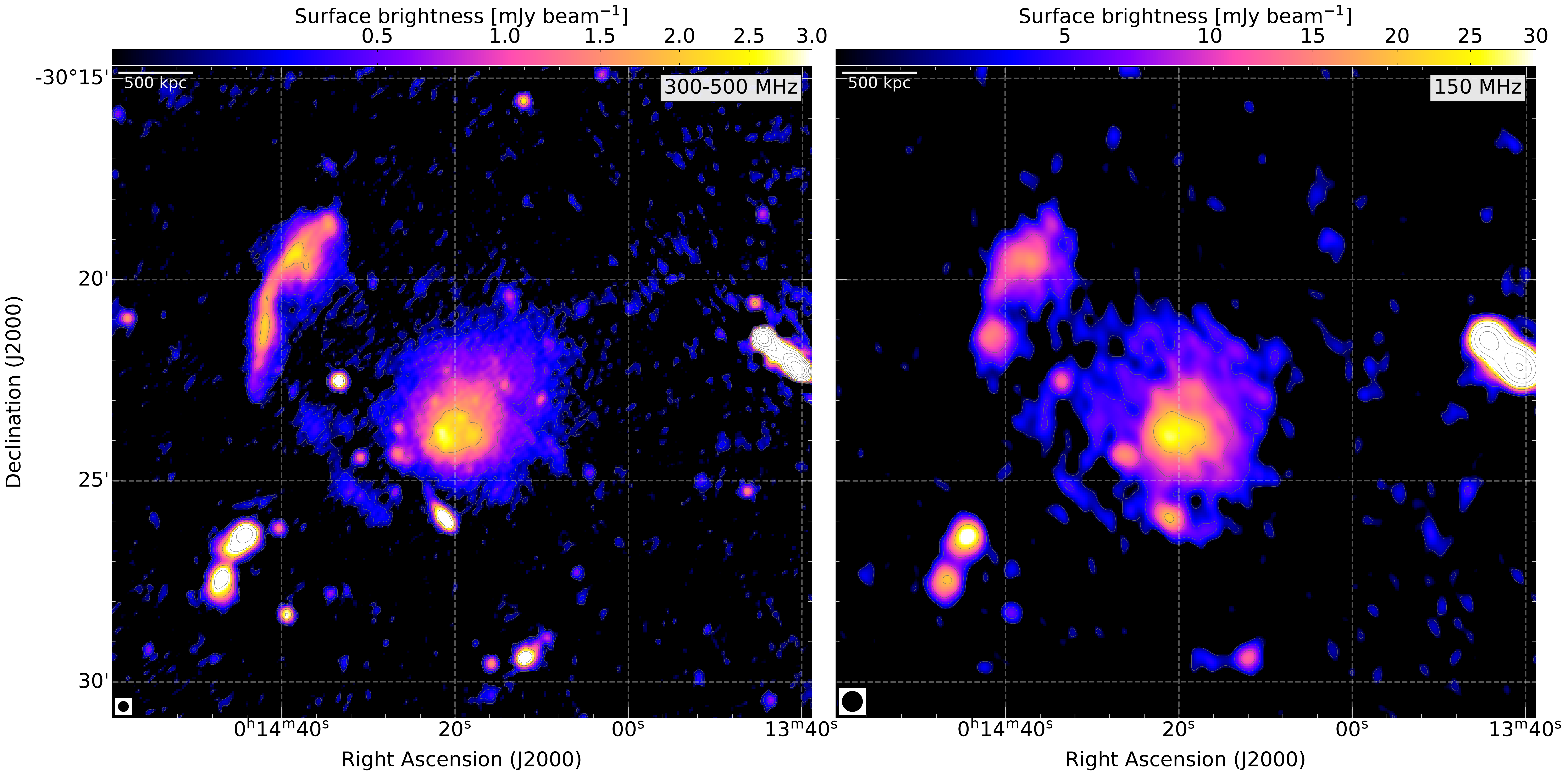}
    \vspace{-0.6cm}
 \caption{uGMRT Band\,3 (left) image at ${15\arcsec}$ resolution and GMRT 150\,MHz (right) narrow band image at $30\arcsec$ resolution. The image properties are given in Table\,\ref{imaging}, IM16 (left-panel) and IM21 (right-panel). Contour levels are drawn at $[1,2,4,8,\dots]\,\times\,3.5\,\sigma_{{\rm{ rms}}}$. In both images there is no region below $-3.0\,\sigma_{{\rm{ rms}}}$. The halo and the main relic R1 are recovered well at both frequencies. Images are created with {\tt Briggs} weighting with {\tt robust=0}. The beam size is indicated in the bottom left corner of the each image.}
      \label{fig2}
  \end{figure*}     
  
%####################################  

We also processed the archival 150\,MHz, 235\,MHz, and 325\,MHz narrow-band GMRT data with the SPAM pipeline. The 235\,MHz and 325\,MHz images are already published \citep{Venturi2013,Paul2019}, therefore not shown in the paper and are used only for flux density measurements. These observations were performed in August  2009 and August 2011 at 150\,MHz and 235\,MHz, respectively. 

\subsection{VLA}
The cluster has been observed with the Karl G. Jansky Very Large Array (VLA) in L and S-band with B\&A, C\&B, and, D\&C  configurations \citep{Pearce2017}. Since we used the fully calibrated VLA L and S band data presented in \cite{Pearce2017}, for a detailed description of the observations and data reduction, we refer the reader to \cite{Pearce2017}. To summarize, for each configuration, 3C147 and 3C138 were observed as  primary calibrators and J0011-2612 as secondary phase calibrator. 

The data were calibrated for the antenna position offsets, elevation dependent gains, parallel-hand delay, bandpass, and gain variations using 3C147 and 3C138. The gains for all the calibrators were then combined, solving for the J0011-2612 flux density. For polarization calibration, the leakage response was determined using the unpolarized calibrator 3C147 and the cross-hand delays were calibrated using 3C138. The absolute position angle was corrected using the polarized calibrator 3C138. The resulting calibrated data were averaged by a factor of 4 in frequency per spectral window (for all three configurations) followed by several round of self-calibrations.

The uncertainty in the radio flux density measurements was estimated as 
\begin{equation}
\Delta S =  \sqrt {(f \cdot S)^{2}+{N}_{{\rm{ beams}}}\ (\sigma_{{\rm{rms}}})^{2}},
\end{equation}
where $f$ is an absolute flux density calibration uncertainty, $S$ is the flux density, $\sigma_{{\rm{ rms}}}$ is the RMS noise, and $N_{{\rm{beams}}}$ is the number of beams. We assume absolute flux density uncertainties of 10\,\% for uGMRT/GMRT \citep{Chandra2004} data, and 2.5\,\% for the VLA data.

%########################################################################
%  Section 4: Results
%######################################################################## 
\section{Results: Radio continuum images}
\label{results}
In Fig.\,\ref{fig1} and \,\ref{fig2}, we show the resulting  Band\,4 and Band\,3, uGMRT radio continuum images of Abell\,2744. To study the radio emission on different varying spatial scales, we show Band\,4 images at different resolutions. These images are created with different uv-tapers.  At Band\,4, with a restoring beam width of $5\arcsec$ we achieved a noise level of $7\mu\,\rm Jy\,beam^{-1}$. The published rms values (at about $16\arcsec$ resolution) of the Abell 2744 field  are $\sigma_{\rm rms, \,610\,MHz}=100\,\mu\,\rm Jy\,beam^{-1}$ \citep{Paul2019},  $\sigma_{\rm rms,\,325\,MHz}=900\,\mu\,\rm Jy\,beam^{-1}$ \citep{Orru2007}, and $\sigma_{\rm rms,\,325\,MHz}=150\,\mu\,\rm Jy\,beam^{-1}$ \citep{Venturi2013}. Our new uGMRT  Band\,4 and Band\,3 images are about a factor of six and three, respectively, deeper than the published GMRT images. The sources are labeled as in \cite{Pearce2017} and extending the list. The main relic (R1) and the halo are detected at all resolutions. The main properties of the diffuse sources in the cluster are summarized in Table\,\ref{Tabel:Tabel2}.

The largest linear scale (LLS) of the main relic is about 1.5\,Mpc at 385\,MHz, similar to what measured at 1.5\,GHz. However, the thickness of the relic increases toward lower frequencies, in particular for the northern part of the relic, which ranges from 270\,kpc to 460\,kpc at 3\,GHz and 385\,MHz, respectively. A linear component of the diffuse emission (R1-A) extending away from the upper northern part of the relic is also visible in the high resolution image (Fig.\,\ref{fig1} top-left). The other known fainter relics, namely R2, R3, and R4, are recovered in the low resolution Band\,3 and Band\,4 images, see Fig.\,\ref{fig1} bottom panels. As apparent in Fig.\,\ref{fig1} top right, the relic R3 seems to be composed of two parts: R3a and R3b. Moreover, there is a fine filament in R3a, also seen in the VLA 1.4 GHz images  \citep{Pearce2017}. The full emission from R3 is not detected fully at 385 and 150\,MHz (also R2) due to the low sensitivity of these two observations in recovering very low surface brightness emission compared to Band\,4 and VLA (1-4 MHz) observations.

The halo is recovered well at 150\,MHz and uGMRT Band\,3 and Band\,4 images. The halo emission is best seen in low resolution images. The LLS of the halo is about 2.5\,Mpc at 675\,MHz. The halo is more extended than detected with the VLA above 1\,GHz, namely 2.1\,Mpc \citep{Pearce2017}. The radio surface brightness is high at the center and decreases toward the outer regions. The halo also shows some small scale surface brightness variations (see Fig.\,\ref{fig1} top right.) 

In the uGMRT high-resolution image (Fig.\,\ref{fig1} top-left), the halo apparently shows a square-shaped morphology. Moreover, the innermost brightest region shows a wide angle cone (similar to the Bullet cluster core). There are also at least 28 discrete ($> 3.5\sigma$) sources detected in the halo region, excluding a head-tail radio galaxy to the south (source C). All discrete sources embedded within the halo are clearly visible in the Band\,4 high resolution image. The combined flux density of the 28 discrete sources is about $\rm 12\,mJy$ at 675\,MHz, extracted from a $10\arcsec$ radio image created using {\tt Briggs} weighting with {\tt robust=0}. Including these sources, the measured flux density of the entire halo is $\rm 112\pm10\,mJy$.  We note that this value is slightly different than the one reported in Table\,\ref{Tabel:Tabel2} due to different imaging parameters and resolution.We emphasize that the flux densities of the halo reported in Table\,\ref{Tabel:Tabel2} do not include contributions from sources A, C, G, H, I, J, K, and L. Out of $\rm 12\,mJy$, the total contribution from sources A, G, H, I, J, K, and L is about $\rm 9\,mJy$ at 675\,MHz, which means the only $\sim3\%$ of total halo flux density resides in the rest of the 21 discrete sources. 

At moderate resolution, our 550-850\,MHz images show ``streams'' of radio emission connecting the northern part of the main relic R1 to the halo emission (Fig.\,\ref{fig1} top-right and bottom-left). The relic R2 is also clearly connected to the halo. 

A low resolution Band\,3 image is shown in the left panel of Fig.\,\ref{fig2}. The very low surface brightness emission seen in the Band\,4 images could not be recovered at the achieved sensitivity at Band\,3. We note that the total on-source time at Band\,4 data was about 24 hours while only 5 hours at Band\,3. The GMRT low resolution 150\,MHz image is shown in the right panel of Fig.\,\ref{fig2}. The morphology of the halo is quite similar at 150\,MHz,  385\,MHz and 675\,MHz.  

\begin{figure*}
    \centering
    \includegraphics[width=1.0\textwidth]{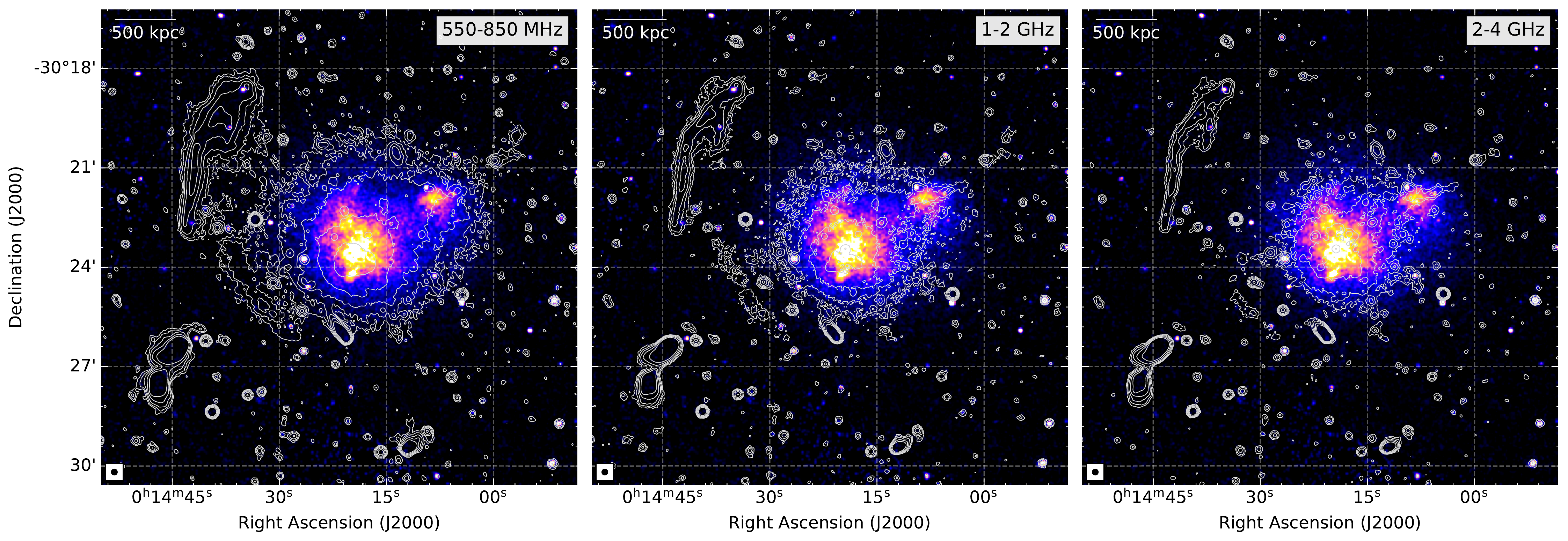}
 \caption{\textit{Chandra} ($0.5-2.0\rm\,keV$ band) X-ray image overlaid with radio contours at 675\,MHz, 1.5\,GHz, and 3.0\,GHz. All radio images are created at a common resolution of $10\arcsec$.  Here, left, middle, and right panels correspond to IM10, IM5, and IM1, respectively. The halo emission is apparently more extended toward low frequencies, in particular to the northeast direction. Contour levels are drawn at $[1,2,4,8,\dots]\,\times\,3.5\,\sigma_{{\rm{ rms}}}$. In all three images there is no region below $-3.0\,\sigma_{{\rm{ rms}}}$. The beam size is indicated in the bottom left corner of the each image.}
      \label{RXimage}
  \end{figure*}

The \textit{Chandra} X-ray image overlaid with radio contours at different frequencies is shown in Fig.\,\ref{RXimage}. The radio emission from the halo extends further to the northeast where the X-ray emission is fainter. The radio and X-ray peaks coincide. The innermost halo emission traces the Bullet-like feature \citep{Owers2011} visible in the X-ray to the south of the main core remarkably well. Moreover, the halo seems to be more extended at low frequencies, in particular in the northeast and northwest direction. The LLS of the halo is at 675\,MHz, 1.5\,GHz and 3\,GHz is 2.5\,Mpc, 2.1\,Mpc, and 1.7\,Mpc, respectively.

%########################################################################
%  Section 5: Spectral analysis 
%######################################################################## 
\section{Spectral analysis}
\label{spectral_anaylsis}

To study the spectral properties of the diffuse emission sources, we combined our new uGMRT (300-850 MHz) observations with those previously presented at 1-4\,GHz \citep{Pearce2017}. We also use the legacy GMRT data at 325\,MHz, 235\,MHz, and 150\,MHz to study the integrated spectra of the halo and R1.  

%%%%%%%%%%%%%%%%%%%%%%%%%%%%%%%%%%%%%%%%%%%%%%%%%%%%%%%%%%%%%%%%%
%  Fig 5: Radio and X ray comparison as a function of frequency 

%%%%%%%%%%%%%%%%%%%%%%%%%%%%%%%%%%%%%%%%%%%%%%%%%%%%%%%%%%%%%%%%%

%%%%%%%%%%%%%%%%%%%%%%%%%%%%%%%%%%%%%%%%%%%%%%%%%%%%%%%%%%%%%%%%%
% Table 2- spectral properties of the diffuse emission sources 
%%%%%%%%%%%%%%%%%%%%%%%%%%%%%%%%%%%%%%%%%%%%%%%%%%%%%%%%%%%%%%%%%
\setlength{\tabcolsep}{6pt}
\begin{table*}[!htbp]
\caption{Properties of the diffuse radio sources in the cluster Abell\,2744.}
\centering
\begin{threeparttable} 
\begin{tabular}{*{10}{c}}
\hline \hline
\multirow{1}{*}{Source} & \multicolumn{2}{c}{VLA (1-4\,GHz)} & \multicolumn{2}{c}{uGMRT (300-850\,MHz)} & \multicolumn{2}{c}{GMRT}&\multirow{1}{*}{$\rm LLS^{\dagger}$} & \multirow{1}{*}{$\alpha$}$^{\dagger\dagger}$& \multirow{1}{*}{$P_{1.5\,\rm GHz}$} \\
 \cline{2-7} 

& $S_{\rm3.0\,GHz}$ & $S_{\rm1.5\,GHz}$&$S_{\rm675\,MHz}$&$S_{\rm385\,MHz}$&$S_{\rm235\,MHz}$&$S_{\rm150\,MHz}$ &&&\\
  & (mJy) & (mJy) & (mJy) & (mJy)&  (mJy) & (mJy) &(Mpc) &&($10^{24}\,\rm W\,Hz^{-1}$) \\
  \cline{2-3} \cline{4-5}\cline{6-7}
  \hline  
R1 & $5.2\pm0.4$ & $12.3\pm1.0$&$30.1\pm1.9$&$70\pm6.8$& $98\pm10$&$161\pm22$&$\sim1.5$ &$-1.17\pm0.03$&$3.93\pm0.03$ \\ 
R2&$0.70\pm0.14$ &$1.9\pm0.2$&$4.0\pm1.0$&$7.0\pm3$& $-$& $-$&$\sim1.1$&$-1.19\pm0.05$&$0.61\pm0.08$\\ 
R3  &$0.55\pm0.07$ &$1.2\pm0.1$&$2.6\pm0.4$&$-$& $-$&$-$&$\sim1.1$&$-1.10\pm0.05$&$0.38\pm0.08$\\ 
R4  &$0.30\pm0.07$ &$0.7\pm0.1$&$5.0\pm0.8$&$3.1\pm0.7$& $-$&$-$& $\sim0.2$&$-1.14\pm0.04$&$0.22\pm0.08$\\ 
Halo&$16.4\pm1.0$ &$42.9\pm2.1$&$105\pm9$&$185\pm16$&$290\pm35$&$526\pm52$& $\sim2.5$&$-1.14\pm0.05$&$13.6\pm0.04$\\ 
\hline 
\end{tabular}
\begin{tablenotes}[flushleft]
\footnotesize
\item{\textbf{Notes.}} Flux densities of relics were extracted from $15\arcsec$ resolution images at 235 MHz, 385 MHz, 675 MHz, 1.5 GHz and 3 GHz, corresponding to images IM3, IM7, IM13, IM17, and IM19, respectively. For imaging properties see Table\,\ref{imaging}. The flux density of the halo is measured from $20\arcsec$ images, namely IM4, IM8, IM14, IM18, IM20, and IM22. All 150 MHz flux densities are measured from IM22. The regions where the flux densities were extracted are indicated in the left panel of Fig.\,\ref{fig6}. The flux density values are measured above $3\sigma$ noise level. Absolute flux density scale uncertainties are assumed to be 10\% for the uGMRT/GMRT data and 2.5\% for the VLA L- and S-band data. $^{\dagger}$The LLS measured at 675\,MHz; $^{\dagger\dagger}$ the integrated spectral index obtained by a single power-law fit. 
\end{tablenotes}
\end{threeparttable} 
\label{Tabel:Tabel2}   
\end{table*}  

%######################%######################
\subsection{Integrated spectra}
\label{int}
%######################%######################

To measure flux densities,  we created images using uniform weighting and a uv-cut of $0.2\,\rm k\lambda$. Here, $0.2\,\rm k\lambda$ is the shortest well sampled baseline of the VLA S-band data. This uv-cut was applied to the VLA L-band, uGMRT Band\,3 and Band\,4 data. The same uv-cut and weighting scheme are used for making spectral index and curvature maps described Sect.\,\ref{index_maps}.  All possible measures were taken to ensure the recovery of the entire flux from the halo and relics at each frequency. 

For integrated spectral analysis, we created images at two different resolutions, namely $15\arcsec$ and $20\arcsec$. Due to different uv-coverages of the GMRT and VLA data, the resulting images have marginally different resolution. Therefore, we smoothened the GMRT and VLA images to the same resolution, i.e., $15\arcsec$ and $20\arcsec$ using the {\tt CASA} task {\tt imsmooth}. The $15\arcsec$ images corresponding to IM3, IM7, IM13, and IM17 and IM19 in Table\,\ref{imaging} were used to measure the flux density of all four relics in the field.  This resolution was chosen to properly exclude the contamination from the halo and other unrelated sources. For the halo we rather used $20\arcsec$ images namely IM4, IM8, IM14, IM18, and IM20 as this resolution allows us to recover the low surface brightness emission in the outermost regions of the halo at the highest signal-to-noise ratio, thus allowing us to measure the true flux density of the halo. We note that the flux density of the halo changes only marginally between $20\arcsec$ and $40\arcsec$ resolution images at 675\,MHz. Similar trends are seen at other observed frequencies. Since 150\,MHz GMRT data do not allow to image at a similar resolution, the flux density values at this particular frequency are extracted from the $30\arcsec$ image, namely IM22 in Table\,\ref{imaging}.  All radio flux densities, unless stated otherwise, include emission above $3\sigma_{\rm rms}$. 
%######################
\subsubsection{Halo}
\begin{figure*}
    \centering
         \includegraphics[width=0.99\textwidth]{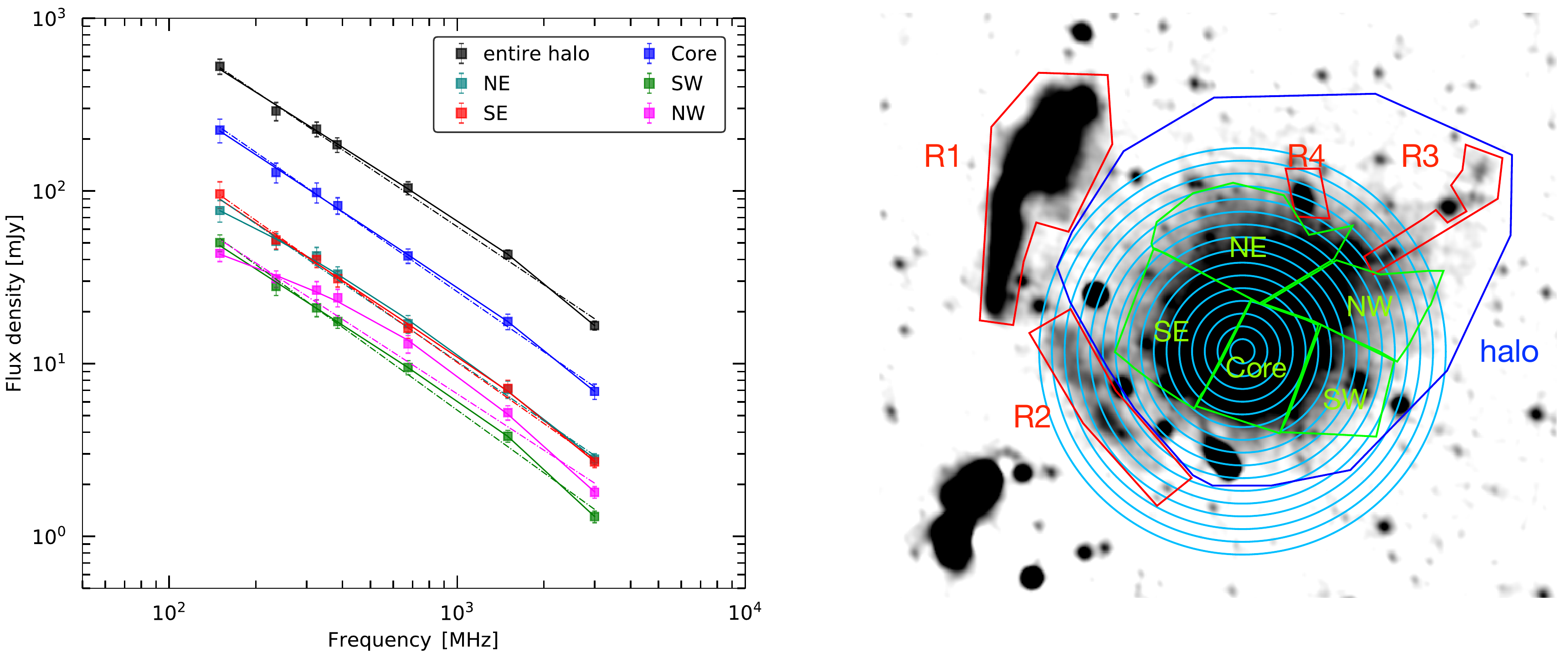}  
 \caption{\textit{Left}: Integrated radio spectra of the entire halo and its subregions. The soild and dash-dotted lines show fitted power-law and double power-law spectra, respectively. The radio spectrum of the halo is  described by a power-law  between 150\,MHz and 3\,GHz. Unlike the entire halo, its subregions show different spectra also featuring high frequency spectral steepening. \textit{Right}: The flux density of the entire halo was extracted from the blue region, labeled as halo. Green regions were used to extract the flux density from the halo subregions. The blue annuli are used to compute the radio and X-ray surface brightness profiles. These annuli have widths of $15\arcsec$, similar to the radio beam size.}
\label{fig6}
\end{figure*}

%######################
By combining our flux density measurement from 150\,MHz to 3\,GHz, we obtained the integrated radio spectrum of the halo. The resulting spectrum is shown in the left panel of Fig.\,\ref{fig6}. The regions used for extracting the integrated flux densities are shown in the right panel of Fig.\,\ref{fig6}. The contaminations from other point sources and relics R2, R3, and R4 were manually measured and then subtracted from the total halo flux density. We note that there are around 28 faint discrete sources embedded in the halo region at 675\,MHz. Most of these sources are also visible at 1.5 and 3\,GHz but at 380 and 150\,MHz we only identified around 9 and 5 sources, respectively. We emphasize that the total flux density of the halo do not include flux contribution from sources A, C, G, H, I, J, K, and L. As mentioned in Sect.\,\ref{results}, the rest of the discrete sources only contribute to about 3\% to the total halo flux density which is almost negligible. 

Our new 150\,MHz, 385\,MHz, and  675\,MHz flux density values of the halo are consistent with the high frequency data points. The halo flux density at 150\,MHz is consistent with those reported by \cite{George2017} using Murchison Widefield Array. Moreover, our uGMRT 385\,MHz data point is in line with the VLA P-band \citep{Orru2007} but different from the value obtained by \cite{Venturi2013}. To know the reason for the difference in the flux density measurements at 325\,MHz, we also processed the archival 325\,MHz GMRT data. From the final image, created using  uniform weighting, we measure the flux density of  $228\pm21$\,mJy at 325\,MHz. This value is consistent with the rest of the data points but significantly lower than those reported by \cite{Venturi2013}, namely $323\pm26$\,mJy. To exactly compare our values with \cite{Venturi2013}, we also imaged 325\,MHz data at $35\arcsec$ resolution with {\tt Briggs} weighting scheme ({\tt robust=0}). The resulting image shows a flux density of $\sim 300$\,mJy, including all discrete sources (except C) embedded in the halo region and R3, R4. This value is comparable to values given in \cite{Venturi2013}, indicating that the difference in flux density at 325\,MHz is simply due to different weighing schemes and contributions from unrelated sources.

%%%%%%%%%%%%%%%%%%%%%%%%%%%%%%%%%%%%%%%%%%%%%%%%%%%%%%%%%%%%%%%%%
%  Fig.6 : integrated  spectra

%%%%%%%%%%%%%%%%%%%%%%%%%%%%%%%%%%%%%%%%%%%%%%%%%%%%%%%%%%%%%%%%%

\setlength{\tabcolsep}{12pt}
\begin{table}
\caption{Reduced Chi-square values for the model fitted in the left panel of Fig.\,\ref{fig6}.}
\centering
\begin{threeparttable} 
\begin{tabular}{ c c c c c }
 \hline  \hline  
 \multirow{1}{*}{}& \multicolumn{2}{c}{$\chi ^{2}_{\rm red}$} \\
 \cline{2-3}  
&\multirow{1}{*}{power-law} & \multirow{1}{*}{double power-law}  \\ 
\hline  
entire halo &$0.67$&$0.93$\\
NW &0.58&0.12\\ 
SE &$0.10$&0.26\\
Core&$0.30$&0.24\\ 
SW &$0.12$&0.06 \\ 
NE &$0.92$&0.02\\
\hline 
\end{tabular}
\end{threeparttable} 
\label{fit_int}   
\end{table}

We first consider the total halo emission. The integrated spectrum of the halo is shown in the left panel of Fig.\,\ref{fig6}. At 3 GHz, the thermal Sunyaev-Zeldovich (SZ) effect could cause a reduction in the measured flux density of the halo. To  check this, we estimate the possible corrections for the SZ-effect at 3\,GHz. The SZ-decrement is predicted assuming a gas density and temperature distribution using a $\beta$-model, based on \cite{Limousin2016}, with a reference temperature of  $T_{\rm X}=10\,\rm keV$. We also apply an inner uv-cut of $0.2\,\rm k\lambda$ for the SZ calculation to match the uv-coverage of our radio observations. When integrating within a region corresponding to the halo, we find a total SZ decrement of about $ -0.2\,\rm mJy$ at 3\,GHz. We conclude that the SZ decrement is about 2\% of the total halo flux density at 3\,GHz, and therefore have a negligible impact on the halo spectrum. 

We fit the integrated spectrum with two models, namely power-law and broken power-law. In Table\,\ref{fit_int}, we compile the reduced chi-square ($\chi ^{2}_{\rm red}$) values. The integrated spectrum of the halo in Abell\,2744 follows a single power-law spectrum between 150\,MHz and 3\,GHz. This is supported by a lower $\chi ^{2}_{\rm red}$ in the case of the single power-law  ($\chi ^{2}_{\rm red}=0.67$) with respect to the double-power law ($\chi ^{2}_{\rm red}=0.93$). The halo has a spectral index of $-1.14\pm0.04$ between 150\,MHz and 3\,GHz. This value is consistent with that reported by \cite{George2017}. The majority of radio halos are reported to show a power-law spectrum between 150 MHz and 1.4 GHz \citep[e.g.,][]{Shimwell2014,Hoang2019,Rajpurohit2020a,Luca2021}. Other than the Abell 2744 halo, integrated spectra above 1.4\,GHz are available only for five halos, namely in Coma cluster \citep{Thierbach2003}, 1RXS\,J0603.3+4214 \citep{Rajpurohit2020a}, MACS\,J0717.5+3745 \citep{Rajpurohit2021b}, Bullet cluster \citep{Shimwell2014}, and Abell S1063 \citep{Xie2020}. Out of these five halos, a power-law spectrum has been found for two of them: 1RXS\,J0603.3+4214 and Bullet cluster. A high frequency spectral steepening has been detected in the other three clusters: Coma cluster,  Abell S1063, and MACS\,J0717.5+3745. In addition, except for the halo in 1RXS\,J0603.3+4214, all show a steep integrated spectrum ($\leq-1.4$). 

The integrated radio spectra of halos provide important information about the underlying particle acceleration mechanism. In turbulent re-acceleration models, a high frequency spectral steepening is expected \citep{Brunetti2001,Cassano2005,Brunetti2014,Brunetti2016}. It remains unclear why out of the above mentioned six halos (including Abell 2744), half is well fitted by a power-law while the other half shows high frequency steepening. To extract more information from our data, we also look into the integrated spectra from subregions of the halo to see whether the subregions also follow a power-law  or not. We divide the halo into five subregions, namely NE, SE, Core, NW, and SW. These regions are depicted in the right panel of Fig.\,\ref{fig6}. 

The resulting integrated radio spectra are shown in the left panel of Fig.\,\ref{fig6}. Intriguingly,  unlike the halo, its subregions show a steepening toward high frequencies and can be described by a double power-law, see Table\,\ref{fit_int}. We note that the integrated spectrum extracted from the combined green subregions (see the right panel of Fig.\,\ref{fig6}) also follows a power-law but is not shown in Fig.\,\ref{fig6}. It is surprising that the entire halo is characterized by a power-law while its subregions show high frequency spectral steepening, especially NE and NW. The Core region is the dominating contributor to the total halo flux density and can be still described by a single power-law. It is puzzling that the combination of all spectra (power-law and double power-law) results in the overall power-law like spectrum, although it is probably due to the fact that the more luminous part (like the cluster core) can be essentially described by a power-law. It is plausible that an overall power-law spectrum may be due to the complex superposition of different components. We discuss this further in Sect.\,\ref{IRX}.

%######################
\subsubsection{R1, R2, R3 and R4}
\begin{figure*}
    \centering
          \includegraphics[width=0.99\textwidth]{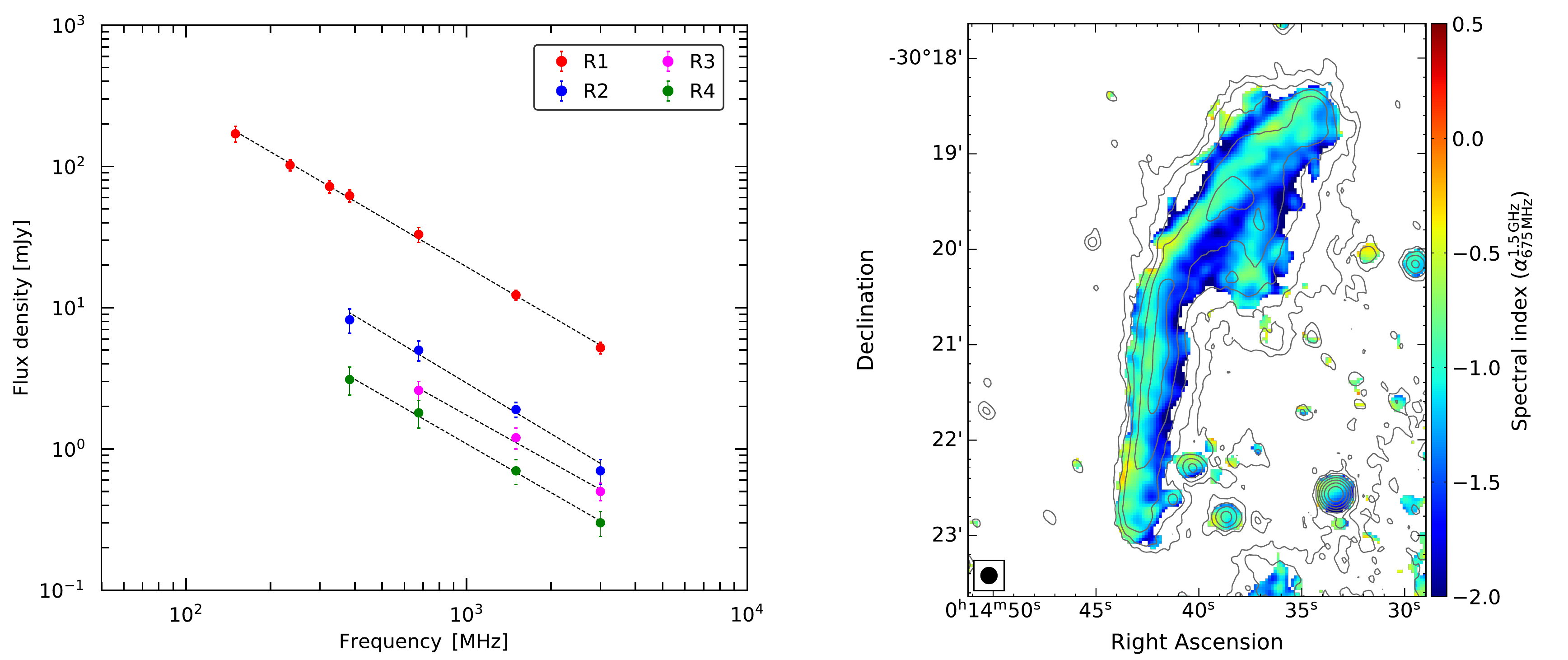}
 \caption{\textit{Left}: Integrated radio spectra of relics R1, R2, R3, and R4. Dashed lines are the fitted power-law with indices given in Table\,\ref{Tabel:Tabel2}. Regions used to obtained the integrated flux densities of all four relics are shown in the right panel of Fig.\,\ref{fig6a}. \textit{Right}: Spectral index map of the main relic R1 between 675\,MHz and 1.5\,GHz at 10$\arcsec$ resolution (corresponding to  IM6 and IM11 in Table\,\ref{imaging}), showing a clear spectral index gradient across the entire relic. Contour levels are drawn at $[1,2,4,8,\dots]\,\times\,3.5\,\sigma_{{\rm{ rms}}}$, and are from the uGMRT Band\,4 image. }
\label{fig6a}
\end{figure*}

%######################
As shown in the left panel of Fig.\,\ref{fig6a}, the main relic R1 follows a simple power-law between 150\,MHz and 3\,GHz. The integrated spectral index of the main relic is $\alpha= -1.17\pm0.03$. This value is consistent  with the values reported by \cite{Pearce2017} and \cite{Paul2019}. 

The other three fainter relics are not detected at a good signal to noise ratio below 385\,MHz. We thus do not include flux density values below 385\,MHz. The spectra of R2 and R4 can be also described by a power-law between 385\,MHz and 3\,GHz, see the left panel of Fig.\,\ref{fig6a}. The integrated spectral indices of R2, R3, and R4 are $-1.19\pm0.05$, $-1.10\pm0.06$, and $-1.14\pm0.05$, respectively. Since R3 is not detected fully at 385\,MHz, its integrated spectral index is obtained only between 675\,MHz and 3\,GHz. 

%%%%%%%%%%%%%%%%%%%%%%%%%%%%%%%%%%%%%%%%%%%%%%%%%%%%%%%%%%%%%%%%%
%  Fig.7 : regions and the spectral index map of the main relic 

%%%%%%%%%%%%%%%%%%%%%%%%%%%%%%%%%%%%%%%%%%%%%%%%%%%%%%%%%%%%%%%%%

\cite{Pearce2017} reported that R3 possesses one of the flattest spectra observed in any relic with an integrated spectral index value of $-0.63$ in the range 1.5 and 3\,GHz. The new spectral index value of R3 is consistent with the typical spectral index of relics. Our integrated spectral indices of R2 and R4 are also significantly flatter compared to those reported by \cite{Pearce2017}. In summary, we find that for all four relics, the slopes of the integrated spectra are between $-1.10$ to $-1.20$.

There exists a relation between radio power and LLS of known relics \citep{Nuza2017}. The ``elongated'' and ``roundish'' relics show a slightly different behavior: elongated relics have greater radio power than roundish relics. \cite{Pearce2017} found that R1, R2, and R3 fall into the known elongated relics group while R4 to that of roundish relics. At 675\,MHz, the LLS of R4 is $\sim 200\,\rm kpc$ which is significantly larger than measured at 1.5\,GHz, namely 50\,kpc \citep{Pearce2017}. For R4, we measure that the radio power is $P_{\rm 1.5\,GHz}=0.22\pm0.08$, see Table\,\ref{Tabel:Tabel2}. With these new values of the LLS and radio-power, R4 also fits nicely in the group of elongated relics.

A radio shock in the ICM, i.e., a shock front with electron acceleration and its subsequent radio emission can be considered (quasi-)stationary if the cooling time of electrons with energies corresponding to the lowest frequency in the observation is much shorter than the timescale on which the shock strength or geometry changes. For those stationary radio shocks, the integrated spectrum, $\alpha_{\rm int}$, is steeper by 0.5 than the injection index, $\alpha_{\rm inj}$,
\begin{equation}
\alpha_{\rm int}=\alpha_{\rm inj}-0.5.
\label{int_inj}
\end{equation}
According to diffusive shock acceleration (DSA) in the test-particle regime, the injection index is related to the Mach number of the shock as
\begin{equation}
\mathcal{M}=\sqrt{ \frac{2\alpha_{\rm inj}-3}{2\alpha_{\rm inj}+1 }}.
\end{equation}
If radio relics are generated according to the stationary radio shock scenario, they are expected to show a power-law with spectral index $-1$ or steeper. However, this (quasi-)stationary condition may not be fulfilled for spherical shocks \citep{Kang2015} or shocks in the presence of turbulent medium \citep{Paola2021}.

The integrated spectra of most of the relics follow the stationary radio shock scenario, at least according to the integrated spectrum. There are  a few exceptions, for example, the relics in Abell\,2256 \citep{vanWeeren2012b,Trasatti2015} and Abell\,3667 \citep{Hindson2014}. The integrated spectral index values of all four relics in Abell\,2744 are in agreement with a stationary radio shock.

%%%%%%%%%%%%%%%%%%%%%%%%%%%%%%%%%%%%%%%%%%%%%%%%%%%%%%%%%%%%%%%%%
%  Fig 8- radio color-color plot of the relic

\begin{figure*}
    \centering
         \includegraphics[width=0.47\textwidth]{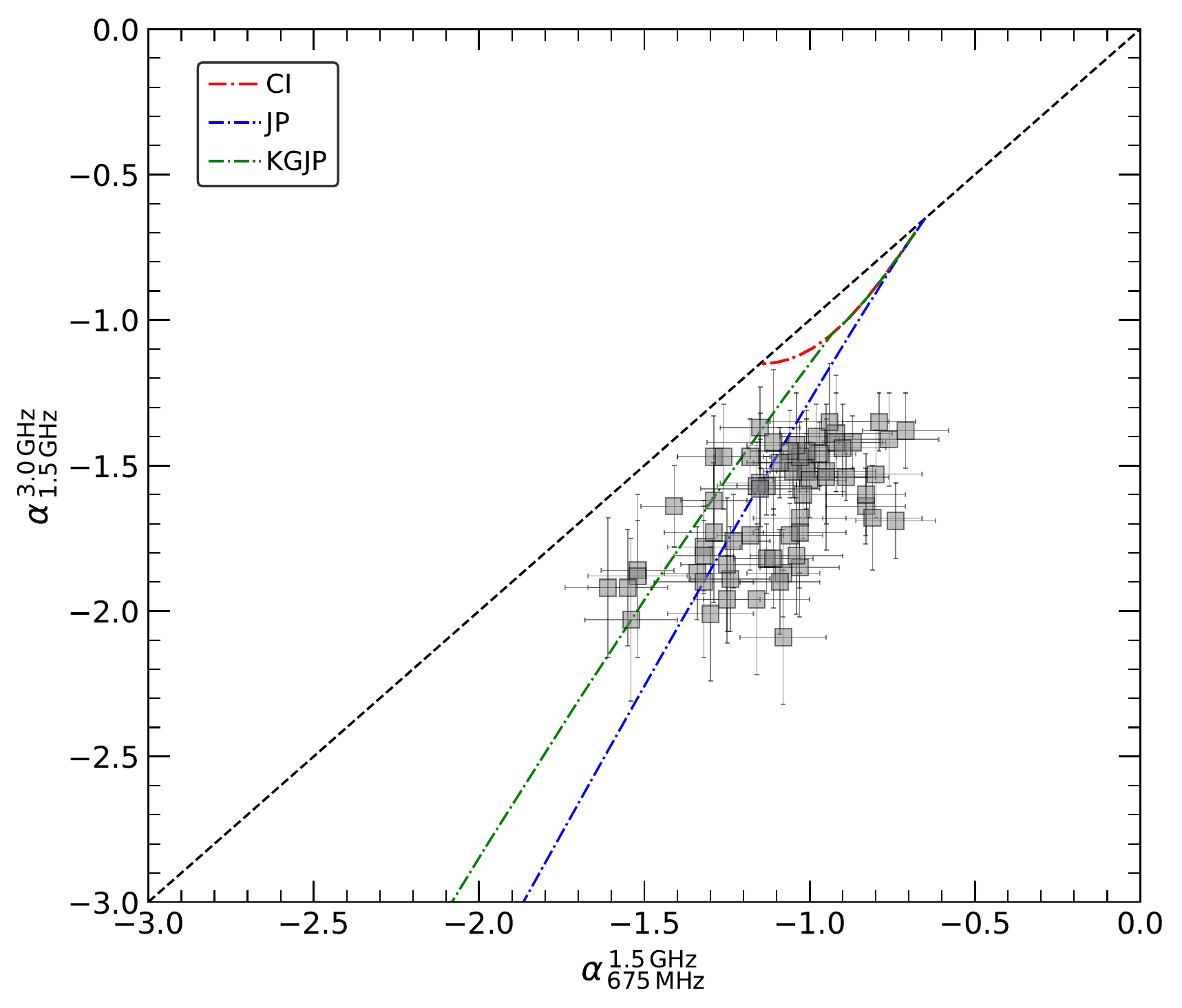} 
             \includegraphics[width=0.47\textwidth]{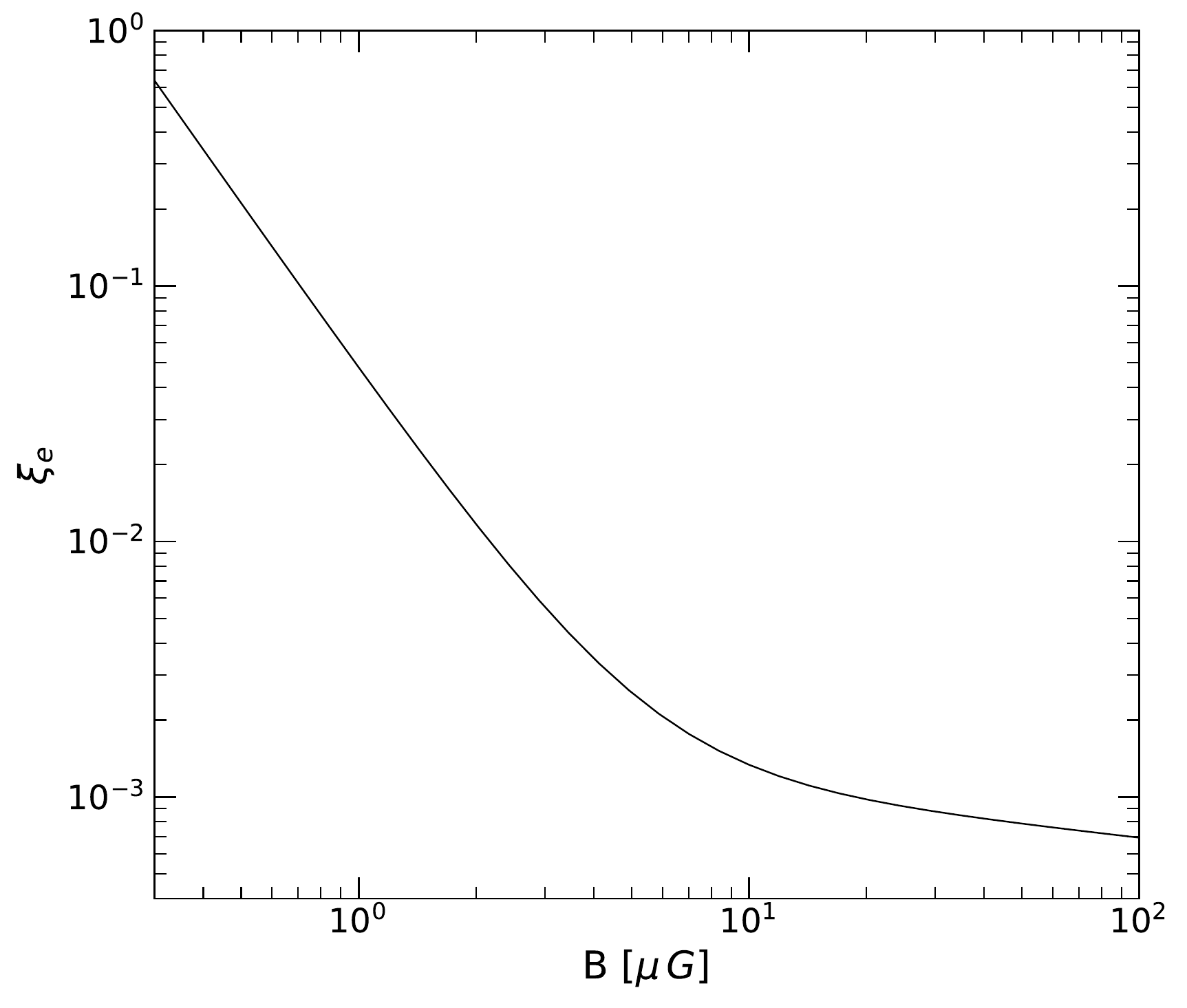} 
 \caption{\textit{Left}: Radio color-color plot of the main relic R1, superimposed with the JP (blue dash-dotted line), KGJP (green dash-dotted), and CI (red dash-dotted) spectral aging models adopting $\alpha_{\rm inj}=-0.67$. The observed data points seems to be relatively consistent with the JP model, suggesting that the relic is very likely seen close to edge-on. To extract spectral index values, we create square-shaped boxes with a width of $10\arcsec$, corresponding to a physical size of about 45\,kpc. \textit{Right}: Particle acceleration efficiency of the main relic R1 as a function of magnetic field. The values are estimated using the Mach number of $\mathcal{M}=3.6$ ($\alpha_{\rm int}=-1.17)$ and $\Psi({\cal M},T_{\rm d})=0.193$. The required acceleration efficiency is about 1\% or below, which can be achieved by accelerating electrons from the thermal pool if the magnetic field is sufficiently strong and a large fraction of the shock shows high Mach number.}
\label{figcc}
\end{figure*}   
%%%%%%%%%%%%%%%%%%%%%%%%%%%%%%%%%%%%%%%%%%%%%%%%%%%%%%%%%%%%%%%%%

%%%%%%%%%%%%%%%%%%%%%%%%%%%%%%%%%%%%%%%%%%%%%%%%%%%%%%%%%%%%%%%%%
%Fig9- spectral index map of the halo

\begin{figure*}
\centering
 \includegraphics[width=0.48\textwidth]{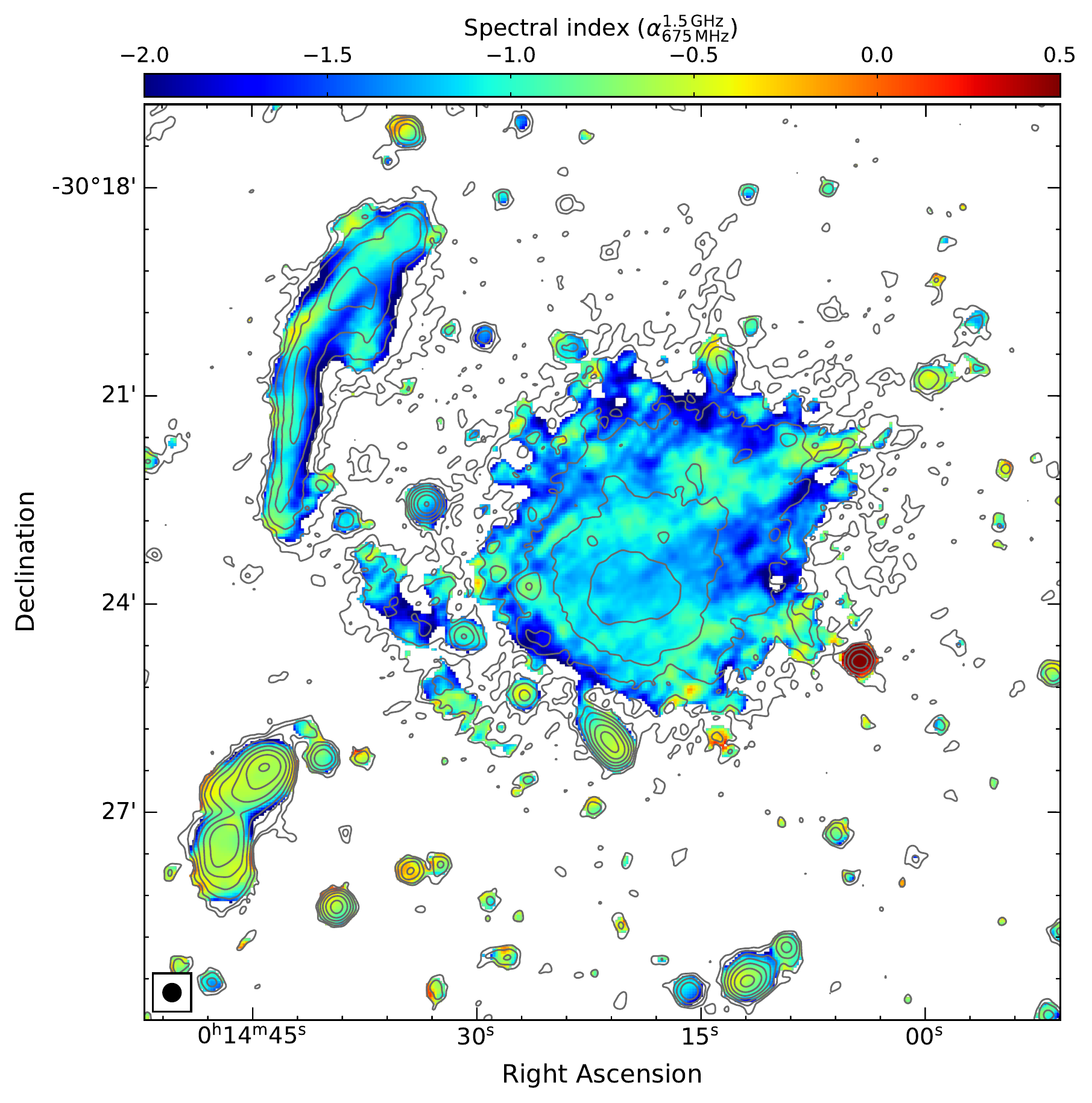}
  \includegraphics[width=0.48\textwidth]{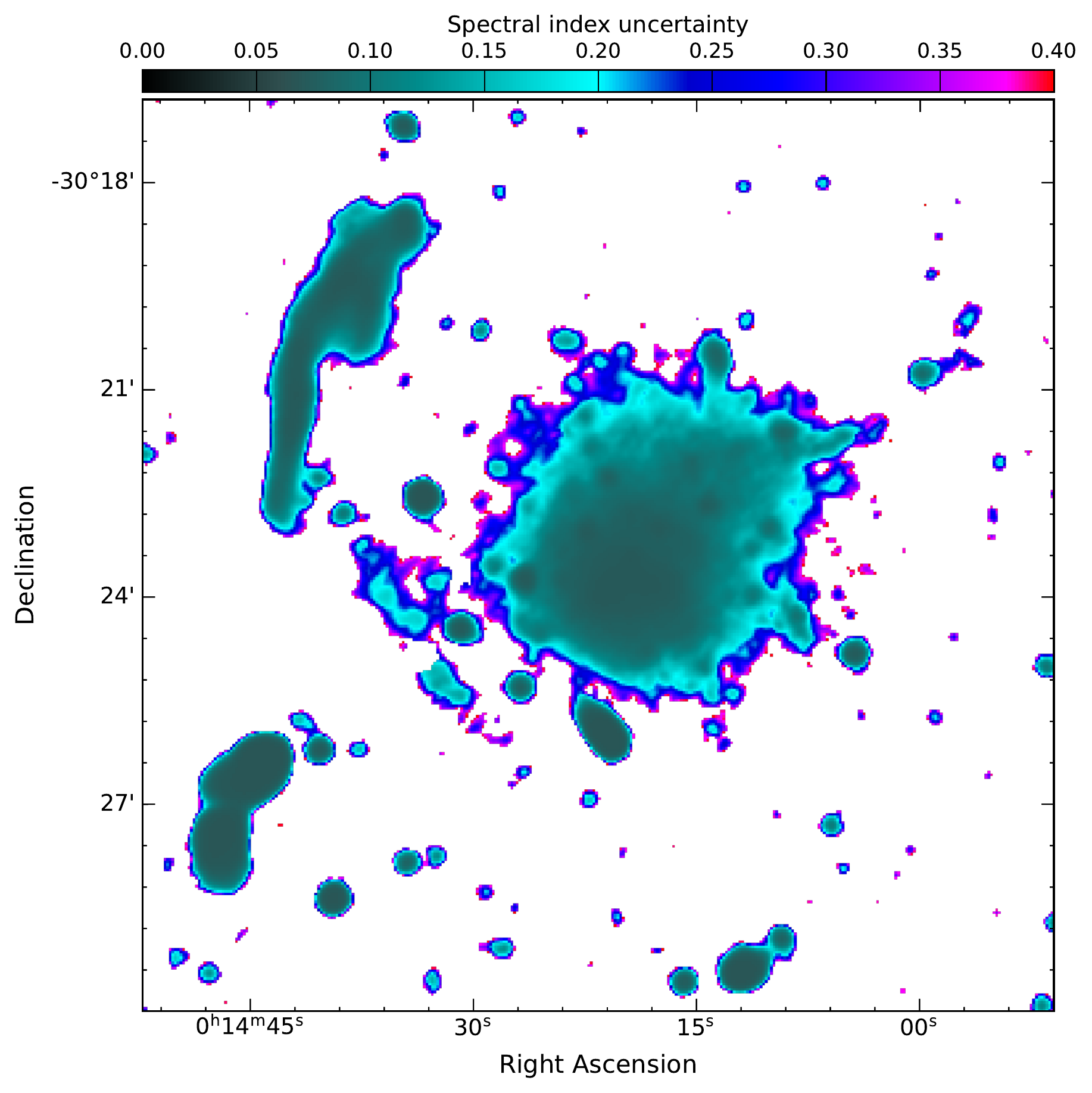}\\
  \vspace{0.5cm}
\includegraphics[width=0.48\textwidth]{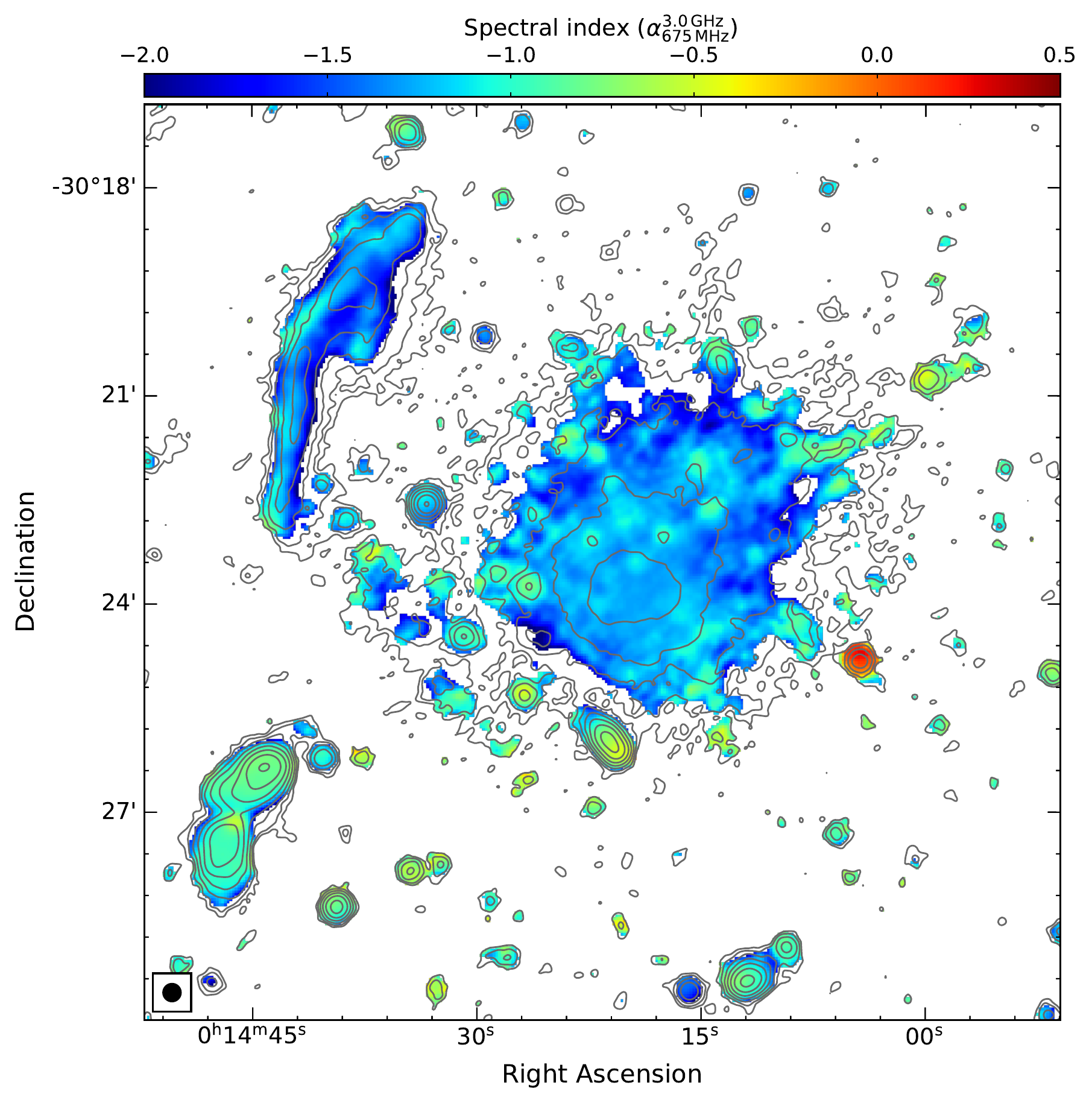} 
  \includegraphics[width=0.48\textwidth]{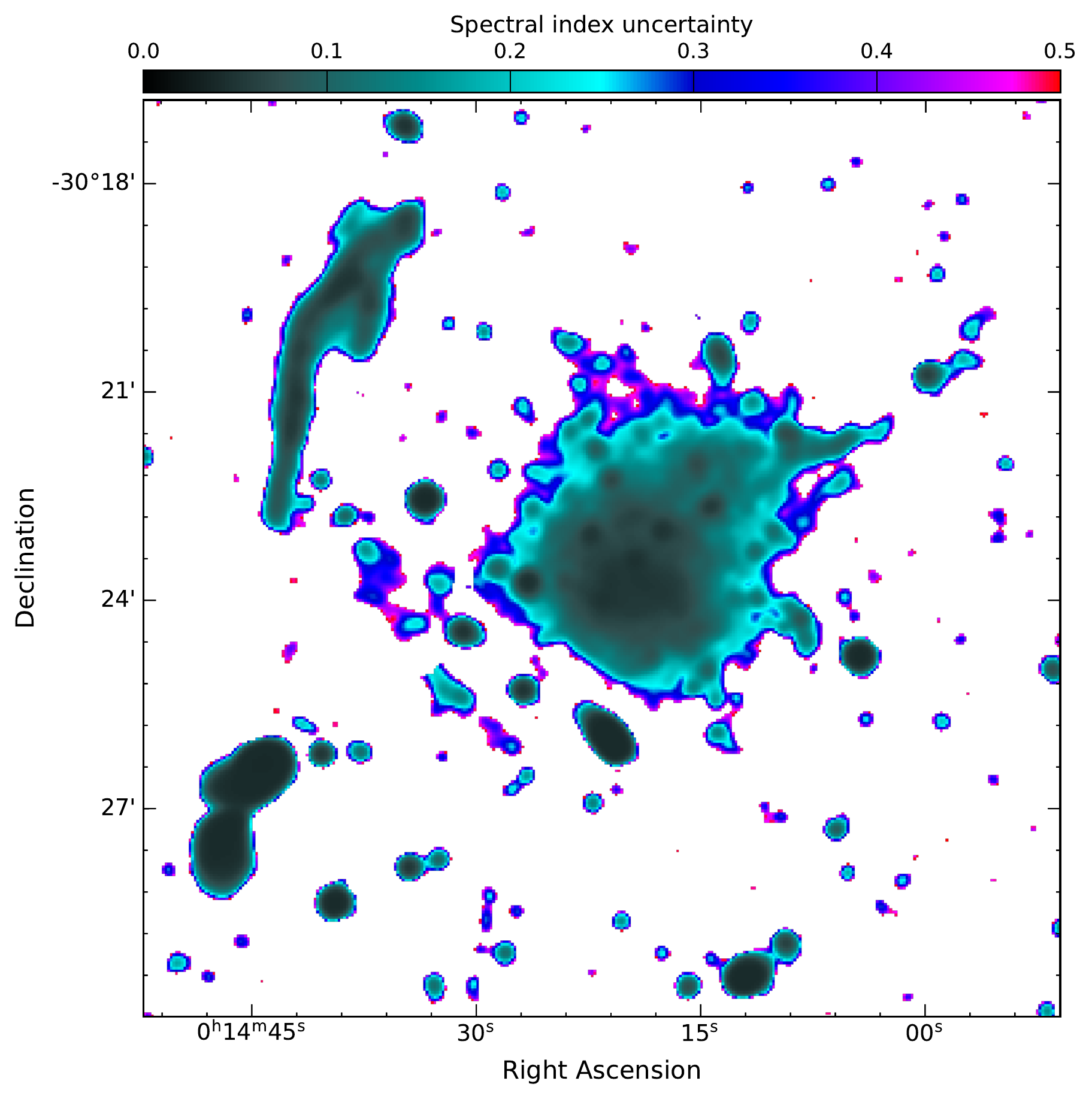}
  \caption{\textit{Left}: Spectral index maps of the halo between 675\,MHz and 1.5\,GHz (top) and 675\,MHz and 3.0\,GHz (bottom) at $15\arcsec$ resolution. The image properties are given in Table\,\ref{imaging}, IM3, IM7, and IM13. Contour levels are drawn at $[1,2,4,8,\dots]\,\times\,3.5\,\sigma_{{\rm{ rms}}}$, and are from the uGMRT Band\,4 image. These maps show the presence of steeper spectral indices in outermost regions of the halo, in particular to the north. \textit{Right}: Corresponding spectral index uncertainty.}
\label{fig3}
\end{figure*}

%%%%%%%%%%%%%%%%%%%%%%%%%%%%%%%%%%%%%%%%%%%%%%%%%%%%%%%%%%%%%%%%%

%%%%%%%%%%%%%%%%%%%%%%%%%%%%%%%%%%%%%%%%%%%%%%%%%%%%%%%%%%%%%%%%%
%Fig. 10- histograms 

\begin{figure*}
\centering
 \includegraphics[width=0.47\textwidth]{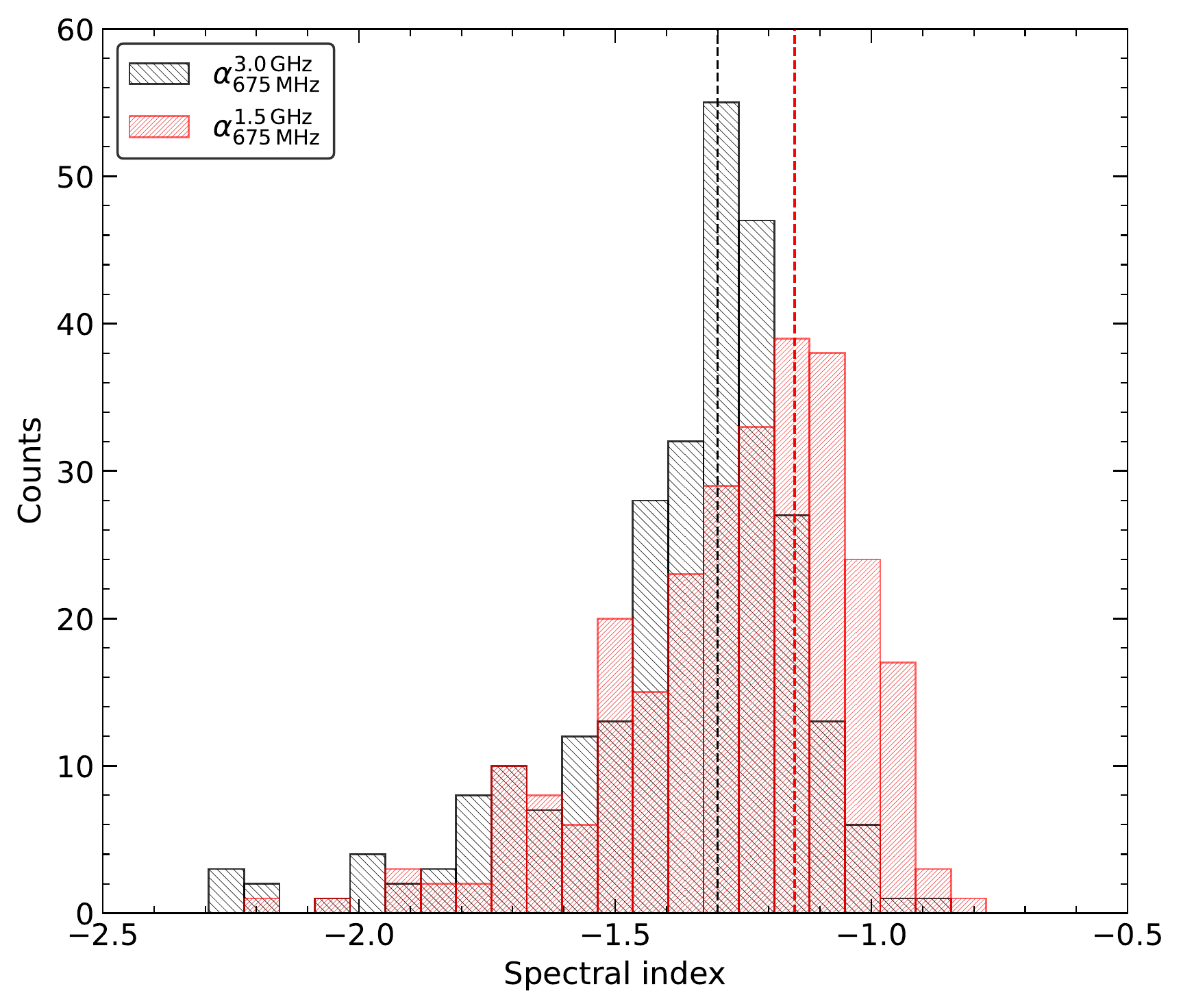}
\includegraphics[width=0.47\textwidth]{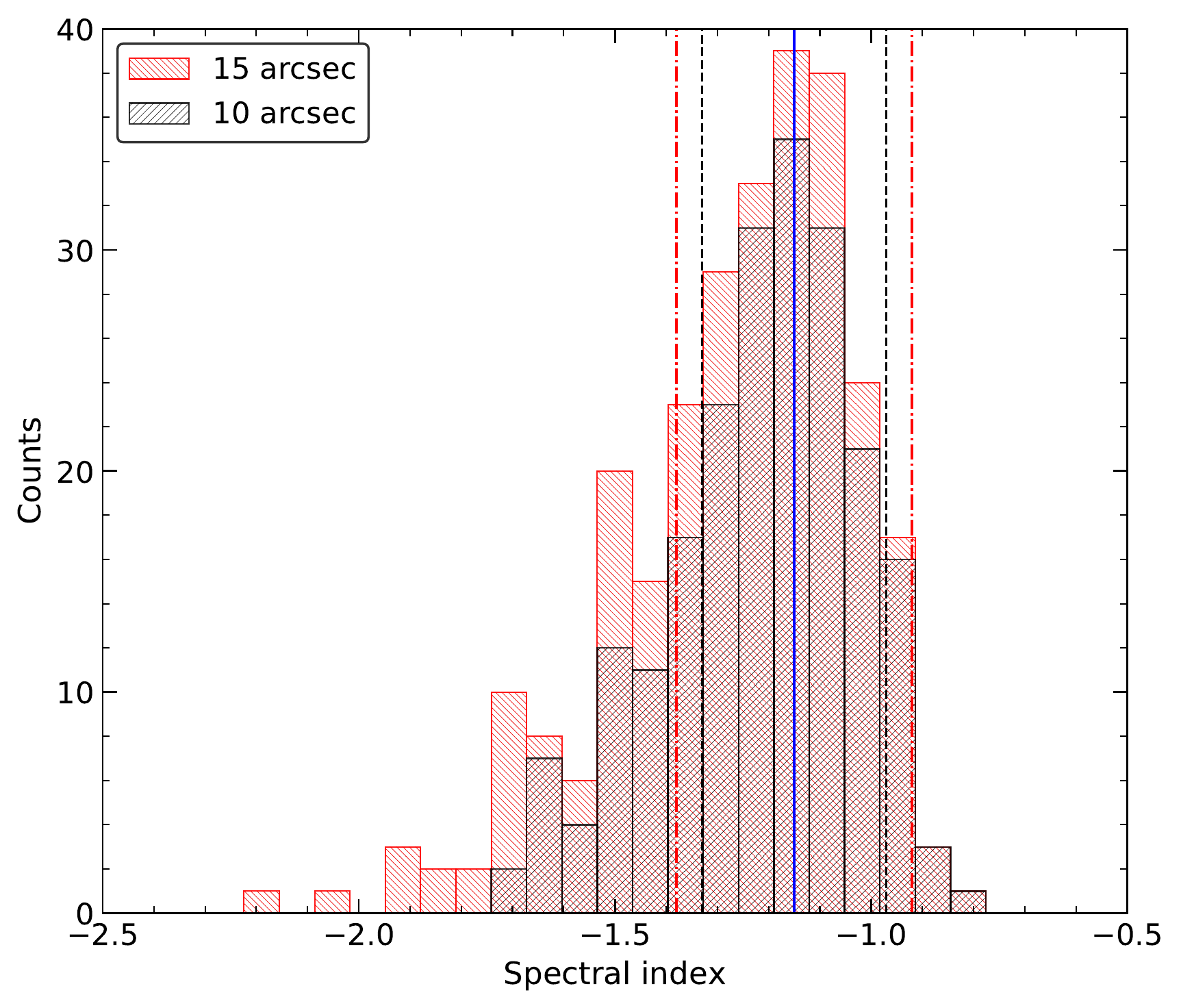} 
  \caption{\textit{Left}: Histogram of the spectral index distribution across the halo in Abell 2744 for different sets of frequencies. The low frequency values were obtained between 675\,MHz and 1.5\,GHz (in red) while the other one between 675\,MHz and 3\,GHz (in black). The spectral indices were extracted from $15\arcsec$ boxes, corresponding to a physical size of about 67\,kpc. The dashed lines represent the median spectral index values. \textit{Right}: Histogram of spectral index values obtained from $10\arcsec$ (black) and $15\arcsec$ (red) resolution radio maps created between 675\,MHz and 1.5\,GHz. The plot suggest that the distribution of the mean spectral index and its standard deviation have a little impact when adopting different resolutions. The solid blue line represents the median value, $ \langle \alpha \rangle =-1.15$, at both $10\arcsec$ and $15\arcsec$ resolution. The dash-dotted (red) and dashed (black) lines represent the standard deviation around the median values at $10\arcsec$ ($\sigma_{10\arcsec}=0.19$) and $15\arcsec$ ($\sigma_{15\arcsec}=0.24$), respectively.}
\label{halo_index}
\end{figure*}

%%%%%%%%%%%%%%%%%%%%%%%%%%%%%%%%%%%%%%%%%%%%%%%%%%%%%%%%%%%%%%%%% 

Recent wideband studies of relics in the 1RXS\,J0603+4214 (aka ``the Toothbrush'')  and MACS\,J0717.5+3745 show that these two relics and subregions show the same broadband power-law spectra with an integrated spectral index of $-1.16\pm0.04$ \citep{Rajpurohit2020a,Rajpurohit2020b,Rajpurohit2021a}. The comparison of these observations with simulations outlined that the integrated radio spectra of relics are dominated by the high end of the distribution of Mach numbers in the shock surface \citep{Wittor2019,Rajpurohit2020a,Paola2021,Wittor2021}. The integrated spectral index of $-1.17\pm0.03$ for the main relic in Abell\,2744 is consistent with that found for the Toothbrush and  MACS\,J0717.5+3745 relics. We emphasize that the radio power of the Toothbrush and MACS\,J0717.5+3745 relics are about 26 and 6 times, respectively, higher than the R1 relic in Abell\,2744. This also suggests that radio-derived Mach numbers may be always dominated by high Mach number shocks regardless of the radio power of relics.  A recent study by \cite{Wittor2021} further strengthen this.  

\subsubsection{Acceleration efficiency for R1 }

For the majority of radio relics, the acceleration of electrons from the thermal pool requires unrealistically large acceleration efficiencies to explain the observed high radio powers of relics, in particular, if X-ray observations does actually provide a reliable measure for the actual strength, see \citet{Botteon2020a} and see \citet{Wittor2021} for a discussion of the reliability of shock strengths derived from X-ray observations. Such efficiencies are hard to reconcile with DSA and it has been argued that the existence of a pre-existing electron population may have been re-accelerated by an earlier episode of shock acceleration, thus increasing the radio power \citep{Kang2012,Pinzke2013}. There are only a few relics where the observed radio brightness can be well explained by acceleration of electrons from the thermal pool via DSA, for example, the relic in Abell\,2249 \citep{Locatelli2020}. 

To explain the observed radio power of the main relic R1, \cite{Eckert2016} estimated the efficiency of shock acceleration from DSA of thermal particles. Based on the earlier X-ray and radio-derived Mach numbers of $\mathcal{M}_{\rm X-ray}=1.7$ and  $\mathcal{M}_{\rm radio}=2.1$, respectively, they concluded that the efficiency needed to account for the observed emission was a factor $\sim 10^2-10^3$ higher than what such weak shocks could achieve based on DSA. 

\cite{Hoeft2007} derived the relation between the radio power $L_{\nu, \rm obs}$ and the physical properties of the downstream plasma for a stationary radio shock and CRe injected with a power-law energy distribution at the shock front
\begin{eqnarray}  
  L_{\nu, \rm obs} 
  & = &  
  C \cdot   
  \xi_{e} \cdot
  \frac{A}{\rm Mpc^2} \cdot 
  \frac{n_{\rm e,\,d}}{\rm 10^{-4} \,cm^{-3}} \cdot 
  \left( \frac{T_{\rm d}}{\rm 7\,keV} \right)^{\frac{3}{2}} \cdot 
  \left( \frac{\nu}{1.4 \, \rm GHz} \right)^{\alpha}  
  \nonumber
  \\
  && \quad 
  \cdot
  \left( \frac{B}{\rm \mu G} \right)^{-1 - {\alpha}} \cdot
  \left( \frac{B_{\rm CMB}^2}{B^2} + 1 \right)^{-1} \cdot
  \Psi({\cal M},T_{\rm d}),
  \label{eff}
\end{eqnarray} 
where $\xi_{e}$ is the fraction of the kinetic energy dissipated at the shock front channeled into the acceleration of electrons from the thermal pool,  $A$ is the surface area of the relic, $n_{\rm e,\,d}$ is the downstream electron density, $T_{\rm d}$ is the downstream electron temperature, $B$ is magnetic field strength in the emitting region, $B_{\rm CMB}$ is the field strength equivalent to the Cosmic Microwave Background energy density, $ B_{\rm CMB} =  3.24 \,(1+z)^{2}\,\mu G$, and $\Psi({\cal M}, T_{\rm d})$ comprises all Mach number dependencies, most importantly the slope of the electron energy spectrum at injection. We note that this function depends weakly on the downstream temperature. We have revised the constant $C$. Applying several improvements to the formalism, most importantly, a power law is assumed for the electron momentum distribution (instead of the energy distribution as assumed in the original work), the isotropic pitch angle distribution is explicitly integrated, and the chemical mixture of hydrogen and helium in the ICM is taken into account, we found  $\rm C = 1.28 \times 10^{27} \, W \, Hz^{-1}$.  To compute the expected radio luminosity, the electron density and temperature can be estimated from X-ray observations. Following \citet{Eckert2016}, we adopt for the downstream region of R1 the properties $n_{\rm e,\,d}=3\times10^{-4}\,\rm cm^{-3}$ and $T_{\rm d}=12$\,keV. For the area of the shock front we adopt $\pi/4\,{\rm LLS}^2 = 1.8\,\rm Mpc^2$. Other parameters used are $\alpha = -1.17$, $ B_{\rm CMB}=5.54\,\mu G$, and $\nu\rm  = 1.5\,GHz$. The integrated spectral index corresponds to a Mach number of 3.57, this leads to $\Psi({\cal M},T_{\rm d})=0.193$.

Unfortunately, neither the acceleration efficiency nor the magnetic field strength in the emitting volume can be tightly constrained from observations. However, since the radio power at 1.5\,GHz  is known from observations (see Table\,\ref{Tabel:Tabel2}) the necessary acceleration efficiency can be given as function of the magnetic field strength, see the right panel of Fig.\,\ref{figcc}.  

For a magnetic field strength of a few $\mu \rm G$ or stronger, the energy fraction, $\xi_e$, channeled into the acceleration of electrons, could have a value of 1\,\% or lower (see the right panel of Fig.\,\ref{figcc}), which might possibly be achieved by accelerating electrons from the thermal pool via DSA. The relic R1 in Abell\,2744 does therefore not require a re-acceleration scenario if a large fraction of the shock front has a strength as obtained from the integrated radio spectrum and if the magnetic field is sufficiently strong. It should be noted, however, that the relic R1 shows a low luminosity; it is in this regard more similar to the relic in Abell\,2249 \citep{Locatelli2020} than to brighter relics.

%##########################
\subsection{Spectral index and curvature maps}
\label{index_maps}
%##########################

\subsubsection{Relic R1}

Radio relics are expected to show a clear spectral index gradient toward the cluster center \citep{vanWeeren2010,Gennaro2018,Rajpurohit2020a}. Such a spectral index gradient reflect the aging of the relativistic electron population while the shock front propagates outwards \citep{vanWeeren2010} or a variation of the Mach number across the shock surface \citep{Skillman2013}. 

The main relic in Abell 2744 is known to show a clear spectral index gradient \citep{Orru2007,Pearce2017,Paul2019}. Our new uGMRT Band\,4 data allow us to create an spectral index map of the relic R1 between 675\,MHz and 1.5\,GHz at higher resolution and with better sensitivity.  The right panel of Fig.\,\ref{fig6a} displays the $10\arcsec$ spectral index map of the relic. A clear spectral gradient toward the cluster center is visible across the entire relic. The injection spectral index at the eastern edge of the relic varies mainly in the range $-0.67\pm0.06$ to $-1.0\pm0.06$ between 675\,MHz and 1.5\,GHz. These values are consistent with those reported by \cite{Paul2019} between 235\,MHz and 610\,MHz but significantly flatter than those found between 1.5\,GHz and 3\,GHz \citep{Pearce2017}.  The difference may be caused by the fact that in  the 1.5 and 3\,GHz spectral index map, the width of the relic is close to the beam size, see \cite{Rajpurohit2018} for a discussion. At low frequencies, the downstream profile typically gets wider, and therefore low frequency spectral index maps are better suited for measuring the injection spectral index \citep{Rajpurohit2018,Rajpurohit2020a}.

For R1, the observed variation in the injection index suggests that there are also regions where the injection index is steeper, which in turn implies the Mach number varies across the shock front. Similar variation in the injection index is reported for the Toothbrush relic \citep[see Fig.\,9 of ][]{Rajpurohit2018} and Sausage relic \citep[see Fig.\,8 of ][]{Gennaro2018}. These variations seem to be consistent with simulations which show that relics are produced by a distribution of Mach numbers \citep{Skillman2013,Wittor2019,Paola2021}. 

 Furthermore, these simulations show that the integrated spectral index, which is measured across the whole relic or its subregions, mirrors the high value tail of the Mach number distribution \citep{Rajpurohit2020a,Paola2021, Wittor2021}. On the other hand, the injection index is measured locally and, hence, it resolves the spectral variations across the relic. The mean injection index measured across the eastern edge of R1 is $-0.90$ implying a shock of Mach number of 2.4 while the integrated spectral index suggest a shock of Mach number 3.6.  We note that similar trends are reported for the Toothbrush, Sausage, and MACSJ0717.5+3745 relics \citep{Hoang2017,Rajpurohit2020a,Rajpurohit2021a}. The comparison of the injection and integrated spectral indices yields that the integrated one is biased toward higher Mach numbers. As the majority of Mach numbers is weaker than the Mach number inferred from the integrated spectral index, the majority of locally measured injection spectral indices is steeper. Therefore, the average injection index is more sensitive to the steeper spectral indices, that are produced by the more numerous weaker shocks.

All merger shock-models predict an increasing curvature in the downstream regions of relics. However, the curvature analysis is available only for five relics: the Toothbrush relic, the eastern relic in 1RXS\,J0603.3$+$4214, the northern relic in CIZA\,J2242.8$+$5301 (aka ``the Sausage"), the southern relic in CIZA\,J2242.8$+$5301, and MACSJ0717.5+3745 \citep{Gennaro2018,Rajpurohit2020a,Rajpurohit2021a}. To check if R1 shows any spectral curvature, we use radio color-color plots \citep{Katz1993,Rudnick1994,vanWeeren2012a,Gennaro2018,Rajpurohit2020a,Rajpurohit2021a}. The radio color-color plots are sensitive to electron populations injected at the shock as well as projection effects, thus provide crucial information about the physical processes operating at the shock. The resulting plot is shown in the left panel of Fig.\,\ref{figcc}. The relic shows a clear negative curvature, as expected from merger shock models.  

Radio relic spectral properties are reported to be affected by projection effects. Projection can be found in the color-color distribution assuming that locally standard models, for example, Jaffe-Perola \citep[JP;][]{Jaffe1973}, continuous injection \citep[CI; ][]{Pacholczyk1970}, and KGJP \citep{Komissarov1994}  can be applied \citep[see][for a detailed discussion]{Rajpurohit2020a}. The color-color plots are also sensitive to the relic's viewing angle \citep{Rajpurohit2021a}. If a relic is observed perfectly edge-on, the spectral shape is expected to follow the JP model, implying that the shock is parallel to the line of sight. Hence, a spectrum of the single spectral age is observed for each line of sight, the case of a perfectly aligned shock front. In contrast, if the relic is inclined along the line of sight, we expect the spectral shape to be consistent with the KGJP model as the spectrum represents a superposition of particle populations of different ages, which are distributed along the line of sight \citep{Rajpurohit2020a}.

In the left panel of Fig.\,\ref{figcc}, we overlay the data points with the JP, CI, and KGJP models, using $\alpha_{\rm inj}=-0.67$. This injection index was adopted because the integrated spectral index of $-1.17$ corresponds to an injection index of about $-0.67$. The observed spectral distribution is at least inconsistent with CI and KGJP models. The overall shape is relatively consistent with the JP model,  suggesting that the shock is very likely aligned parallel to the line of sight. We note that there are no data points in the range $-0.5$ to $-1.2$ between 1.5 and 3.0\,GHz. We emphasize that image resolution may play a critical role as spectral indices are vulnerable to smoothing effects, for example, between 1.5 and 3\,GHz we are not necessarily measuring the actual injection index as the width of the relic may be close to the beam size. Moreover, as the spectral index values are extracted from $10\arcsec$ resolution radio maps, this also steepens the actual spectral index \citep{Rajpurohit2018}.

%%%%%%%%%%%%%%%%%%%%%%%%%%%%%%%%%%%%%%%%%%%%%%%%%%%%%%%%%%%%%%%%%
%Fig. 11: curvature map of the halo

\begin{figure*}
    \centering
    \includegraphics[width=0.48\textwidth]{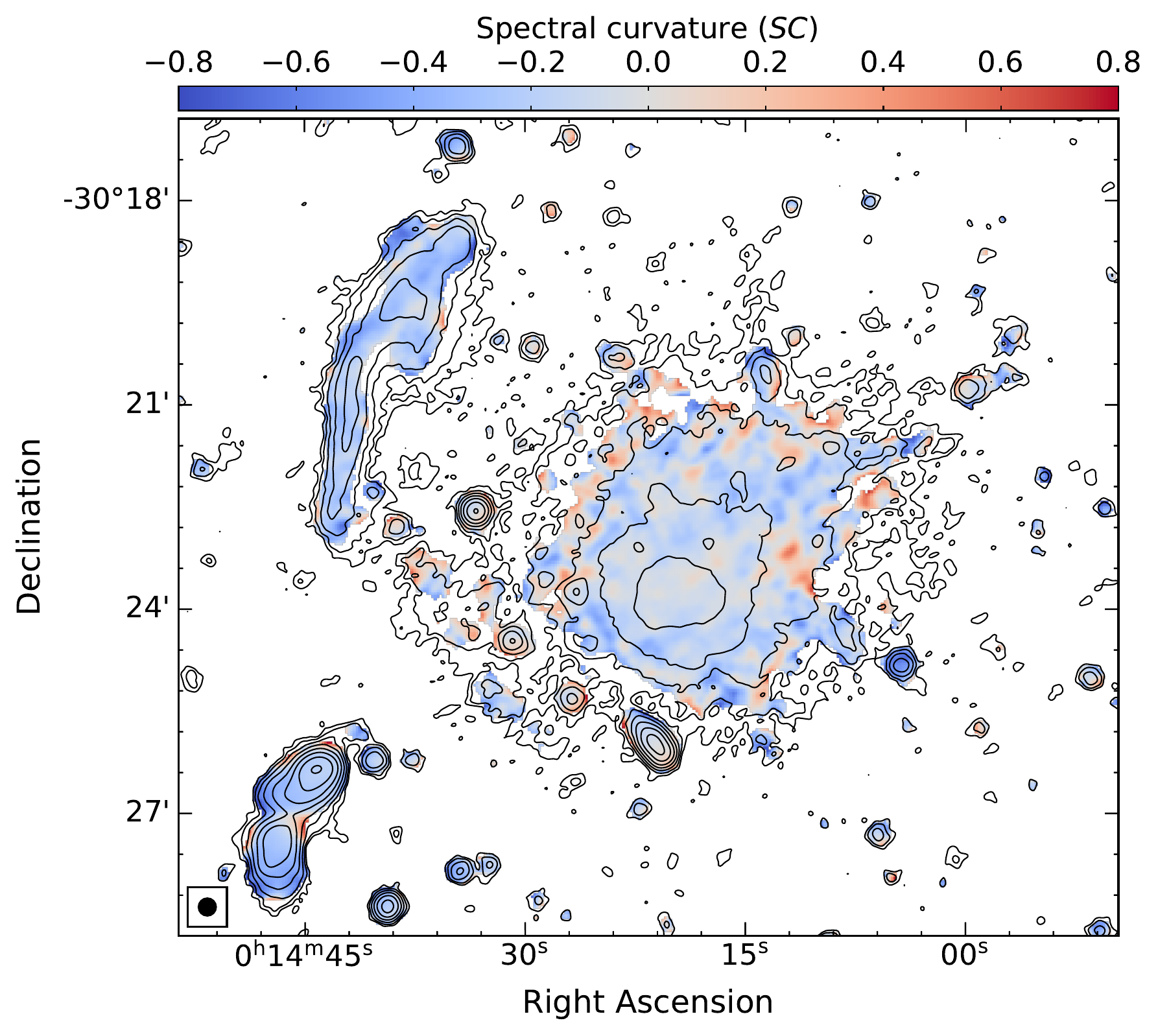} 
     \includegraphics[width=0.48\textwidth]{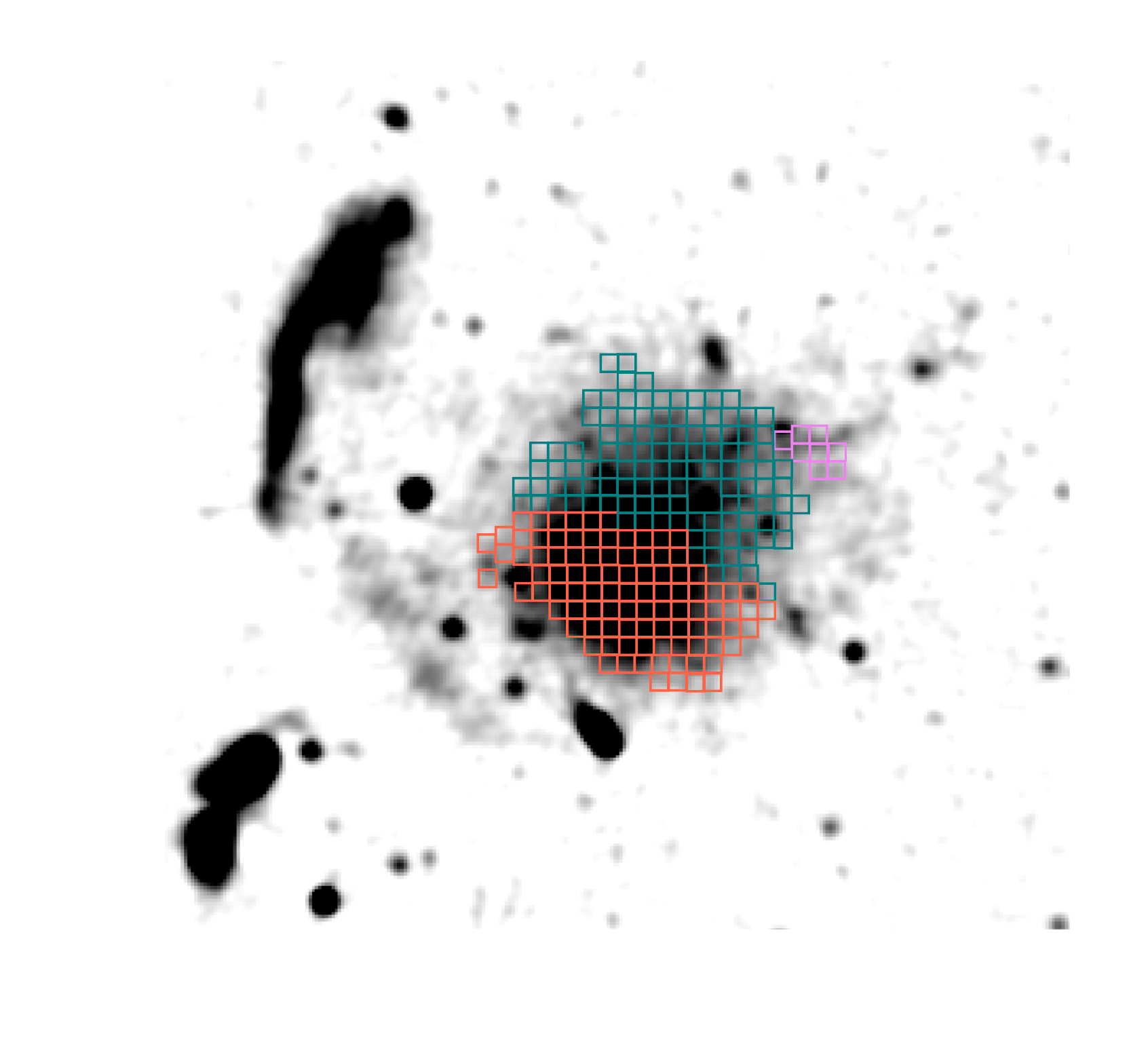} 
 \caption{\textit{Left}: Three frequency spectral curvature map of the halo at $15\arcsec$ resolution, created using 675\,MHz, 1.5\,GHz, and 3\,GHz radio maps.  The image properties are given in Table\,\ref{imaging}, IM3, IM7, and IM13. The halo shows a hint of spectral curvature. The SC is negative for a convex spectrum. Contour levels are drawn at $[1, 2, 4, 8, . . . ] \times 3.5\,\sigma_{\rm rms}$, and are from the uGMRT Band\,4 image.  \textit{Right}: Regions where the radio spectral index values were extracted between 675\,MHz and 3\,GHz. Each box has a width of $15\arcsec$ corresponding to a physical size of about 67\,kpc. The same regions are used to obtained $\alpha-I_{\rm X}$ correlation in the left panel of  Fig.\,\ref{figindex}.}
\label{halocc}
\end{figure*}  
%%%%%%%%%%%%%%%%%%%%%%%%%%%%%%%%%%%%%%%%%%%%%%%%%%%%%%%%%%%%%%%%%

For the main relic R1, from the spatially resolved $10\arcsec$ spectral index map between 675\,MHz and 1.5\,GHz, we measure an injection index that in some regions is as flat as $-0.67$. The derived value corresponds to a shock of Mach number 3.6 (using Eq.\,\ref{int_inj}). We measured an integrated spectral index of $-1.17\pm0.3$ for the main relic, suggesting a shock of Mach number $\mathcal{M}=3.6_{-0.2}^{+0.3}$. This value is in agreement with the one derived from the injection spectral index. We do not find any discrepancy in the Mach numbers obtained from the radio integrated and injection spectral indices for R1.

Using \textit{XMM-Newton} data, \cite{Eckert2016} reported the detection of a weak shock front with $\mathcal{M}={1.7^{+0.5}_{-0.3}}$ via the surface brightness jump at the eastern edge of the R1 relic. However, based on the X-ray temperature jump \cite{Hattori2017} found a significantly higher Mach number shock, namely $\mathcal{M}=3.7\pm0.4$.  They speculated that the Mach number obtained from the temperature jump is associated with the strong shock heating of the ICM along an X-ray filament at that location. There is clearly a large discrepancy in the shock Mach number derived from the X-ray surface brightness and temperature jump. Interestingly, the X-ray derived Mach number from the temperature jump is consistent with the radio-derived Mach number. If the Mach number obtained from the temperature jump represents the true shock strength, R1 will be among the few relics where the radio and X-ray Mach numbers agree remarkably well with each other, namely $\mathcal{M}=3.6$.      
  
%##########################
\subsubsection{Halo}
\label{halo_maps}
%##########################
The observed patterns of spectral index fluctuations provide important information about the physical processes in halos and the balance of acceleration processes as well as synchrotron and inverse Compton losses. 

In \cite{Pearce2017}, high-frequency spectral index maps (at $15\arcsec$ and $30\arcsec$ resolutions) of the halo were derived using the VLA 1.5 and 3\,GHz data. Recently, \cite{Paul2019} presented low-frequency (235-610 MHz) spectral index maps of the halo at $25\arcsec$ resolution. The new uGMRT Band\,4 data allow us to study the spectral index distribution between 675\,MHz and 3\,GHz with improved sensitivity and better resolution.  

We created maps for two frequency sets: $\rm 675\,MHz{-}1.5\,GHz$ and $\rm 675\,MHz{-}3\,GHz$. The resulting maps are shown in Fig. \,\ref{fig3}. It is evident that the halo shows the presence of localized regions in which the spectral index is significantly different from the average. This is generally expected for turbulent re-accelerations models \citep{Brunetti2014}. In the inner regions of the halo and toward subcluster 5, the spectral index is relatively flat. In addition, the regions with flat spectral indices appear to follow the Bullet-like X-ray morphology. Regions with flatter spectral indices are indicative of the presence of more energetic particles or a higher local magnetic field strength. They are generally influenced by ongoing mergers. The ICM temperature in innermost regions also show  variations from about  4\,keV to 10\,keV \citep{Pearce2017}. 

 %%%%%%%%%%%%%%%%%%%%%%%%%%%%%%%%%%%%%%%%%%%%%%%%%%%%%%%%%%%%%%%%%
% Fig. 12 - color-color plot of the halo
\begin{figure*}
    \centering
    \includegraphics[width=0.47\textwidth]{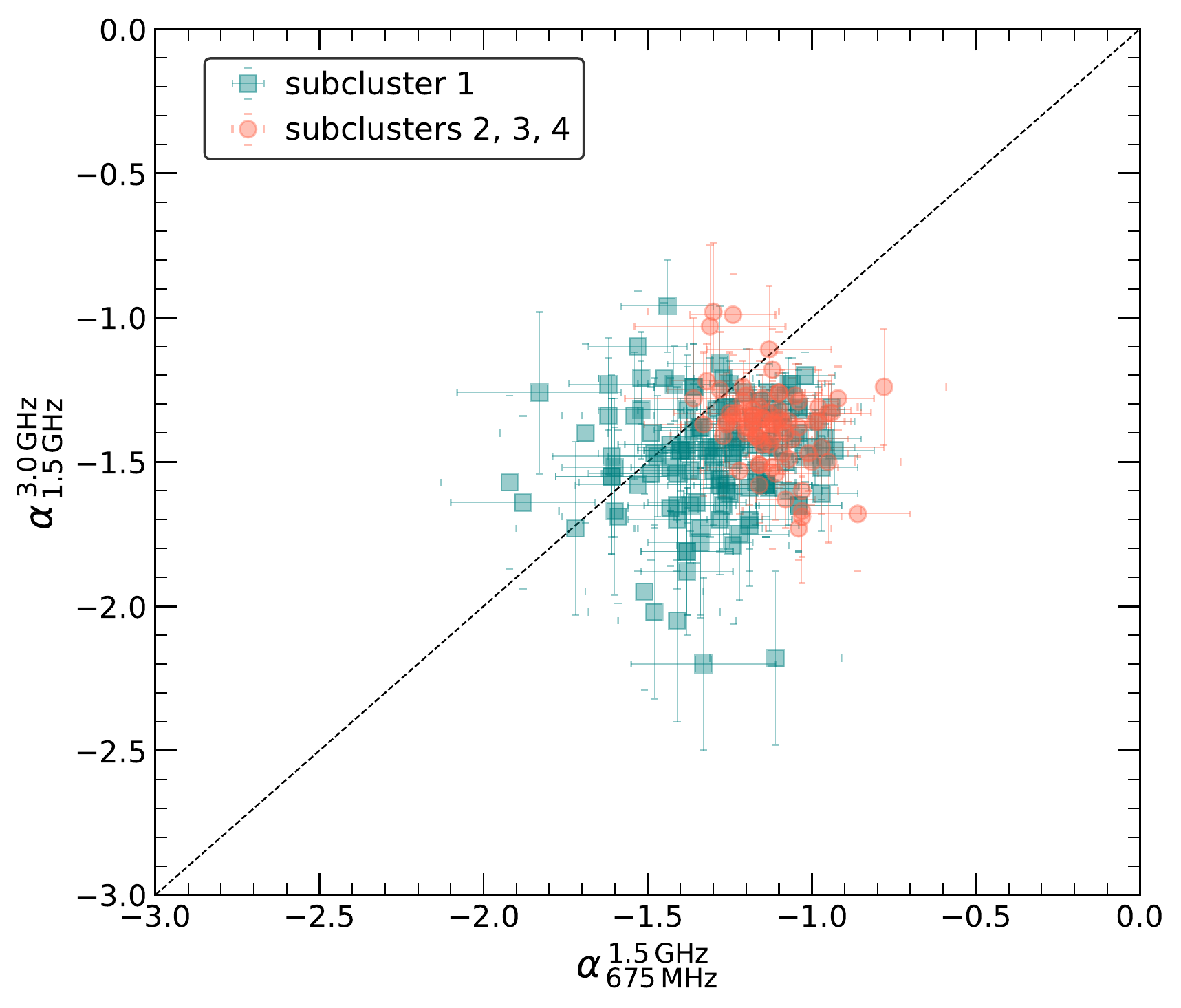} 
    \includegraphics[width=0.51\textwidth]{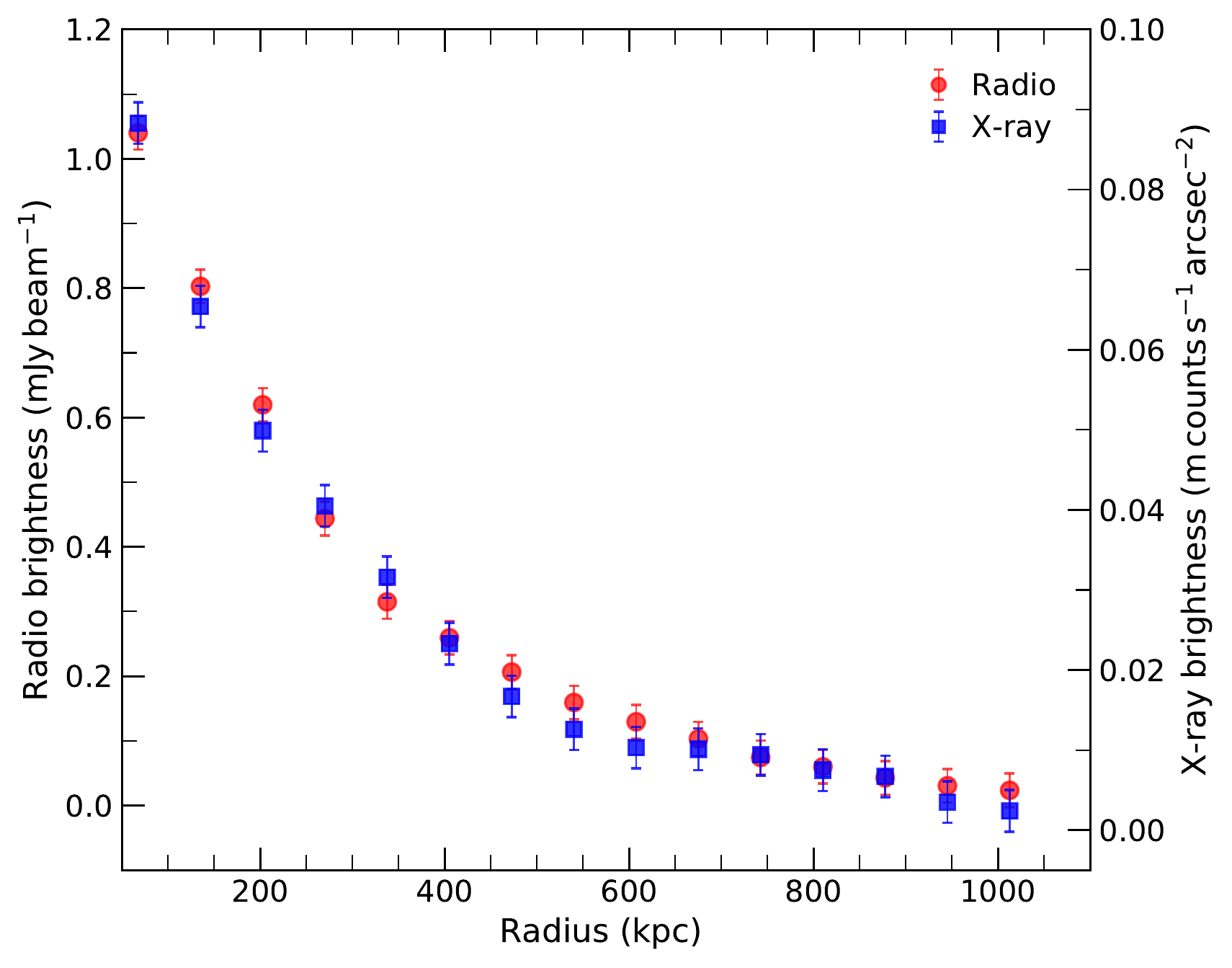}
 \caption{\textit{Left}: Radio color-color plot of the halo, showing a complex spectral curvature distribution. The orange data points are extracted from the southern part of the halo while turquoise one from the northern part. The curvature distribution is seems different in the northern and southern parts of the halo. Regions used for extracting the spectral index values are are shown in the right panel of Fig.\,\ref{halocc}. \textit{Right}: Comparison between the radio (red) and the X-ray (blue) surface brightness profiles for the Abell\,2744 halo. The X-radio and X-ray profiles are remarkably similar.}
\label{halocc1}
\end{figure*}  
%%%%%%%%%%%%%%%%%%%%%%%%%%%%%%%%%%%%%%%%%%%%%%%%%%%%%%%%%%%%%%%%%

Moreover, the spectral index gets steeper in the outermost regions, in particular in the northern part of the halo. According to the re-accelerations models, such steeper indices are expected in the simplified situation where the acceleration rate is constant due to the decline in the magnetic field with radius for $B^2<<B_{\rm CMB}^2$  \citep{Brunetti2001}. The spectral index trends are similar for indices derived between 675\,MHz and 1.5\,GHz and between 675\,MHz and 3\,GHz, see Fig. \,\ref{fig3}. 

\cite{Pearce2017} studied the spectral variations across the radio halo  between 1.5 and 3\,GHz. They found that the spectral distribution is more or less uniform across the halo with $ \langle \alpha \rangle = -1.37 $ and standard deviation of $\sigma= 0.28$. In the left panel of Fig. \,\ref{halo_index}, we show the histogram of the spectral index distribution from 675\,MHz-1.5\,GHz and 675\,MHz-3\,GHz, extracted from $15\arcsec$ resolution maps. Between 675\,MHz-1.5\,GHz, the spectral index distribution shows the median spectral index of $\langle\alpha_{\rm 675\,MHz}^{\rm 1.5\,GHz}\rangle=-1.15$ and a standard derivation of $\sigma_{15\arcsec}=0.18$. This value is consistent with the integrated spectral index of the halo, namely $-1.14\pm0.05$. Between 675\,MHz and 3\,GHz, we measure a median value of $\langle\alpha_{\rm 675\,MHz}^{\rm 3.0\,GHz}\rangle=-1.31$ and $\sigma_{15\arcsec}=0.24$, which is comparable with those reported by \cite{Pearce2017} between 1.5 and 3\,GHz.

%%%%%%%%%%%%%%%%%%%%%%%%%%%%%%%%%%%%%%%%%%%%%%%%%%%%%%%%%%%%%%%%%
%Fig. 13- radio-Xray correlation

\begin{figure*}
    \centering
     \includegraphics[width=1.00\textwidth]{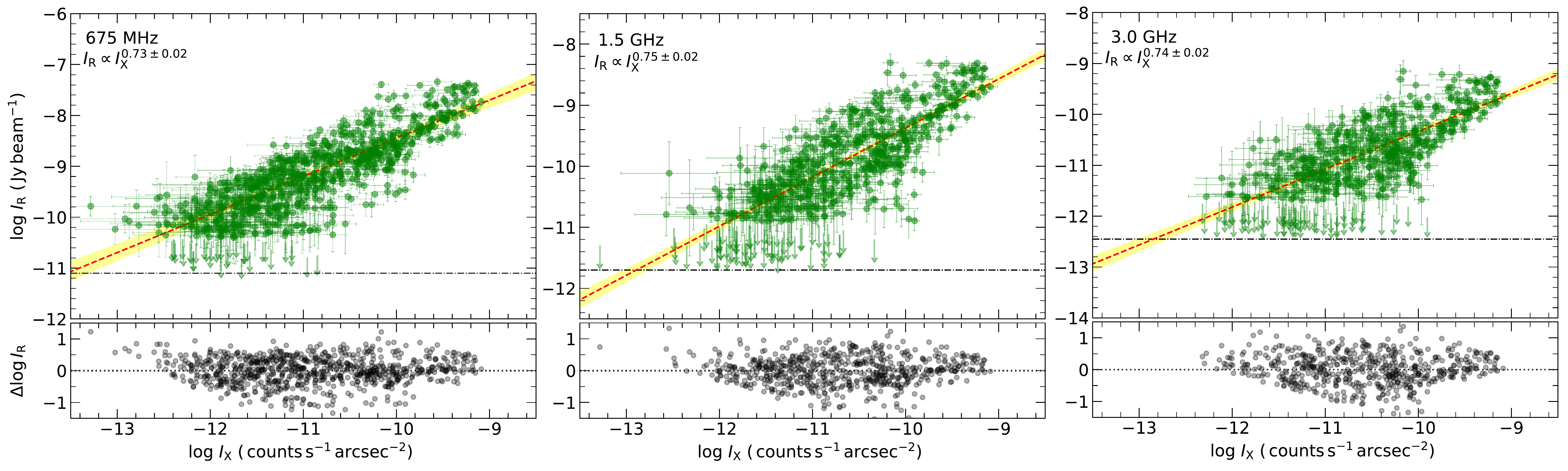}
           \vspace{-0.3cm}
 \caption{\textit{Left}: $I_{\rm R}{-}I_{\rm X}$ relation of the halo in Abell\,2744, extracted in square boxes with width of $10\arcsec$ (about 45\,kpc). The X-ray surface brightness is extracted from the \textit{Chandra} ($0.5-2.0$\,keV band) image smoothed with a Gaussian FWHM of $3\arcsec$. The radio surface brightness is extracted from radio maps at $10\arcsec$ resolution. The {\tt Linmix} best-fit relations are indicated by red dashed lines. Green circles depict cells where the radio and X-ray surface brightness is above $3\sigma$ level. The upper limits (arrows) represent cells with data points below $2\sigma$ radio noise level. The horizontal black dash-dotted lines indicate the $1\sigma$ in the radio maps. Yellow show samples from the posterior distribution. The best-fit are reported with the corresponding 95\% confidence regions. The lower panel shows the residuals of log$\, I_{\rm R}$ and log$\,I_{\rm X}$ with respect  to the {\tt Linmix} best fit line. The radio brightness strongly correlates with the X-ray at all three frequencies. The best-fitting slopes at 675\,MHz, 1.5\,GHz and 3\,GHz are $0.73\pm0.02$, $0.75\pm0.02$, and $0.74\pm0.02$, respectively. }
\label{fig4}
\end{figure*}
%%%%%%%%%%%%%%%%%%%%%%%%%%%%%%%%%%%%%%%%%%%%%%%%%%%%%%%%%%%%%%%%% 

 At $15\arcsec$, we measure the median spectral index uncertainty  of 0.15 and 0.10 between 675\,MHz and 1.5\,GHz and 675\,MHz and 3.0\,GHz, respectively.  If the variations in the spectral index are the result of measurement uncertainties, the median uncertainty is expected to be comparable to the standard deviation. The median error between 675\,MHz-3.0\,GHz is about 2.4 times the standard deviation, hinting at a small-scale spectral index fluctuation across the halo as found for radio halos in Abell 2255 \citep{Botteon2020}, Abell 520 \citep{Vacca2014,Hoang2019}, and MACS\,J0717.5$+$3745 \citep{Rajpurohit2021b}. In the latter case, it was suggested that the large scatter around the average spectral index is due to strong inverse Compton (IC) losses because of the high redshift of that cluster. At high redshift, IC losses are strong and the electron acceleration could produce synchrotron emission preferentially in regions with $B\sim B_{\rm CMB}$, whereas the acceleration process is quenched where turbulence is lower. We emphasize that the other two clusters are located at a relatively low redshift. The fluctuations in the spectral index across the halo are expected in turbulent re-acceleration models caused by magnetic field variations and different (re-)acceleration efficiency within the halo \citep[e.g.,][]{Brunetti2001,Petrosian2001,Brunetti2007}. In contrast, the secondary models predict a uniform spectral index distribution due to continuous generation of  relativistic electrons through the collision between relativistic protons and thermal protons in the ICM \citep[e.g.,][]{Pfrommer2008,Ensslin2011}.

 %%%%%%%%%%%%%%%%%%%%%%%%%%%%%%%%%%%%%%%%%%%%%%%%%%%%%%%%%%%%%%%%%
% Table 3: IX-IR correlation bets-fit parameters  
%%%%%%%%%%%%%%%%%%%%%%%%%%%%%%%%%%%%%%%%%%%%%%%%%%%%%%%%%%%%%%%%%

\setlength{\tabcolsep}{8pt}
\begin{table*}
\caption{{\tt Linmix} fitting slopes and Spearman ($r_{s}$) and Pearson ($r_{p}$) correlation coefficients of the data for Figs.\,\ref{fig4} and \ref{fig4_sub}.}
\centering
\begin{threeparttable} 
\begin{tabular}{ c  c  c  c c c  c  c  c  c }% c | c}
 \hline  \hline  
\multirow{1}{*}{} & \multirow{1}{*}{$\nu$} &  \multicolumn{4}{c|}{$3\sigma$} & \multicolumn{4}{c}{$2\sigma$}\\
 \cline{2-10}
&& $b$ &   $\sigma_{\rm int}$ &   $r_{s}$   & $r_{p}$ & $b$ & $\sigma_{\rm int}$ & $r_{s}$ & $r_{p}$\\
  \hline  
&675\,MHz& $0.67\pm0.02$ & $0.10\pm0.01$ & 0.86 & 0.83 & $0.73\pm0.02$ &$0.13\pm0.01$ &0.87&0.85 \\ 
Halo&1.5\,GHz & $0.71\pm0.03$&$0.13\pm0.01$&0.81&0.77&$0.75\pm0.02$&$0.15\pm0.01$&0.84&0.78 \\ 
&3.0\,GHz& $0.66\pm0.03$&$0.12\pm0.01$&0.77&0.75&$0.74\pm0.02$&$0.16\pm0.02$&0.81&0.77 \\
  \hline  
&$675\,\rm MHz$& $0.53\pm0.02$& $0.05\pm0.01$& $0.82$& $0.78$& $0.65\pm0.02$& $0.08\pm0.01$& $0.86$& $0.78$\\
Northern&$1.5\,\rm GHz$& $0.54\pm0.04$& $0.07\pm0.01$& $0.72$& $0.71$& $0.68\pm0.04$& $0.11\pm0.01$& $0.77$& $0.81$\\ 
&$3.0\,\rm GHz$& $0.42\pm0.04$& $0.05\pm0.01$& $0.82$& $0.71$& $0.60\pm0.04$& $0.10\pm0.01$& $0.75$& $0.64$\\ 
\hline
&$675\,\rm MHz$& $0.72\pm0.02$& $0.07\pm0.01$& $0.92$& $0.90$& $0.77\pm0.02$& $0.08\pm0.01$& $0.94$& $0.90$\\
Southern&$1.5\,\rm GHz$& $0.82\pm0.02$& $0.05\pm0.01$& $0.88$& $0.75$& $0.82\pm0.02$& $0.10\pm0.01$& $0.92$& $0.81$\\ 
&$3.0\,\rm GHz$& $0.70\pm0.06$& $0.09\pm0.01$& $0.84$& $0.80$& $0.75\pm0.03$& $0.10\pm0.01$& $0.90$& $0.85$\\ 
\hline 
\end{tabular}
\end{threeparttable} 
\label{fit}   
\end{table*}

%%%%%%%%%%%%%%%%%%%%%%%%%%%%%%%%%%%%%%%%%%%%%%%%%%%%%%%%%%%%%%%%% 

To check the effect of the cell size on the median spectral index and standard deviation using two different resolutions, $10\arcsec$ and $15\arcsec$. As shown in the right panel of Fig.\,\ref{halo_index}, between 675\,MHz and 1.5\,GHz the mean spectral index remains the same at both resolutions, namely $-1.15$. The standard deviations at $10\arcsec$ and $15\arcsec$ are 0.19 and 0.24, respectively. This indicates that the distribution of the mean value and its standard deviation have little effect when adopting different resolutions.        

We also investigate if the halo shows any sign of spectral curvature. The spectral curvature (SC) map was derived as:
\begin{equation}
{\rm{SC}} = -\alpha_{\rm low}+\alpha_{\rm high}.
\end{equation}
We use the $15\arcsec$ resolution maps at 675\,MHz, 1.5\,GHz and 3\,GHz. The low frequency spectral index map is created between 675\,MHz and 1.5\,GHz and the high frequency map between 1.5\,GHz and 3\,GHz. The SC is negative for a convex spectrum. The curvature map of the halo is shown in the left panel of Fig.\,\ref{halocc}. The halo shows localized regions with a clear curvature. 

We also performed the radio color-color analysis for the halo. The regions used for extracting spectral indices are shown in the right panel of Fig.\,\ref{halocc} and the resulting color-color plot in the left panel of Fig.\,\ref{halocc1}. The majority of data points lie below the power-law line, a clear sign of negative curvature. No curvature is expected in the secondary models of halo formation because cosmic-ray protons do not lose significant amount of energy, therefore the spectral index remains constant producing a power-law without any curvature \citep[e.g., ][]{Dolag2000,Pfrommer2008}. The number of radio halos where spectral curvature has been measured is still extremely limited. Only for two radio halos has such an analysis been carried out, namely MACS\,J0717.5+3745 \citep{Rajpurohit2021b} and 1RXS\,J0603.3+4214 \citep{Rajpurohit2020a}.  The halo in MACS\,J0717.5+3745 shows a significant curvature while no curvature is found in the 1RXS\,J0603.3+4214 halo. The curvature distribution across the halo in Abell\,2744 is quite different from the MACS\,J0717.5+3745 halo. Moreover, the curvature distribution seems different in the northern and southern parts of the halo. 

It is worth noting the difference in the spectral properties of the halos in Abell 2744, 1RXS\,J0603.3+4214, and MACS\,J0717.5+3745. The 1RXS\,J0603.3+4214 halo shows a power-law spectrum, a remarkably uniform spectral index distribution, and no spectral curvature \citep{vanWeeren2016a, Rajpurohit2020a} while the one in MACS\,J0717.5+3745 shows a curved spectrum, strong spectral index fluctuations, steeper spectral indices and curvature in the outermost regions \citep{Rajpurohit2021b}. The present halo, on the other hand shows an overall power-law spectrum with small-scale spectral index fluctuations, steeper spectral index in the outermost regions and a hint of spectral curvature. We conclude that an overall power-law spectrum observed in radio halos does not necessary imply a uniform spectral index distribution. In fact, an overall power-law may be also observed in combination of quite a variety of spectral indices and curvature distributions.

%########################################################################
%  Section 6: X-ray and radio correlations 
%######################################################################## 
\section{X-ray and radio correlations}
\label{IRX}
The similarity between X-ray and radio morphologies of halos in galaxy clusters indicates a connection between the energetics of the nonthermal (magnetic fields and relativistic electrons) and thermal components of the ICM. The morphology of the radio halo is remarkably similar to the X-ray emission in Abell\,2744, Fig.\,\ref{RXimage}

To examine the profiles of the X-ray and radio emission of the halo, we divided the halo into regions of concentric circular annuli centered on the X-ray surface brightness peak (see the right panel of Fig.\,\ref{fig6}). We used the $15\arcsec$ 675\,MHz radio image. The mean surface brightness and standard deviations were estimated within concentric rings. The discrete sources  (both from radio and X-ray images) and the relics R2, R3, and R4 were masked out and were excluded from the analysis. In the right panel of Fig.\,\ref{halocc1} we show the resulting radial profiles. Despite the fact that the X-ray emission is characterized by the presence of several substructure, the radio and X-ray brightness matches quite well, as also reported by \cite{Govoni2001a}. Both, radio and X-ray brightness falls rapidly up to a radii of 600\,kpc and flatter profile is seen at a radius $> 600\,\rm\,kpc$.

%%%%%%%%%%%%%%%%%%%%%%%%%%%%%%%%%%%%%%%%%%%%%%%%%%%%%%%%%%%%%%%%% 
%Fig. 14- radio-Xray correlation subregions

\begin{figure*}
    \centering
     \includegraphics[width=1.00\textwidth]{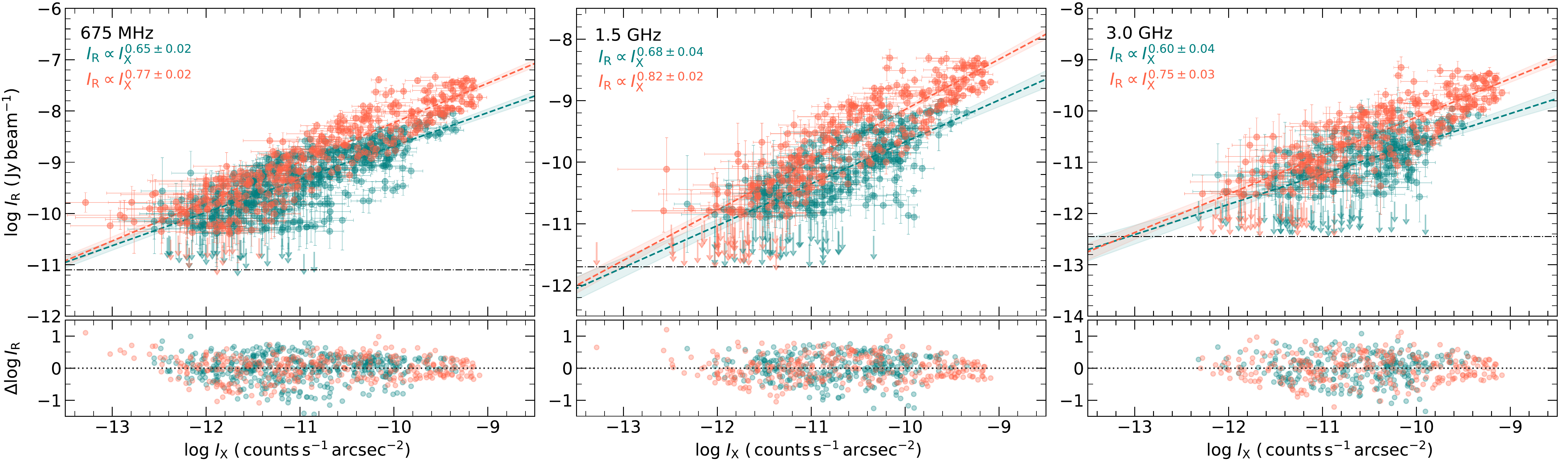}
           \vspace{-0.3cm}
 \caption{\textit{Left}: $I_{\rm R}{-}I_{\rm X}$ relation for the northern (turquoise) and southern (orange) parts of the halo. The upper limits (arrows) represent cells with data points below $2\sigma$ radio noise level. The horizontal black dash-dotted lines indicate the $1\sigma$ in the radio maps. The turquoise and orange dashed lines correspond to the best-fit obtained separately for the northern and southern, respectively, parts of the halo. The best-fits are reported with the corresponding 95\% confidence regions. The correlation slope is indeed different for the northern and southern regions of the halo: the southern part of the halo is steeper than the northern part at all three frequencies.}
\label{fig4_sub}
\end{figure*}

%##########################
\subsection{Spatial correlation between X-ray and radio brightness}
\label{IR-IX}
%##########################

For several halos, a point-to-point analysis of radio and X-ray surface brightness has been performed mostly at 1.4\,GHz. Most of the halos are reported to show a linear or sublinear correlation at 1.4\,GHz \citep{Govoni2001a,Govoni2001b,Shimwell2014,Rajpurohit2018,Hoang2019,Cova2019,Xie2020,Botteon2020,Rajpurohit2021b,Luca2021}. This relationship is generally described by a power law:
\begin{equation}
\rm {log}\, \it I_{\rm R}=a+ b\,\rm {log}\,\it I_{\rm X},   
\end{equation}
where the slope $b$ determines whether the thermal components of the ICM (i.e., the thermal gas density and temperature) declines (if $b<1$) faster than the nonthermal components (i.e., the magnetic fields and relativistic particles) or vice-versa (if $b>1$).

High resolution, sensitive uGMRT, VLA, and \textit{Chandra} X-ray data, allow us to perform a detailed investigation of the interplay between the thermal and nonthermal components of the ICM. We make use of  675\,MHz, 1.5\,GHz and 3\,GHz radio images created at $10\arcsec$ resolution, with uniform weighting scheme and a uvcut of $0.2\,\rm k\lambda$. The \textit{Chandra} X-ray image was smoothed with a Gaussian FWHM of $3\arcsec$. We construct a grid covering the entire halo region, excluding R2, R3, R4, and discrete sources. The width of each cell is $10\arcsec$ (45\,kpc). To retain good SNR, we include those areas where the radio and X-ray surface brightness is above the $3\sigma$ level. However, regions where the radio surface brightness is above  $2\sigma$ are included as an upper limit. The radio brightness is expressed in $\rm Jy\,beam^{-1}$ and X-ray in $\rm counts\,s^{-1}\,arcsec^{-2}$.

Fig.\,\ref{fig4} shows the point-to-point comparison between the X-ray and radio brightness at 675\,MHz, 1.5\,GHz, and 3\,GHz. The halo shows a clear positive correlation at all three frequencies: higher radio brightness is associated with higher X-ray brightness, in agreement with \cite{Govoni2001a}. We use the {\tt Linmix}\footnote{\url{https://linmix.readthedocs.io/en/latest/src/linmix.html}} package \citep{Kelly2007} to determine the best-fitting parameters to the observed data. {\tt Linmix} uses a Bayesian hierarchical approach to linear regression considering measurement uncertainties on both axes. It also incorporates nondetection on y-variable (upper limit) and allows the estimation of intrinsic random scatter in the regression. The correlation strength was measured by the Spearman and Pearson correlation coefficients.

%%%%%%%%%%%%%%%%%%%%%%%%%%%%%%%%%%%%%%%%%%%%%%%%%%%%%%%%%%%%%%%%%
%Fig.15 - spectral index vs temp and Ix relations

\begin{figure*}
    \centering     
     \includegraphics[width=1.0\textwidth]{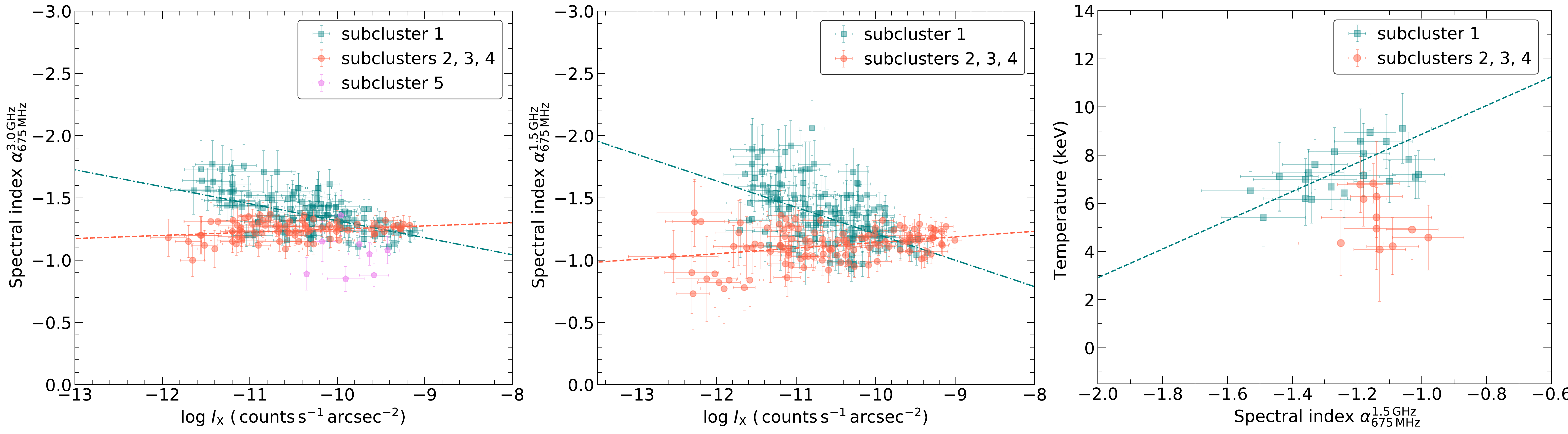}
 \caption{\textit{Left}: $\alpha{-}I_{\rm X}$ relation of the halo in Abell\,2744 between 675\,MHz and 3\,GHz. Regions where the spectral indices were extracted are shown in the right panel of Fig.\,\ref{halocc}. \textit{Middle}: $\alpha{-}I_{\rm X}$ between 675\,MHz and 1.5\,GHz. The halo shows evidence of two different components with possibly different evolutionary stage. {\tt Linmix} best-fit relations are indicated by dashed lines obtained separately for two different regions. In Table\,\ref{fit_index}, we summarize the best-fit parameters. \textit{Right}: $\alpha{-}T_{\rm X}$ relation for the Abell 2744 halo. The spectral index values were extracted in the same regions as used for the X-ray temperature measurements.}
\label{figindex}
\end{figure*} 
%%%%%%%%%%%%%%%%%%%%%%%%%%%%%%%%%%%%%%%%%%%%%%%%%%%%%%%%%%%%%%%%%

In Table\,\ref{fit}, we summarize the best-fit slopes and corresponding correlation coefficients for each radio frequency considered here. The $I_{\rm R}$ and $I_{\rm X}$ are strongly correlated at 675\,MHz, 1.5\,GHz and 3\,GHz. We find a sub-linear slope of $b_{\rm 675\,MHz}=0.73\pm0.02$, $b_{\rm 1.5\,GHz}=0.75\pm0.02$, and $b_{\rm 3\,GHz}=0.74\pm0.02$ at 675\,MHz, 1.5\,GHz, and 3\,GHz, respectively. A sublinear slope is also found for the radio halos in Abell 2319 \citep{Govoni2001a}, Coma cluster \citep{Govoni2001a}, MACS\,J0717.5$+$3745 \citep{Rajpurohit2021b}, MACS\,J1149.5$+$2223 \citep{Luca2021}, and Abell 2163 \citep{Feretti2001}.  

As shown in Fig.\,\ref{fig4}, the number of data points are different at 675\,MHz, 1.5\,GHz and 3\,GHz, indicating that we are sampling different regions. We note that the range of X-ray surface brightness is smaller at 3\,GHz, suggesting the total halo region at 3\,GHz is smaller or more central. Therefore, to compare correlations in the same region, we performed a new fitting of the correlations at 675\,MHz and 1.5\,GHz by including only the boxes used at 3\,GHz. This results in the correlation slopes of $b_{\rm 675\,MHz} = 0.77\pm0.02$ and $b_{\rm 1.5\,GHz}= 0.78\pm0.02$. While these values are marginally steeper than those listed in Table\,\ref{fit}, it suggests that the constant and sublinear slope found at 675\,MHz, 1.5\,GHz and 3\,GHz is independent of regions.

Intriguingly, the slope in the Abell\,2744 halo is remarkably uniform, namely $0.73\pm0.02$, $0.75\pm0.02$, and $0.74\pm0.02$ at 675\,MHz, 1.5\,GHz, and 3\,GHz, respectively. Recently, we investigated the $I_{\rm R}{-}I_{\rm X}$ relation  for the  MACS\,J0717.5$+$3745 halo at different frequencies \citep{Rajpurohit2021b}. Unlike Abell 2744 halo, the correlation slope in the MACS\,J0717.5$+$3745 halo gets steeper toward high frequencies implying a high frequency spectral steepening. As discussed in Sect\,\ref{alpha_temp}, the halo consists of at least two main components (northern and southern; see Fig.\,\ref{subcomponents}), therefore we also fit the data separately to check if these two components show different behavior in the $I_{\rm R}$ and $I_{\rm X}$ relation. The resulting plot is shown in Fig.\,\ref{fig4_sub} with corresponding best-fitting slopes. Compared to the northern part, the radio emission across the southern part of the halo is strongly correlated with the X-ray surface brightness, see Table\,\ref{fit}. In addition, the correlation slope is different for the northern and southern parts, suggesting different cluster dynamics in these regions. 

The correlation slope of the southern part of the halo is clearly significantly  steeper than the northern part at all three frequencies. It implies that along the denser regions (subclusters 2 and 3) the radio emissivity is enhanced more than the X-ray emissivity. The correlation slope is still relatively uniform as a function of frequency for both the northern and southern parts. The uniform correlation slope in the Abell 2744 halo is may be due to the fact that at 3\,GHz, we are not sampling regions with low X-ray brightness.

Our slope at 1.5 GHz is significantly flatter than reported  previously by \cite{Govoni2001a}, namely $b_{\rm 1.4\,GHz}=0.99\pm0.05$. We note that \cite{Govoni2001a} used a Least-Square fitting which may give different results. To check this, we also fit the 1.5\,GHz data using Least Squares regression. We emphasize that this regression method is less powerful than {\tt Linmix} since it  does  not  account  for  measurement uncertainties in both variables and does not provide an estimate of the intrinsic scatter. Using Least-Square fitting, we obtained a  slope of $0.80\pm0.03$, still inconsistent with \cite{Govoni2001a}.

It is worth noting that \cite{Govoni2001a} used rather low resolution radio ($50\arcsec$) and X-ray ($30\arcsec$) images. We thus repeated the fitting using {\tt Linmix} at $25\arcsec$ and $50\arcsec$ resolution images. We created two new grids covering the halo. Each cell size in these grids is similar to the beam of the radio image at these two resolutions. We obtained a slope of $b_{\rm 25\arcsec,\,1.5\,GHz}=0.81\pm0.02$ and $b_{\rm 50\arcsec,\, 1.5\,GHz}=0.82\pm0.02$. This indicates that the correlation slope changes slightly when using low resolution images. The change in the slope is negligible when going from $20\arcsec$ ($b_{\rm 25\arcsec,\,1.5\,GHz}=0.81\pm0.02$) to $50\arcsec$ ($b_{\rm 50\arcsec,\, 1.5\,GHz}=0.82\pm0.02$) but the slope is slightly flatter at higher resolution, namely at $10\arcsec$ ($b_{\rm 10\arcsec,\, 1.5\,GHz}=0.75\pm0.02$).  The same trends are noticed at 675\,MHz and 3\,GHz. The change in image resolution thus cannot explain the difference in the slope obtained by \cite{Govoni2001a} and ours. Lastly, we fit the low resolution data points using Least Squares regression and also  included R2, R3, and R4 regions embedded in the halo. This results in a slope of $0.96\pm0.03$ which is consistent with \cite{Govoni2001a}.  We conclude that the $I_{\rm R}{-}I_{\rm X}$ correlation slope for the halo in Abell 2744 is indeed sublinear.

The correlation slopes provide information about the different acceleration models and the distribution of magnetic fields  \citep[e.g.,][]{Dolag2000,Govoni2001a,Storm2015}. The radio emissivity due to synchrotron emission, $j_{\rm R}$,  depends on the number density of relativistic electrons and the magnetic field strength
\begin{equation}
    j_{\rm R}\propto N_{0}\cdot B^{(\delta+1)/2} \cdot  \nu^{-(\delta-1)/2},
    \label{Er}
\end{equation}
where $B$ is the magnetic field strength, $\nu$ is the frequency, and $\delta$  (with $\alpha=\frac{1-\delta}{2}$) is the slope of the electron energy distribution. For a typical radio spectral index of $-1$ then $j_{\rm R}\propto n_{\rm CRe}\cdot B^2 \cdot  \nu^{-1}$ (where $n_{\rm CRe}$ is density of radio-emitting electrons).

The X-ray emissivity, $j_{\rm X}$ depends on the thermal gas density, $n_{e}^{2}$, and  temperature, $T_{\rm X}$ as
\begin{equation}
 j_{\rm X} \propto n_{e}^{2} \cdot T^{1/2}_{\rm X}.
\end{equation}
A linear relation between $I_{\rm R}$ and $I_{\rm X}$ implies that the magnetic field and relativistic particles are connected to the thermal electron density as:  
\begin{equation}
n_{\rm CRe}\cdot B^2 \cdot  \nu^{-1} \propto n_{e}^{2} \cdot T^{1/2}_{\rm X}.
\end{equation}
For example, in the case where the energy density of both radio emitting electrons and magnetic field scale with thermal energy (assuming a quasi-isothermal plasma), the radio emissivity scales with the thermal X-ray emissivity as $j_{\rm R}/j_{\rm X}\propto T^{1/2}_{X}$. Therefore, sublinear slopes suggest that the energy density of radio-emitting electrons and/or that of the magnetic field decline less rapidly than the thermal energy density \citep[e.g.,][]{Govoni2001a}. In reacceleration models these trends allow to constrain the way the turbulent energy flux is dampled into relativistic electrons and magnetic fields \citep{Brunetti2014}. A superlinear or linear slope is generally expected in the secondary models, depending on the weak and strong magnetic field, as in this case the radio emissivity scales with the thermal X-ray emissivity as $j_{\rm R}/j_{\rm X}\propto T^{1/2}_{X}\cdot B^2/(B^2 + B_{\rm CMB}^2)$  \citep[e.g.,][]{Dolag2000,Govoni2001a}.

The sublinear $I_{\rm R}{-}I_{\rm X}$ scaling in the Abell\,2744 halo together with the curved spectra in its subregions, small-scale spectral index fluctuations, steeper spectral indices in the outermost regions and spectral curvature disfavors secondary models.

%%%%%%%%%%%%%%%%%%%%%%%%%%%%%%%%%%%%%%%%%%%%%%%%%%%%%%%%%%%%%%%%%
%Figs 16 and 17 - polarization maps

 \begin{figure*}[!thbp]
    \centering
    \includegraphics[width=0.43\textwidth]{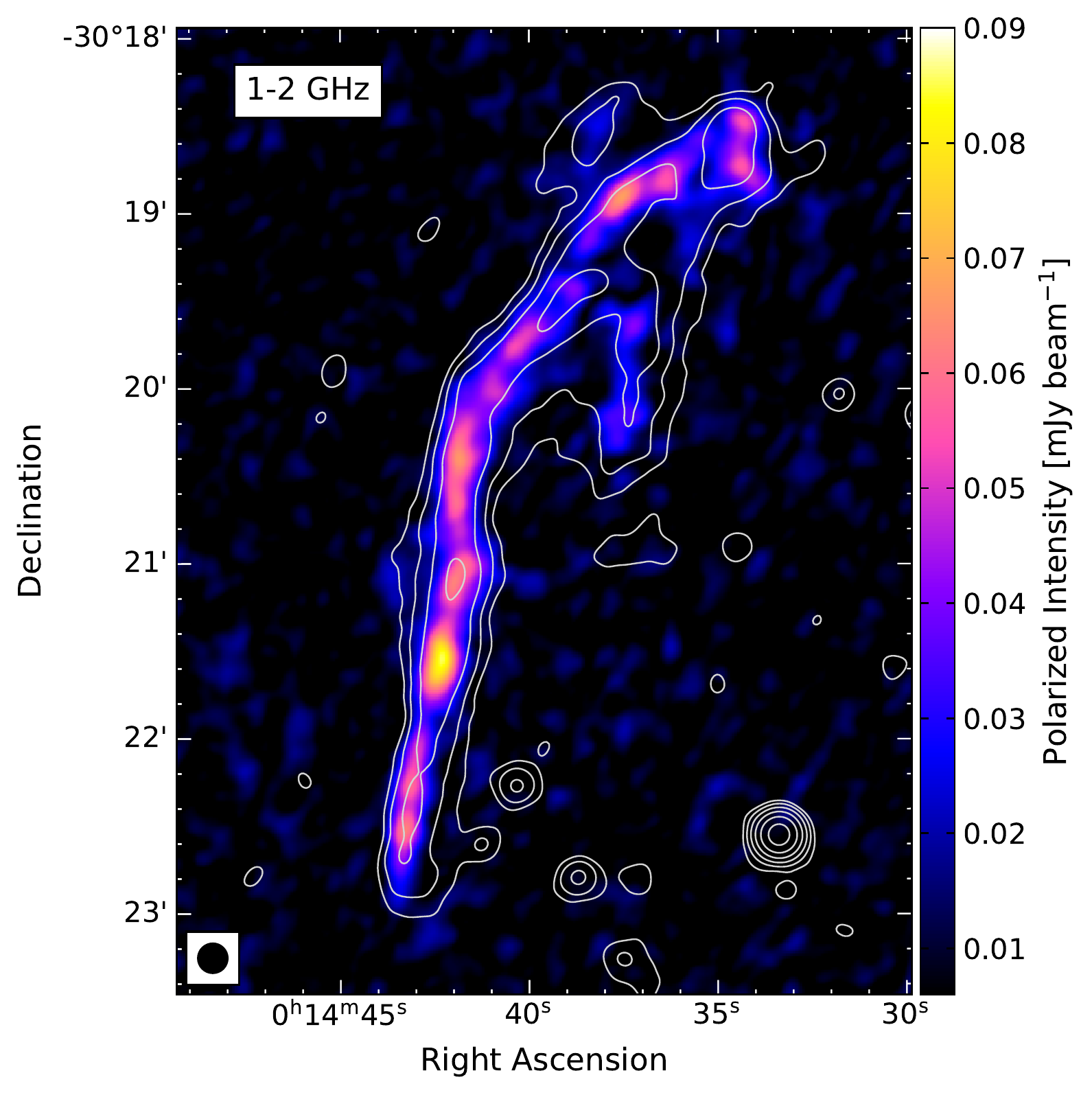}
        \includegraphics[width=0.44\textwidth]{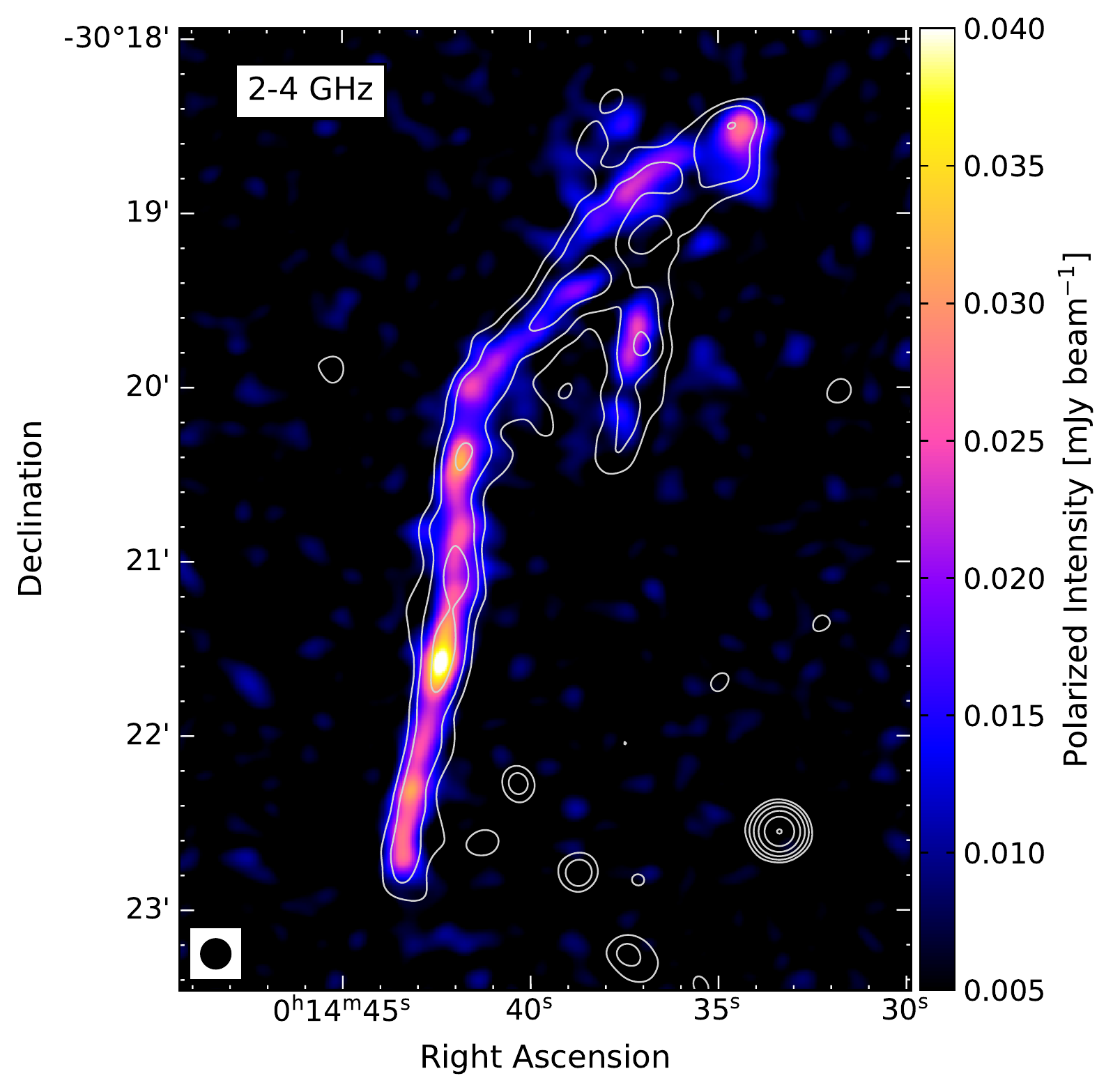}
    \vspace{-0.2cm}
    \caption{Polarization intensity images of the main relic R1 at $10\arcsec$ resolution. The images show that the relic is polarized over its entire length from 1 to 4\,GHz. Contour levels are drawn at $[1,2,4,8,\dots]\,\times\,5\sigma_{{\text{ rms}}}$ and are from the Stokes $I$ images. The beam sizes are indicated in the bottom left corner of the each image.}
      \label{poli}
  \end{figure*}  
  
  \begin{figure*}[!thbp]
    \centering
            \includegraphics[width=0.43\textwidth]{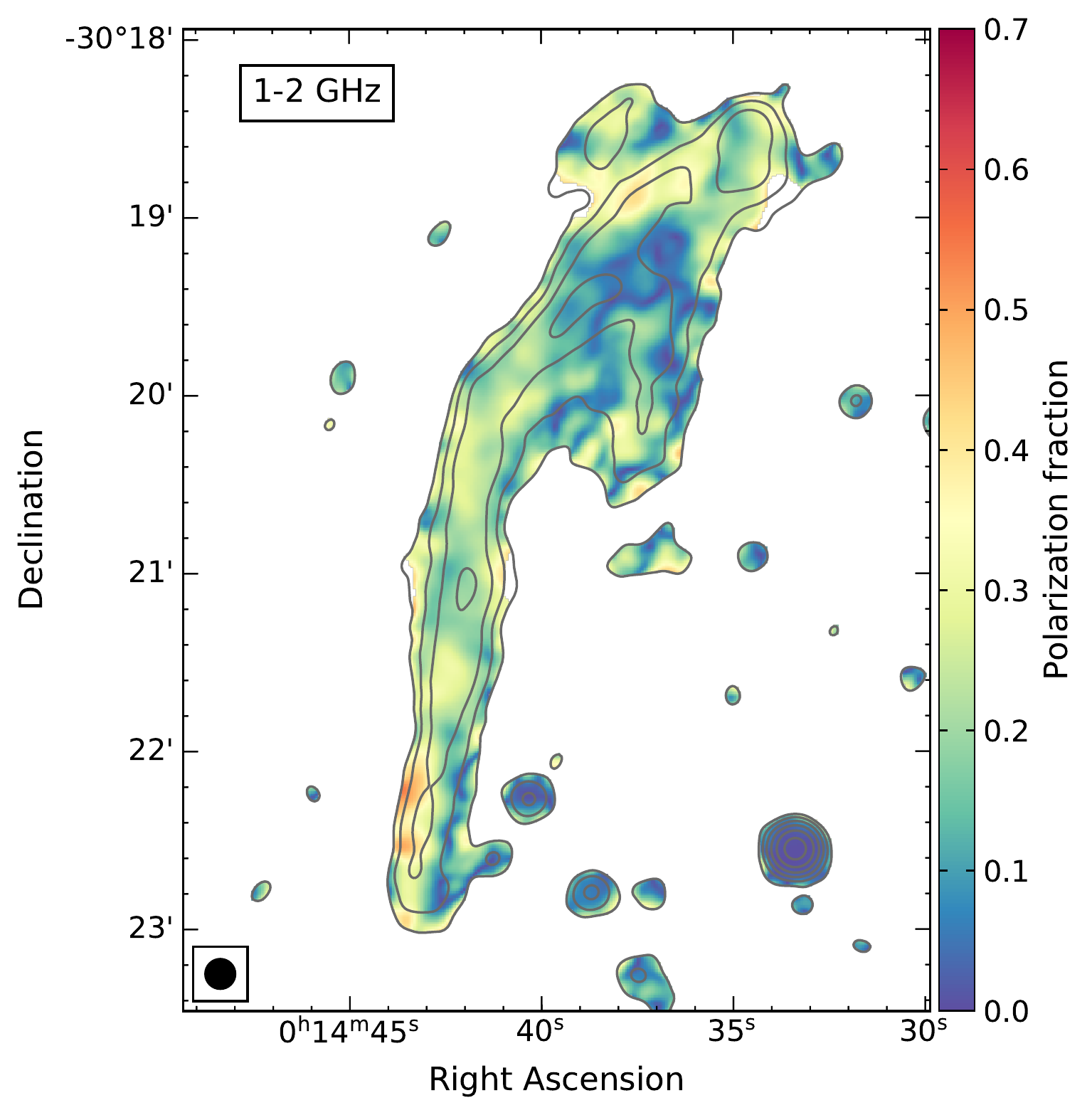}
        \includegraphics[width=0.43\textwidth]{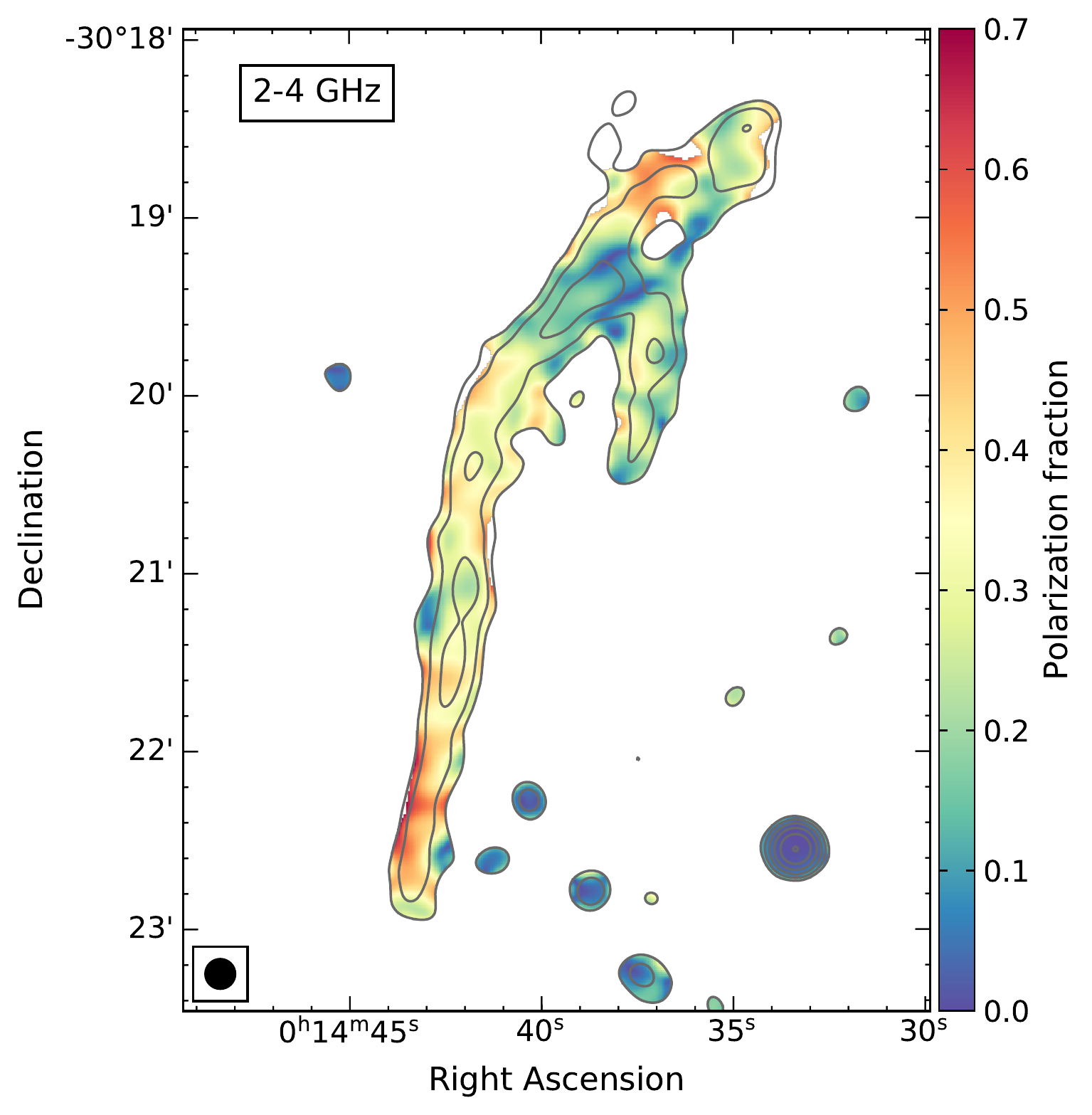}
    \vspace{-0.2cm}
    \caption{Fractional polarization maps of the main relic at $10\arcsec$ resolution. The degree of polarization across the relic decreases when going from S-band (right) to L-band (left). Some local fluctuations in polarization fraction are visible, particularly in the northern part of the relic. Contour levels are same as in Fig.\,\ref{poli}.}
      \label{fract}
  \end{figure*}  
 
%%%%%%%%%%%%%%%%%%%%%%%%%%%%%%%%%%%%%%%%%%%%%%%%%%%%%%%%%%%%%%%%%

%%%%%%%%%%%%%%%%%%%%%%%%%%%%%%%%%%%%%%%%%%%%%%%%%%%%%%%%%%%%%%%%%
% Table 4 : spectral index versus Ix and Temp bets-fit parameters  
%%%%%%%%%%%%%%%%%%%%%%%%%%%%%%%%%%%%%%%%%%%%%%%%%%%%%%%%%%%%%%%%%

\setlength{\tabcolsep}{5pt}
\begin{table}
\caption{{\tt Linmix} best fitting parameter of the data for the right panel of Fig.\,\ref{figindex} ($I_{\rm X}-\alpha$ correlation).}
\centering
\begin{threeparttable} 
\begin{tabular}{ c c c c  c c}
 \hline  \hline  
& entire halo& subcluster 1&  subclusters 2, 3, and 4 \\ 
\hline  
$b_{\rm 675\,MHz}^{\rm 3.0\,GHz}$&$-$& $-0.13$& $0.02$\\
$r_{\rm s,\,\alpha_{675\,MHz}^{1.5\,GHz}}$&$0.30\pm0.08$& $-0.68\pm0.08$& $0.80\pm0.08$\\ 
$r_{\rm s,\,\alpha_{675\,MHz}^{3.0\,GHz}}$&$0.42\pm0.14$& $-0.83\pm0.08$& $0.65\pm0.12$\\ 

\hline 
\end{tabular}
\end{threeparttable} 
\label{fit_index}   
\end{table}
 
 %%%%%%%%%%%%%%%%%%%%%%%%%%%%%%%%%%%%%%%%%%%%%%%%%%%%%%%%%%%%%%%%%
%Fig 18: vector distribution 

  \begin{figure*}[!thbp]
    \centering
            \includegraphics[width=0.47\textwidth]{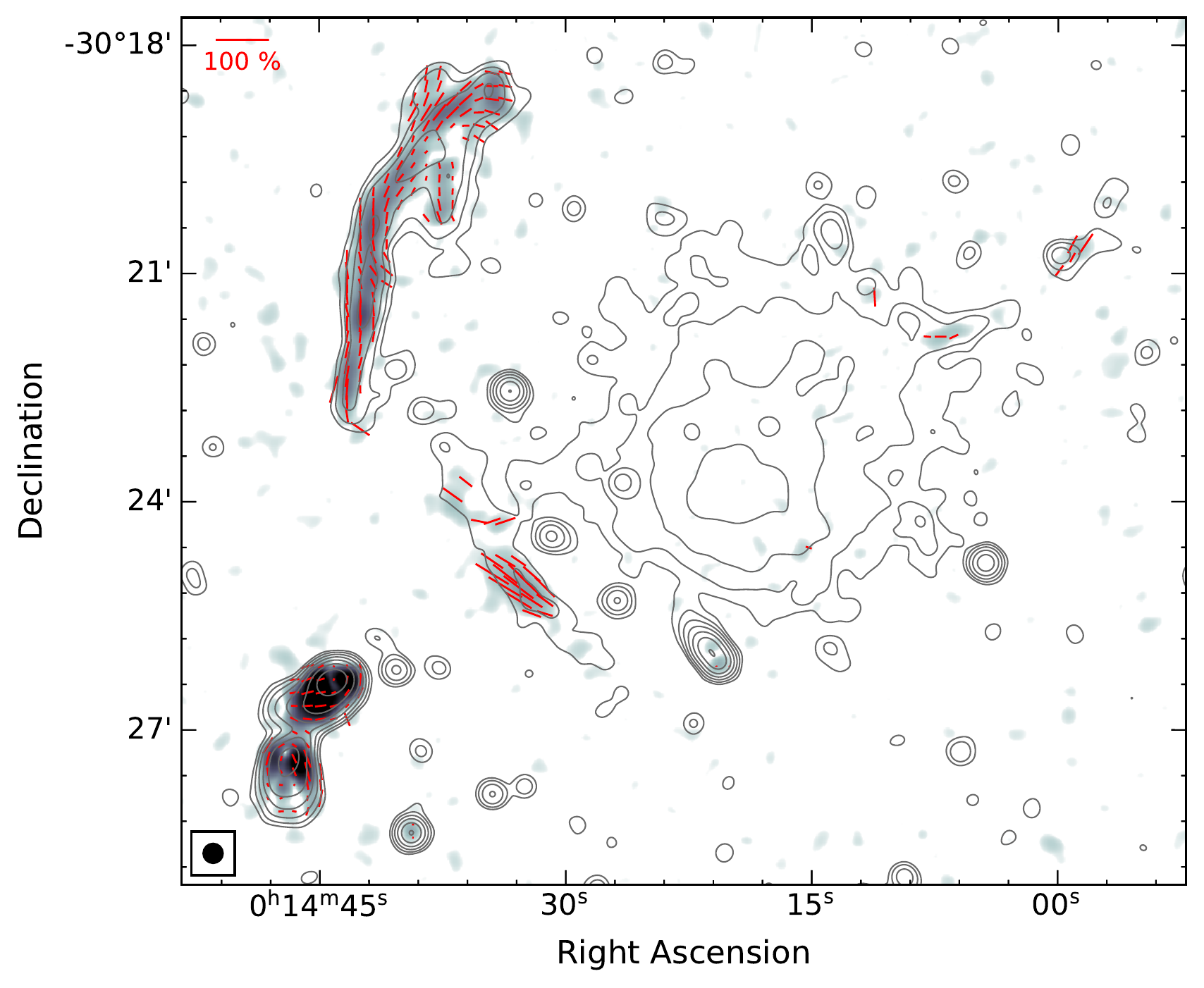}
        \includegraphics[width=0.47\textwidth]{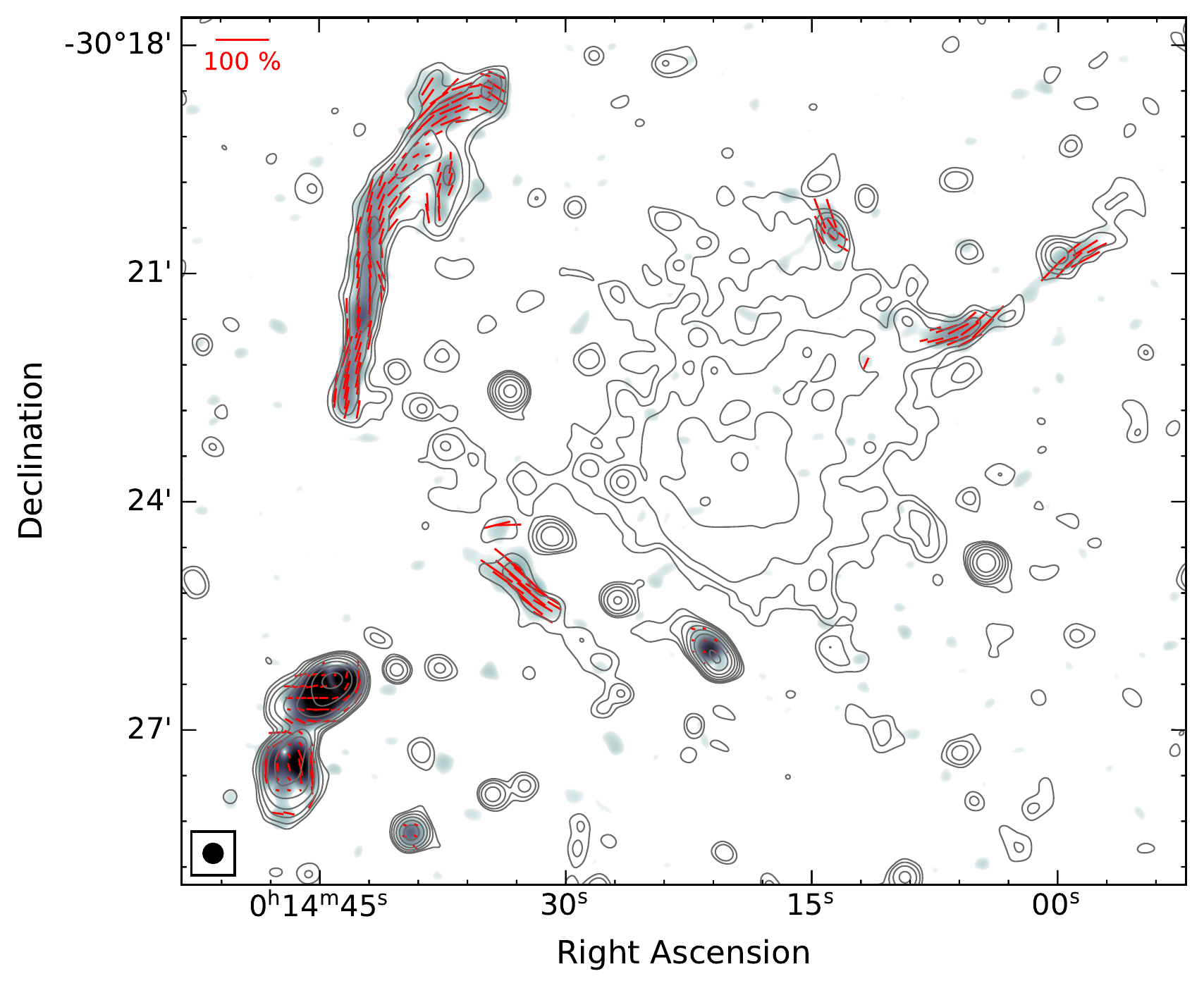}
    \vspace{-0.2cm}
    \caption{VLA L-band (left) and S-band (right) polarization intensity maps at $15\arcsec$ resolution. Red lines represent the magnetic field vectors. Their orientation represents the projected B-field corrected for Faraday rotation and the contribution from the Galactic foreground. For all four relics, the magnetic-field vectors are uniformly disturbed and parallel to the orientation of the radio emission. The vector lengths is proportional to the polarization percentage. No vectors were drawn for pixels below $3\sigma$. Contour levels are drawn at $[1,2,4,8,\dots]\,\times\,5\sigma_{{\text{ rms}}}$ and are from the Stokes $I$ images.}
      \label{fract_low}
  \end{figure*} 

%%%%%%%%%%%%%%%%%%%%%%%%%%%%%%%%%%%%%%%%%%%%%%%%%%%%%%%%%%%%%%%%%

%##########################
%% alpha-I_X
\subsection{Spatial correlation between spectral index and X-ray brightness}
\label{alpha_ICM}
%##########################

We study the point-to-point distribution of the halo spectral index with the thermal brightness. To extract the X-ray surface brightness and spectral indices, we again created a grid of cells, see the right panel of Fig.\,\ref{figcc}. Since the halo is more extended toward lower frequencies, we examine the radio spectral index between $\alpha^{\rm{3\,GHz}}_{\rm{675\,MHz}}$ and $\alpha^{\rm{1.5\,GHz}}_{\rm{675\,MHz}}$. The spectral index values are obtained from $15\arcsec$ resolution radio maps. The regions used for extracting spectral indices between 675\,MHz and 3\,GHz and the X-ray surface brightness are shown in the right panel of Fig.\,\ref{fig6}. We considered only regions where the flux density is above $3\sigma$ in both the maps, and excluded regions contaminated by discrete point sources. 

%%%%%%%%%%%%%%%%%%%%%%%%%%%%%%%%%%%%%%%%%%%%%%%%%%%%%%%%%%%%%%%%%
% Table 5 : polarization properties of relics in Abell 2744
%%%%%%%%%%%%%%%%%%%%%%%%%%%%%%%%%%%%%%%%%%%%%%%%%%%%%%%%%%%%%%%%%

\setlength{\tabcolsep}{30pt}
\begin{table*}[!htbp]
\caption{Polarization properties of all four relics in the cluster Abell\,2744.}
\centering
\begin{threeparttable} 
\begin{tabular}{ c | c | c | c | c   }% c | c}
\hline \hline
\multirow{1}{*}{Source} & \multicolumn{2}{c|}{VLA} & \multirow{1}{*}{RM range} & \multirow{1}{*}{$<\sigma_{\rm RM}>$} \\
 \cline{2-3} 
&S-band & L-band  &   \\
  \cline{2-3} 
  &$P_{{\rm{3.0\,GHz}}}$ & $P_{{\rm{1.5\,GHz}}}$   &     \\
 &( \%) & (\%) & ($\rm rad\,m^{-2}$) & ($\rm rad\,m^{-2}$)   \\
  \cline{2-3} \cline{3-5}
  \hline  
R1 & $30\pm4$ & $19\pm2$&$+1$ to $+19$& $ 5\pm1 $\\ 
R2 & $34\pm4$ & $27\pm3$&$+4$ to $+16$& $4\pm1$\\ 
R3 & $32\pm3$ & $13\pm2$&$-2$ to $+18$& $6\pm2$\\ 
R4 & $25\pm2$ & $8\pm1$&$-$& $-$ \\ 
\hline 
\end{tabular}
\begin{tablenotes}[flushleft]
\footnotesize
\item{\textbf{Notes.}} {The mean fractional polarization for the main relic R1 is measured from $10\arcsec$ resolution image. We note that our S-band polarization value of  R1 is slightly higher than those obtained by \cite{Pearce2017} because we measured polarization fraction from Faraday corrected S-band images. Since the polarized emission from the fainter relics R2, R3, and R4 are recovered well in low resolution images therefore, we report the mean fractional polarization measured from $15\arcsec$ resolution maps.}
\end{tablenotes}
\end{threeparttable} 
\label{polarization}   
\end{table*}

The resulting  $\alpha-I_{\rm X-ray}$ distribution for the halo in Abell\,2744 is shown in Fig.\,\ref{figindex}. The halo apparently shows two distinct trends. To check the significance of any correlation, we fit the data assuming a relation of the form:

\begin{equation}
    \alpha=a+b\,\rm {log}\,\it I_{\rm X}.   
\end{equation}

The fitting results are given in Table\,\ref{fit_index} and demonstrate that there is no significant correlation between these two quantities. Given that a positive and negative correlation have been observed for the halo in Abell 2255 \citep{Botteon2020} and MACS\,J0717.5$+$3745 \citep{Rajpurohit2020b}, respectively,  the lack of a significant correlation indicates that this radio halo may be in a different evolutionary state or it may be a multicomponent halo. To check this second possibility, we also fit the data separately. Interestingly, the halo clearly follows two different trends: one section of the halo shows a positive correlation, while the other shows a negative correlation (see Fig.\,\ref{fit_index}). This is the first time that such distinct correlation trends are detected in a radio halo, with a statistical measure of the significance of the correlation.

It is worth stressing that the strong/weak lensing analysis of Abell\,2744 reveals a complex merger, involving at least four subclusters \citep{Golovich2019}. Subcluster 1 is elongated along the east-west direction, see Fig.\,\ref{subcomponents}. The X-ray data reveal an additional small northwest ICM component, close to the location of R3 (Fig.\,\ref{subcomponents}). Interestingly, regions following the anticorrelation belong to subcluster 1. A similar correlation has also been recently observed for the halo in MACS\,J0717.5$+$3745 \citep{Rajpurohit2020b}. The observed anticorrelation in the northern part of the halo implies that the faint X-ray regions show a steeper spectral index. This suggests that there is a steepening in the outermost regions of the halo. The spectral steepening in the outermost part of the northern part of the cluster is also evident in the spectral index maps shown in Fig.\,\ref{fig3}. Moreover, the data points that fall into the positive correlation belong to the subclusters 2, subcluster 3, and subcluster 4 (Fig.\,\ref{subcomponents}). In the next section, we discuss the presence of these two distinct components in detail. 

%%%%%%%%%%%%%%%%%%%%%%%%%%%%%%%%%%%%%%%%%%%%%%%%%%%%%%%%%%%%%%%%%
%Fig 19- Faraday spectra

  \begin{figure*}[!thbp]
    \centering
            \includegraphics[width=1.00\textwidth]{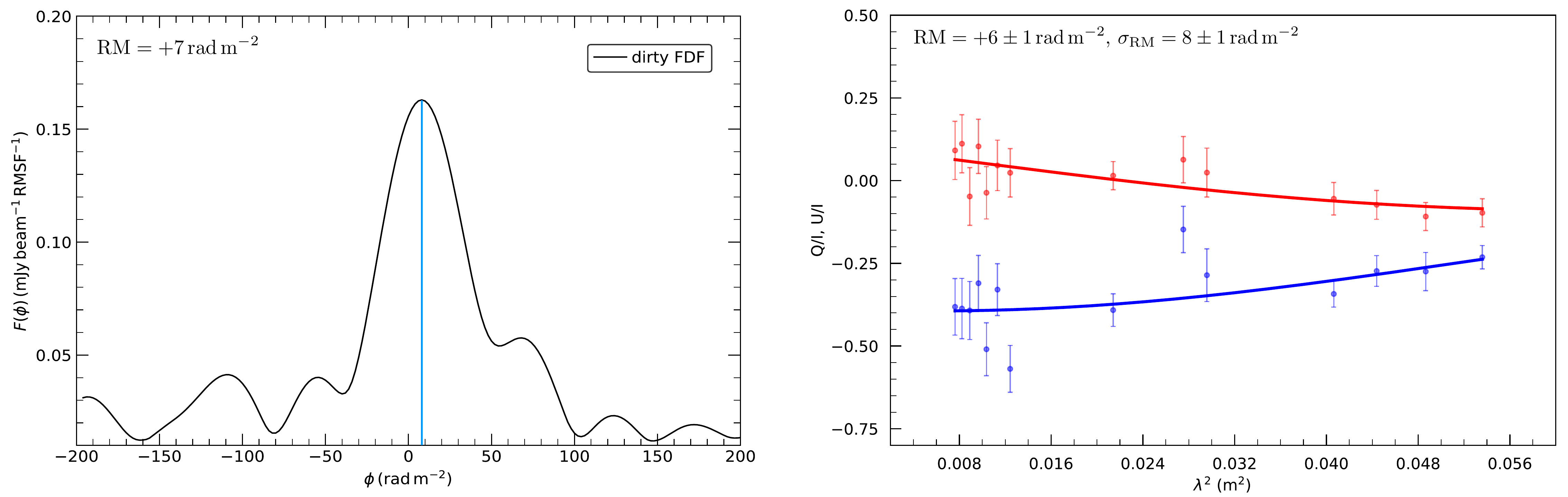}
    \vspace{-0.2cm}
    \caption{\textit{Left}: Faraday spectrum of one of the regions within R1 obtained using RM-synthesis. The blue line indicates the peak RM. \textit{Right}: Corresponding QU-fitting spectrum fitted with a single RM component and an exponential external depolarization term. The fractional Stokes $Q$ (blue) and $U$ (red) with the dots showing the observed  data points and the solid lines the best-fit model from the QU-fitting. Both RM-synthesis and QU-fitting spectra are consistent with a single RM component that has a low value of Faraday dispersion measure, indicating that the relic R1 is located far outside of the cluster or toward the observer.}
      \label{RMspectra}
  \end{figure*} 
%%%%%%%%%%%%%%%%%%%%%%%%%%%%%%%%%%%%%%%%%%%%%%%%%%%%%%%%%%%%%%%%%

%##########################
\subsection{Spatial correlation between spectral index and ICM temperature}
\label{alpha_temp}
%##########################
 We also compare the ICM temperature with the spectral index. We use the \textit{Chandra} temperature map presented in \cite{Pearce2017}. The resulting plot is shown in the right panel of Fig.\,\ref{figindex}. We again fit the data using {\tt Linmix} to search for a possible correlation between spectral index and temperature. The Spearman correlation coefficient is $0.18$. As found by \cite{Pearce2017}, the relation between $\alpha-T_{\rm X}$  does not show any significant correlation. 

Since in the $\alpha-I_{\rm X}$ relation, the halo shows two distinct trends, we also fit the data separately. As seen in the right panel of  Fig.\,\ref{figindex}, in the northern part of the halo (subcluster 1) the flat spectral index regions seem to be associated with higher temperatures. For this region, we find that the Spearman correlation coefficient is $r_{\rm s}=0.56$, indicating a moderate correlation between the spectral index and ICM temperature. Regions associated with the southern part of the halo (subclusters 2, 3, and 4) show relatively uniform spectral indices, namely between $-1.0$ to $-1.25$. The temperature in these regions varies from 3 to 7\,keV.

To the best of our knowledge, the point-to-point correlation between the spectral index and ICM temperature has been quantitatively reported for one more halo, namely Abell\,2255 \citep{Botteon2020}. They found a hint of mild anticorrelation between the X-ray temperature and spectral index for the Abell\,2255 halo. The moderate positive correlation between the spectral index and the ICM temperature for the northern part of the halo supports the idea that a fraction of the gravitational energy, which is dissipated during cluster mergers in  heating  the  thermal  plasma,  is converted into  re-acceleration of relativistic particles and amplification of the magnetic field  \citep{Orru2007}. 

The presence of two different clear trends in the $\alpha-I_{\rm X}$ and  $\alpha-T_{\rm X}$ relations provide strong evidence that the Abell 2744 halo is a two-component halo. The clear separation between these two components may  trace different evolutionary stages of the same phenomenon. Abell 2744 is known to show a very complicated merger dynamics \citep{Merten2011,Owers2011,Medezinski2016,Golovich2019}. Both optical and X-ray analyses point to a violent major merger in the central region along the north-south axis with a larger line of sight component.

\cite{Owers2011} suggested that some component of the northern subcluster is directed toward the north-northwest. The tight anticorrelation between the radio spectral index and X-ray surface  brightness and a moderate positive correlation between the spectral index and ICM temperature in the north-northwest direction favors this.  Moreover, \cite{Owers2011} speculated that the small northwestern clump (subcluster 5) is most likely in a post, off-center, core passage phase traveling to the north or north-east. In the $\alpha-I_{\rm X}$ correlation, the data points from subcluster 5 is in line with the detected anticorrelation in the northern part of the halo, which seems to be consistent with the fact that this component may be traveling to the north-east. 

In the southern part of the cluster we detected a positive correlation between the spectral index and X-ray surface brightness. If interpreted with turbulent reacceleration models, this suggests that gas dynamics and consequently both the turbulent and magnetic field properties in the dense regions, namely subclusters 2 and 3, are slightly different from the external (less dense) regions and from the northern part of the halo \citep[e.g.,][]{Brunetti2001,Brunetti2007,Bonafede2018}.

%########################################################################
%  Section 7: Polarization
%######################################################################## 

\section{Polarization and Faraday Rotation } 
\label{polarization_analysis}

\cite{Pearce2017} found that all four relics in Abell\,2744  are highly polarized at 2-4\,GHz and show an aligned B-vectors distribution. They presented S-band polarization maps without correcting for Faraday rotation effect.  We  report here Faraday Rotation Measure (RM) analysis from 1-4 GHz to understand the polarization properties of the relics in the field. For polarization calibration details, we refer to \cite{Pearce2017}.

We create Stokes $IQU$ cubes at $10\arcsec$ resolution. These are only used to study the main relic R1. Because of the radio surface brightness of the components R2, R3, and R4 relative to R1 are very low, these are studied using lower-resolution $15\arcsec$ Stokes $IQU$ cubes. The imaging was performed with {\tt Briggs} weighing and {\tt robust=0.5}. As the output images has slightly different resolutions, all images were smoothed to a common resolution, namely $10\arcsec$ and $15\arcsec$.  Images with poor sensitivity or significant artifact contamination were discarded. The polarization properties are summarized in Table\,\ref{polarization}.

The resulting polarization intensity maps of the main relic R1 are shown in Fig.\,\ref{poli}. The relic is polarized over its entire length in both, the VLA S- and L-band. The morphology of the polarized emission is similar to the total power emission. In Fig.\,\ref{fract}, we show the fractional polarization maps at $10\arcsec$ resolution. The polarization fraction across the relic varies between 2 and 64\%, where it can be measured. The spatially averaged polarization fraction over R1 is about 19\% at 1-2\,GHz and 30\% at 2-4\,GHz, showing that the degree of polarization decreases significantly with increasing wavelength.   

In addition, we observe some local fluctuations in the polarization fraction across the relic, particularly in the northern part of the relic.  These local fluctuations are most likely an outcome of a perturbed ICM at the location of the relic.

The beam depolarization could lead to a significant loss of the observed polarized signal. We also measured the polarization fraction from $15\arcsec$ maps. The resulting values are about 12\% lower than those obtained at $10\arcsec$ resolution. This implies that the intrinsic polarization is very likely higher than reported in Table\,\ref{polarization}.

At L-band (1-2\,GHz), there is tentative evidence to suggest that the high degree of polarization is observed at the eastern edge of the relic (see Fig.\,\ref{fract}). This is the area where we expect to witness the fresh acceleration of ICM electrons. In particular, at the northern part of the relic (excluding R1-A), the polarization fraction decreases as we approach the downstream region. Recently, \cite{DiGennaro2021} found a clear gradient in the polarization fraction across the Sausage relic. They speculated that such a trend is possibly due to projections and/or decrease in the magnetic field anisotropy toward the cluster center. Such trends is expected in the presence of a shocked turbulent ICM (Dom\'{i}nguez-Fern\'{a}ndez et al. submitted). On the other hand, if the magnetic field remains regular at the position of the relic, we would expect a polarization fraction increasing toward the downstream of the shock front.

As shown in Fig.\,\ref{fract_low}, R2 and R3 are also highly polarized in L-band while no polarized emission is detected for R4. The degree of polarization drops quickly for R3 (from 32\% to 13\%) and R4 (from 25\% to 8\%) with increasing wavelength while moderately for R2; see Table\,\ref{polarization}. As previously reported by \cite{Pearce2017}, the B-field vectors are strikingly aligned with the orientation of the emission for all relics and even across R1-A. We do not find any small-scale variation in the B-field orientation across and along the relics in both the L and S-band.  

RM-synthesis \citep{Brentjens2005} was performed on $IQU$ cubes with 131 spectral channels using ``{\tt pyrmsynth}'' code\footnote{\url{https://github.com/mrbell/pyrmsynth}}. The RM-cube synthesizes a range of RM from $-200\,\rm rad\,m^{-2}$ to $+200\,\rm rad\,m^{-2}$ with a bin size of $1\,\rm rad\,m^{-2}$. The Faraday spectrum of one of the regions within R1 is shown in the left panel of Fig.\,\ref{RMspectra}. The RM across the entire relic is relatively uniform and shows a well-defined single peak.  

The mean RM across R1 is about $+8\,\rm rad\,m^{-2}$. Our own Milky Way contributes to extragalactic RM values. The expected Galactic RM at the location of the cluster is $\sim+6\,\rm rad\,m^{-2}$ \citep{Oppermann2012}. Therefore, the detected RM's are consistent with the average Galactic foreground.

To analyze the Faraday rotation properties of the emission and of the intervening ICM, we also applied QU-fitting to the main relic R1. Since the relic has a very low surface brightness, to detect the relic with a sufficiently good signal to noise in Stokes $Q$ and $U$ we imaged L- and S-band data per spectral window rather than using channel wise Stokes $IQU$ images. We divided the relic into 33 square boxes with width of $10\arcsec$. As the fractional polarization varies between 1-4\,GHz, we fit the data with a single RM component and an exponential depolarization term as formulated in \cite{Burn1966}.
\begin{equation}
  P(\lambda^{2})
  = 
  I_{0}(\lambda) \, p_{0} \, 
  e^{2i (\psi_{0}+{\rm{RM}}\lambda^{2})} \,
  e^{-2\sigma_{\rm RM}^{2} \lambda^{4}},
  \label{sigmarm}
\end{equation}
where $P(\lambda^{2})$ is the complex polarization as a function of wavelength, $I_{0}$ the total intensity of the source, and $p_{0}$ is the intrinsic polarization fraction.  $\psi_{\rm 0}(\lambda^2)$ is the intrinsic polarization angle, RM is the rotation measure  and $\sigma_{\rm RM}$ is the standard deviation (known as Faraday dispersion measure) of the Gaussian distributed rotation measures.

We performed QU-fitting using the functionality available as part of the \texttt{RM-tools} suite\footnote{\url{https://github.com/CIRADA-Tools/RM}}. The best-fit is determined by a nested sampling algorithm. The QU-fitted spectrum of one of the regions is shown in the right panel of Fig.\,\ref{RMspectra}. A single component in Faraday space with depolarization term provides a good fit, as also seen in the Faraday spectra obtained from RM-Synthesis. The RM mainly varies from $+1$ to $+19\,\rm rad\,m^{-2}$ across the relic. The $\sigma_{\rm RM}$ is in the range $3-8\,\rm rad\,m^{-2}$. The lower value of Faraday dispersion measure implies that any contribution from the Faraday rotating intervening material is very small and the relic is located far outside the cluster or toward the observer. The $\sigma_{\rm RM}$ values across R2 and R3 are also low, indicating that both of these relics are located in the low density ICM.  

Under a simple hypothesis, Faraday dispersion measure can be used to constrain the magnetic field strength as \citep{Sokoloff1998} 
\begin{equation}
\sigma_{\rm{RM}}=\sqrt{(1/3)}\,0.081\,\langle n_{\rm e}\rangle\,B\,(L\,t/f)^{0.5},
\label{sigmaRM}
\end{equation}
where $\langle n_{\rm e}\rangle$ is the average thermal electron density of the ionized gas along the line of sight in $\rm cm^{-3}$, $B$ is the magnetic field strength in $\upmu$G, and $f$ is the volume filling factor of the Faraday-rotating gas. $L$ is the path length through the thermal gas and $t$ is the turbulence scale, both in pc. We use $t=40\rm\,kpc$ (the RM fluctuations measured from $10\arcsec$ images), $L=2.4$\,Mpc (the total extent of cluster), and  $f=0.5$ \citep[e.g.,][]{Murgia2004}. 

At the location of the main relic, the thermal electron density is about $10^{-4}\rm\,cm^{-3}$ \citep{Eckert2016,Botteon2020a}. The  highest value of $\sigma_{\rm RM}$ across the R1 relic is about $\rm 8\,rad\,m^{-2}$. By inserting all values in Eq.\,\ref{sigmaRM}, we obtained $B\sim0.5\,\upmu \rm G$.  
  
The polarization analysis suggests that R2 and R3 are located at the cluster outskirts. Both sources show a power-law spectrum, we do not find any hint of curved spectrum between 385\,MHz and 3\,GHz. In addition, R2 is reported to trace a shock front \citep{Pearce2017}. Based on the morphology, and the spectral and polarization properties we conclude that R2 and R3 are shock related structures. Shock waves generated by cluster mergers can compress AGN fossil radio plasma or fossil plasma can passively evolve from a switched off AGN,   producing so called radio phoenices or AGN relics. Both of these sources have relatively small sizes because with time radio plasma losses most of the energetic electrons responsible for the radio emission by radiative energy losses \citep{Ensslin2001}. Due to the small size, therefore, R4 could be a radio phoenix or AGN relic. 
  
%########################################################################
%  Section 8: Conclusions
%######################################################################## 
 
\section{Conclusions and summary}
\label{summary}
In this work, we have presented uGMRT (300-850\,MHz) total power and VLA (1-4 GHz) polarization observations of the galaxy cluster Abell\,2744. These observations allowed us to perform a detailed spectral and polarization analysis of the diffuse radio emission sources present in this system. To quantify the connection between the thermal and nonthermal components of the intracluster medium, we also combine the radio data with \textit{Chandra} X-ray observations. In our new images, we detect the previously known radio relics R1, R2, R3, and R4, and the radio halo. From these observations we were able to characterize the spectral properties and the radio-versus-X-ray relations of the central halo emission in unprecedented detail. We summarize the overall results as follows:

\begin{enumerate} 

\item{} The halo emission is more extended than previously reported, namely $\sim 2.5\,\rm Mpc$. The observed total extent of the halo changes with the observing frequency, i.e, it is more extended toward low frequencies.\\

\item{} The radio integrated emission from the entire halo follows a power-law spectrum between 150\,MHz and 3\,GHz and has a slope  of $\alpha=-1.14\pm0.04$. In contrast, subregions show a slightly different spectra and high frequency steepening. This suggests that an overall power-law spectrum can be observed in combination of a variety of different spectra.\\

\item{} The spatially resolved spectral index maps of the halo show the presence of localized regions in which the spectral index is significantly different from the average. The spectral index in the innermost regions is relatively flat and steepens in the outer regions. This is consistent with the turbulent (re-)acceleration models, assuming that there is more dissipated turbulent kinetic power or a higher magnetic magnetic field strength in the innermost halo region.  We find a mean scatter of $0.24$ around the mean spectral index of $-1.15$ between 675\,MHz and 3.0\,GHz, suggesting small-scale spectral index fluctuations across the halo. \\

\item{} The spatially resolved map also reveals spectral curvature across the halo. The spectral shapes inferred from spatially resolved map show regions with a complex curvature distribution.\\

\item{} The radio brightness of the halo strongly correlates with the X-ray brightness at all observed frequencies. The slopes of the correlation remains remarkably uniform  at 675\,MHz, 1.5\,GHz, and 3\,GHz, namely $0.74\pm0.02$. In addition, the correlation slope across the southern part of the halo is significantly steeper than the northern part of the halo. \\

\item{} The point-to-point comparison between the X-ray surface brightness and spectral index across the halo reveals two different trends. We find a strong anticorrelation in the northern part of the halo and a positive correlation in the southern part of the halo. This suggests that the radio halo indeed consists of two major components, with distinct evolutionary signatures.\\

\item{} The point-to-point comparison between the spectral index and the ICM temperature also shows two different trends. This  further strengthens the fact that the halo has at least two different components. We find a moderate positive correlation between these two quantities in the northern part of the halo.\\

\item{}  The integrated radio spectrum of the main relic, R1, follows a simple power-law between 150\,MHz and 3\,GHz. The spectral index for the main relic R1 is $-1.17\pm0.04$. The radio spectra suggest a shock with Mach number 3.6 which is intriguingly consistent with the X-ray shock Mach number ($\mathcal{M}=3.7\pm0.4$) obtained from the temperature jump at the location of the relic, and is disagreement with previous X-ray estimates of the shock strength based on the density jump. If CRe are accelerated from the thermal pool, a plausible acceleration efficiency below 1\% results in the observed radio power of the relic R1 if a large fraction of the shock front has a strength obtained from the integrated spectrum and the magnetic field strength is of a few $\mu\rm G$ or stronger. \\

\item{}  The other three fainter relics also follow a power-law between 385\,MHz and 3\,GHz. We obtained spectral indices of $\alpha_{\rm int,\, R2}=-1.19\pm0.05$, $\alpha_{\rm int,\,R3}=-1.10\pm0.05$, and $\alpha_{\rm int, \,R4}=-1.14\pm0.04$. For all four relics in the Abell 2744 field, the integrated spectral index values suggest shocks with Mach number in the range $3.3-4.5$. This suggests that the integrated radio spectra of relics may be always dominated by high Mach number shocks as found in simulations \citep{Wittor2019,Paola2021}, irrespective of radio brightness.  Our results also suggests that the properly derived integrated spectral indices are in the  $-1.0$ to $-1.2$ range,  implying that a tail in the Mach distribution determines the integrated spectral index. \\

\item{} Relics R1, R2, and R3 are highly polarized between 1-4\,GHz and show a single RM component. The mean RMs in these three relics are very close to the Galactic foreground. The Faraday dispersion is below $\rm 10 rad\,m^{-2}$, indicating very little Faraday-rotating intervening material along the line of sight. This suggest that R1, R2, and R3 are located in regions of low density ICM, toward the observer. 

\end{enumerate} 

The presence of curved spectra in the halo subregions, the spectral index fluctuations, the steep spectral indices in the outermost regions, a hint of spectral curvature, and the sub-linear $I_{\rm R}-I_{\rm X}$ correlation slope is in line with turbulent re-acceleration models for the origin of radio emitting electrons. The high quality data from the uGMRT and the VLA  suggests that we are entering into a new level of complexity in radio halos. Our findings highlight that the combination of two or more components may result in an overall power-law spectrum in radio halos. 

In conclusion, the emerging complexity in the distribution of nonthermal components in Abell\,2744 echoes the richness of dynamical structures highlighted by optical and X-ray studies \citep[e.g.,][]{Owers2011,Merten2011,Hattori2017,Golovich2019}, hints at a multiple-merger scenario for this system. Different episodes of matter accretions, following the infall along the multiple filaments revealed by X-ray and lensing studies \citep[][]{Eckert2016,Jauzac2018}, are likely to explain also the unusual variety of radio structures in  Abell\,2744, making it a remarkable laboratory to study the interplay between thermal and non-thermal energy components of the ICM. 

%########################################################################
%  Acknowledgements
%######################################################################## 

\begin{acknowledgements}
We thank Nicola Locatelli for helpful discussion. KR, FV, and PDF acknowledge financial support from the ERC Starting Grant "MAGCOW", no. 714196. RJvW acknowledges support from the ERC Starting GrantClusterWeb 804208. CJR, MB, EB, and AB acknowledges financial support from the ERC Starting Grant ``DRANOEL'', number 714245. W.F. acknowledges support from the Smithsonian Institution and the Chandra High Resolution Camera Project through NASA contract NAS8-03060. DW is funded by the Deutsche Forschungsgemeinschaft (DFG, German Research Foundation)- 441694982. This research made use of computer facility at IRA Bologna, Italy, Th\"uringer Landessternwarte, Tautenburg, Germany, and  the HPC resources at the Physical Research Laboratory (PRL), India. 

We thank the staff of the GMRT that made these observations possible. GMRT is run by the National Centre for Radio Astrophysics of the Tata Institute of Fundamental Research. The National Radio Astronomy Observatory is a facility of the National Science Foundation operated under cooperative agreement by Associated Universities. The scientific results reported in this article are based in part on observations made by the \textit{Chandra} X-ray Observatory and published previously in \cite{Pearce2017}. Finally, we wish to acknowledge the developers of the following python packages, 
which were used extensively during this project: \texttt{aplpy} 
\citep{Robitaille2012}, \texttt{astropy} \citep{Astropy2013}, 
\texttt{matplotlib} \citep{Hunter2007}, \texttt{numpy} \citep{Numpy2011} and 
\texttt{scipy} \citep{Jones2001}.
\end{acknowledgements}

\bibliographystyle{aa}

\bibliography{ref.bib}

\end{document}